\documentclass[12pt]{article}
\usepackage{epsfig}
\begin{document}
\newcounter{popnr}
\def\fn#1{{\mathop{{\rm #1}}}}
\def\theequation{\thesection.\arabic{equation}}
\renewcommand{\theequation}{\arabic{section}.\arabic{equation}}
\newcommand{\alpheqn}{\setcounter{popnr}{\value{equation}}
                      \addtocounter{popnr}{1}
                      \setcounter{equation}{0}
   
\renewcommand{\theequation}{\arabic{section}.\arabic{popnr}\alph{e
quation}}}
\newcommand{\reseteqn}{\setcounter{equation}{\value{popnr}}
     \renewcommand{\theequation}
     {\arabic{section}.\arabic{equation}}}
\def\be{\begin{equation}}
\def\ee{\end{equation}}
\def\bq{\begin{equation}}
\def\eq{\end{equation}}
\def\bqa{\begin{eqnarray}}
\def\eqa{\end{eqnarray}}
\def\roughly#1{\mathrel{\raise.3ex
\hbox{$#1$\kern-.75em\lower1ex\hbox{$\sim$}}}}
\def\lsim{\roughly<}
\def\gsim{\roughly>}
\def\llgm{\left\lgroup\matrix}
\def\rrgm{\right\rgroup}
\def\vectrl #1{\buildrel\leftrightarrow \over #1}
\def\partrl{\vectrl{\partial}}
\def\gslash#1{\slash\hspace*{-0.20cm}#1}
%\renewcommand{\theequation}{\arabic{section}.\arabic{equation}}

%\begin{flushright}
~
%\end{flushright}

\renewcommand{\thefootnote}{\fnsymbol{footnote}}

\begin{center}
{\bf The Color Dipole Picture of low-x DIS: 
Model-Independent and Model-Dependent Results}%\footnote{Supported by Deutsche 
%Forschungsgemeinschaft, contract number schi 189/6-2}\footnote{email:
%Dieter.Schildknecht@physik.uni-bielefeld.de}
\end{center}
\vspace {0.5 cm}
\begin{center}
{\bf  Masaaki Kuroda\footnote[1]{Email: kurodam@law.meijigakuin.ac.jp}}\\[2.5mm]
Institute of Physics, Meijigakuin University\\ [1.2mm] 
Yokohama, Japan\\ [3mm] 
{\bf Dieter Schildknecht\footnote[2]{Email: schild@physik.uni-bielefeld.de}} \\[2.5mm]
Fakult\"{a}t f\"{u}r Physik, Universit\"{a}t Bielefeld \\[1.2mm] 
D-33501 Bielefeld, Germany \\[1.2mm]
and \\[1.2mm]
Max-Planck Institute f\"ur Physik (Werner-Heisenberg-Institut),\\[1.2mm]
F\"ohringer Ring 6, D-80805, M\"unchen, Germany
\end{center}

\vspace{1 cm}

 \renewcommand{\thefootnote}{\arabic{footnote}}
    \setcounter{footnote}{0}

\baselineskip 18pt

\begin{center}
{\bf Abstract}
\end{center}

We present a detailed examination of the color-dipole picture (CDP) of low-$x$ deep
inelastic scattering. We discriminate model-independent results, not depending
on a specific parameterization of the dipole cross section, from model-dependent
ones. The model-independent results include the ratio of the longitudinal to
the transverse photoabsorption cross section at large $Q^2$, or equivalently
the ratio of the longitudinal to the unpolarized proton structure function, $F_L
(x,Q^2)=0.27 F_2 (x, Q^2)$, as well as the low-$x$ scaling behavior of the
total photoabsorption cross section $\sigma_{\gamma^*p} (W^2,
Q^2)=\sigma_{\gamma^*p} (\eta (W^2, Q^2))$ as $\log (1 / \eta (W^2, Q^2))$ for
$\eta (W^2, Q^2) <1$, and as $1/\eta (W^2, Q^2)$ for $\eta (W^2, Q^2) \gg
1$. Here, $\eta (W^2, Q^2)$ denotes the low-$x$ scaling variable, $\eta (W^2,
Q^2)=(Q^2 + m^2_0) / \Lambda^2_{sat} (W^2)$ with $\Lambda^2_{sat} (W^2)$ being
the saturation scale. The model-independent analysis also implies     \\
$\lim\limits_{W^2\rightarrow\infty, Q^2 {\rm fixed}} \sigma_{\gamma^*p} (W^2,
Q^2) / \sigma_{\gamma p}
 (W^2) \rightarrow 1$ at any $Q^2$ for asymptotically
large energy, $W$. 
Consistency with pQCD evolution determines the underlying gluon distribution
and the numerical value of $C_2 = 0.29$ in the expression for the saturation
scale, $\Lambda^2 (W^2) \sim (W^2)^{C_2}$. 
In the model-dependent analysis, by restricting the mass of
the actively contributing $q \bar q$ fluctuations by an energy-dependent upper
bound, we extend the validity of the color-dipole picture to $x \cong Q^2 / W^2
\le 0.1$. The theoretical results agree with the world data on DIS for $0.036
{\rm GeV}^2 \le Q^2 \le 316 {\rm GeV}^2$.

\vspace{0.5cm}

\section{Introduction}
In terms of the (virtual) forward-Compton-scattering amplitude, deep inelastic 
scattering (DIS) at low values of the Bjorken scaling variable, $x \cong Q^2 / W^2 \ll 1$,
proceeds via forward scattering of massive (timelike) hadronic fluctuations of the photon, much like 
envisaged by generalized vector dominance \cite{Sakurai,Gribov,FR}
\footnote{Compare also ref. \cite{APP} for a recent review and further references.} 
a long time ago. In QCD, the hadronic fluctuations may be described as 
quark-antiquark
states that interact with the nucleon in a gauge-invariant manner as
color-dipole states \cite{Nikolaev,Cvetic}, coupled to the gluon field in the 
nucleon via 
(at least) two gluons \cite{Low}.
This is the color-dipole picture 
(CDP) of low-x DIS. Compare fig. 1.
\begin{figure}[h!]
\centerline{\epsfig{file=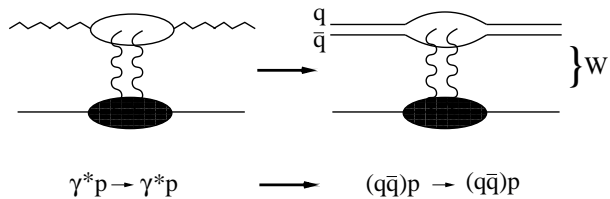,width=8cm} }
\caption{The fluctuation of the photon $\gamma^*$ into a massive $q \bar q$
  color-dipole state and the interaction of the color dipole with the gluon
  field of the nucleon. }
\end{figure}
%Fig.1 

A detailed representation of the experimental results on the photoabsorption cross
section requires an ansatz for the dipole cross section i.e. an ansatz for the
cross section for the scattering of the color-dipole state on the nucleon. Such an ansatz cannot
be formulated entirely free from parameters, just as fit parameters are
required for the related description of the DIS data in terms of the gluon
distribution\footnote{Compare e.g. ref. \cite{DEV}, Chapter 4, and the
bibliography given there.} of the nucleon at low x.

In the first part of the present work, we will show that nevertheless much of
the general features of the DIS experimental data \cite{H1} on the 
photoabsorption cross
section at low x can be derived in the CDP without a detailed
parameter-dependent ansatz for
the dipole-proton interaction 
cross section i.e. model-independently. The general results follow from the very nature of
the $q \bar q$
interaction with the nucleon as the interaction of a color-dipole state. The
model-independent results include the ratio of the longitudinal to the
transverse photoabsorption cross section at low x and 
large $Q^2$ \cite{Ku-Schi}, as well as
the empirically established low-x scaling,the dependence of the photoabsorption
cross section on a single variable $\eta (W^2, Q^2)$ i.e. $\sigma_{\gamma^*p} (W^2, Q^2) =
\sigma_{\gamma^*p} (\eta (W^2, Q^2))$ \cite{DIFF2000}. 
The empirical dependence on $\eta (W^2, Q^2)$, as
$1/\eta (W^2, Q^2)$ for $\eta (W^2,Q^2) \gg 1$, and as $\ln (1/\eta (W^2, Q^2))$ for
$\eta (W^2, Q^2) \ll 1$, is a general feature of the dipole interaction. Here, $\eta (W^2, Q^2)$
denotes the scaling variable, $\eta (W^2, Q^2) \equiv (Q^2 +
m^2_0)/\Lambda^2_{sat} (W^2)$ with $m^2_0 \simeq 0.15 GeV^2$, and 
$\Lambda^2_{sat} (W^2)$ denotes the appropriately defined ``saturation scale''
which rises as a small fixed power, $C_2$ of the square of the $\gamma^* p$ 
center-of-mass energy, $\Lambda^2_{\rm sat} (W^2) \sim (W^2)^{C_2}$.

A detailed model for the dipole cross section will be analyzed and compared
with the world experimental data in 
Sections 3 to 5 of the present paper, and conclusions will be presented in
Section 6.

\section{The CDP: Model-independent Results}

A model-independent prediction of the longitudinal-to-transverse ratio of
the photoabsorption cross section was recently presented \cite{Ku-Schi}. 
Based on the general analysis of the transverse and the longitudinal
photoabsorption cross section in Sections 2.1 and 2.2, we will present a more
detailed account of the underlying argument in Section 2.3. 
After a general discussion on the CDP in
Section 2.4, we will deal with low-x scaling in Section 2.5
and derive the functional dependence of the photoabsorption cross section
on the scaling variable $\eta (W^2, Q^2)$. In Section 2.6, we analyze 
the photoabsorption cross section in the limit of $W^2 \to \infty$ at
fixed values of $Q^2 > 0$. The $\eta (W^2, Q^2)$ dependence implies that
the photoabsorption cross section for $W^2\rightarrow \infty$ at fixed $Q^2>0$
converges towards 
a $Q^2$-independent limit that coincides with $(Q^2 = 0)$ photoproduction.
In Section 2.7, we will show that the consistency of the CDP with DGLAP
evolution \cite{Lipatov} for the sea quark distribution function constrains the energy
dependence of the saturation scale,
$\Lambda^2_{\rm sat} (W^2)$, and of the structure function $F_2 (x
\cong Q^2 / W^2, Q^2)$ for $x < 0.1$.
We will also elaborate on the connection between the CDP and the extraction 
of the gluon distribution of the proton. We compare the gluon distribution
underlying the CDP with the gluon distributions that were extracted from the experimental
data by directly employing the pQCD improved parton picture in the analysis of
the experimental data.

\subsection{The longitudinal and the transverse photoabsorption cross
section at large \boldmath $Q^2$, \unboldmath part I.}

The transverse-position-space representation \cite{Bj}  of the longitudinal and
the transverse photoabsorption cross section \cite{Nikolaev,Cvetic}
\be
\sigma_{\gamma^*_{L,T}} (W^2, Q^2) = \int dz \int d^2 \vec r_\bot
\vert \psi_{L,T} (\vec r_\bot, z (1 - z), Q^2) \vert^2~~\sigma_{(q\bar q)p}
(\vec r_\bot, z (1 - z), W^2) 
\label{2.1}
\ee
summarizes in compact form the structure of the $x \approx Q^2/W^2 \le 0.1$
interaction of a $q \bar q$ pair, originating from a $\gamma^*_{L,T} \to
q \bar q$ transition, with the gluon field of the nucleon. The square of the
``photon wave function'' $\vert \psi_{L,T} (\vec r_\bot, z(1-z), Q^2) \vert^2$
describes the probability for the occurrence of a $q \bar q$ fluctuation of transverse
size, $\vec r_\bot$, of a longitudinally, $\gamma^*_L$, or 
a transversely polarized photon,$\gamma^*_T$, of virtuality $Q^2$.
The variable $z$, with $0 \le z \le 1$, characterizes the
distribution of the momenta between quark and antiquark. In the rest frame of
a $q \bar q$ fluctuation of mass $M_{q \bar q}$, the variable $z$ 
determines \cite{Cvetic}
the direction of the three-momentum of the quark with respect to the photon
direction. The dipole cross section, related to the imaginary part of the
$(q \bar q)p$ forward scattering amplitude, is denoted by $\sigma_{(q \bar q)p}
(\vec r_\bot, z (1-z), W^2)$. For generality, we include a potential dependence
on the ``$q \bar q$-configuration variable'' $z(1-z)$. The 
dipole cross
section depends on the center-of-mass energy, $W$, of the $(q \bar q)p$ scattering 
process \cite{Cvetic, DIFF2000, Forshaw, Ewerz},
since it is a timelike massive $q \bar q$ pair, the photon dissociates or
fluctuates into \footnote{In this respect, we differ from 
ref. \cite{Nikolaev}, where the dipole cross section is assumed to depend on
$x \cong Q^2/W^2$. Compare also the discussion on this point in 
Section 2.4.}. 
The interaction of a massive $q \bar q$ pair with the proton (the integration
over $d^2\vec r_\bot$ corresponding to an integration over fluctuation masses)
depends 
on $W$ and, in particular, is independent of the photon-virtuality,
$Q^2$. This point is inherently connected with the mass-dispersion
relation \cite{Sakurai, Gribov} of generalized vector dominance, 
and it was recently 
elaborated upon
from first principles of quantum field theory in ref. \cite{Ewerz}.

The gauge invariance for the interaction of the $q \bar q$ color
dipole with the color field in the nucleon requires a representation of the 
dipole cross section of the form \cite{Nikolaev, Cvetic}
\be
\sigma_{(q \bar q)p} (\vec r_\bot, z(1-z), W^2) = \int d^2 \vec l_\bot \tilde{\sigma}
(\vec l^{~2}_\bot, z(1-z), W^2) \left(1-e^{-i~ \vec l_\bot \cdot \vec r_\bot}\right),
\label{2.2}
\ee
where the transverse momentum of the gluon absorbed by the dipole state is
denoted by $\vec l_\bot$. In the important limit of a small-size dipole,
$\vec r^{~2}_\bot \to 0$, from (\ref{2.2}) we have
\be
\sigma_{(q \bar q)p} (\vec r_\bot, z (1-z), W^2) = \frac{\pi}{4} 
\vec r^{~2}_\bot \int d \vec l^{~2}_\bot \vec l^{~2}_\bot 
\tilde{\sigma} (\vec l^{~2}_\bot, z (1-z), W^2).
\label{2.3}
\ee
A dipole of vanishing transverse size must obviously 
have a vanishing cross section 
(``color-transparency'') as in (\ref{2.3}), when interacting with the gluon field. The
validity of the approximation (\ref{2.3}) requires
\be
\vec r^{~2}_\bot \vec l^{~2}_\bot < \vec r^{~2}_\bot \vec l^{~2}_{\bot~Max}
(W^2) < 1,
\label{2.4}
\ee
where $\vec l^{~2}_{\bot Max} (W^2)$ characterizes the W-dependent domain of 
$\vec l^{~2}_\bot < \vec l^{~2}_{\bot~ Max} (W^2)$ in which 
$\tilde{\sigma} (\vec l^{~2}_\bot,z(1-z),W^2)$, at 
a given energy $W$, by assumption is appreciably different from zero. 
For the subsequent
discussion, it will be useful to introduce the variables $\vec r^{~\prime}_\bot
= \sqrt{z (1-z)} \vec r_\bot$ and $\vec l^{~\prime}_\bot = 
\vec l_\bot/\sqrt{z(1-z)}$ \cite{MKDS}. In terms of these variables the restriction
(\ref{2.4}) becomes
\be
\vec r^{~\prime 2}_\bot \vec l^{~\prime 2}_\bot < \vec r^{~\prime 2}_\bot
\vec l^{~\prime 2}_{\bot~Max} (W^2) < 1.
\label{2.5}
\ee

The validity of (\ref{2.3}) to (\ref{2.5}) is an 
integral part of the 
CDP. 
The absorption of a gluon of transverse momentum squared
$\vec l_\bot^{~2} < \vec l_{\bot~Max}^{~2} (W^2)$ by a $q \bar q$
fluctuation (unless the absorbed gluon is re-emitted by the absorbing quark)
 increases the mass of the $q \bar q$ fluctuation. At any given 
squared energy, $W^2$, the contributing $q \bar q$ masses, and consequently
the values of $\vec l_\bot^{~\prime 2}$ actively contributing to the cross
section, must be bounded by an upper
limit, since only fluctuations of sufficiently long 
lifetime\footnote{The well-known expression for the lifetime of a 
hadronic fluctuation is given in (2.60) below.} do contribute 
to the Compton forward-scattering amplitude of the CDP.

Color transparency (\ref{2.3}) determines the photoabsorption cross section
(\ref{2.1}) for sufficiently large $Q^2$. This will be elaborated upon next.

We will
consider massless quarks. Inserting the explicit representation of the 
photon wave function in (\ref{2.1}), we find the well-known expression 
$(Q \equiv \sqrt{Q^2})$ \cite{Nikolaev}
\bqa
& & \sigma_{\gamma^*_{L,T}p} (W^2, Q^2) = \frac{3\alpha}{2 \pi^2} \sum_q
Q^2_q Q^2 \cdot \label{2.6} \\
& & \hspace*{-0.5cm} \cdot \left\{ \matrix{ 
4 \int d^2 \vec r_\bot \int dz z^2 (1-z)^2 \cdot K^2_0 (r_\bot \sqrt{z(1-z)} Q)
\sigma_{(q \bar q)p} (r_\bot, z(1-z),W^2),\cr
\hspace{-6cm}    \int d^2 \vec r_\bot \int dz (1-2z (1-z)) z (1-z) \cr
\hspace{4cm}     \cdot K^2_1 (r_\bot
\sqrt{z (1-z)}Q) \sigma_{(q \bar q)p} (r_\bot, z(1-z), W^2).}\right.
\nonumber
\eqa
Here, $r_\bot \equiv | \vec r_\bot |$, and $K_{0,1} (r_\bot \sqrt{z(1-z)} Q)$ 
denotes modified Bessel functions.

A compact and direct way of deriving the large-$Q^2$ behavior of the
cross sections in (\ref{2.6}) makes use of the strong fall-off of the modified
Bessel functions at large values of their argument,
\be
K^2_{0,1} (y) \sim \frac{\pi}{2y} e^{-2y},~~~~~(y \gg 1).
\label{2.7}
\ee
The integral over $\int d^2 \vec r_\bot = 
\pi \int d \vec r^{~2}_\bot$ in (\ref{2.6}) is
accordingly dominated by
\be
r^\prime_\bot Q \equiv r_\bot \sqrt{z(1-z)} Q < 1.
\label{2.8}
\ee
As soon as $\vec r^{~\prime 2}_\bot > 1/Q^2$, the integrand in (\ref{2.6})
yields negligible contributions. 
The interval for $r^\prime_\bot$ defined by the condition (\ref{2.8}) is
contained in the interval (\ref{2.5}), where color transparency is valid, provided
$Q^2$ is sufficiently large, such that
\be
\vec r^{~\prime 2}_\bot < \frac{1}{Q^2} < 
\frac{1}{\vec l^{~\prime 2}_{\bot~Max} (W^2)},
\label{2.9}
\ee
or
\be
Q^2 > \vec l^{~\prime 2}_{\bot~Max} (W^2).
\label{2.10}
\ee
Under this constraint, the photoabsorption cross section (\ref{2.6}) can be
evaluated by inserting the $\vec r^{~2}_\bot \to 0$ expression (\ref{2.3}). One
obtains
\bqa
& & \sigma_{\gamma^*_{L,T}p} (W^2, Q^2) = \frac{3\alpha}{2} \sum_q
Q^2_q Q^2 \cdot \label{2.11} \\
& & \hspace{-0.5cm}   \cdot \left\{ \matrix{ 
\int d \vec r^{~2}_\bot \vec r^{~2}_\bot 
\int dz z^2 (1-z)^2  \cdot K^2_0 (r_\bot \sqrt{z(1-z)} Q)
\int d \vec l^{~2}_\bot \vec l^{~2}_\bot \tilde{\sigma} (\vec l^{~2}_\bot, 
z(1-z),W^2),\cr
\hspace{-6cm}  \frac{1}{4}\int d \vec r^{~2}_\bot  \vec r^{~2}_\bot \int dz (1-2z (1-z))
z (1-z) \cdot \cr 
\hspace{4.2cm}    \cdot K^2_1 (r_\bot
\sqrt{z (1-z)}Q) \int d \vec l^{~2}_\bot \vec l_\bot^{~2} \tilde{\sigma}
(\vec l^{~2}_\bot, z(1-z), W^2).}\right.
\nonumber
\eqa
In terms of the variable $\vec r^{~\prime}_\bot$ from (\ref{2.8}) the photoabsorption cross
section (\ref{2.11}) is given by
\bqa
& & \sigma_{\gamma^*_{L,T}p} (W^2, Q^2) = \frac{3\alpha}{2} \sum_q
Q^2_q Q^2 \cdot \nonumber\\
& & \hspace*{-1cm} \cdot \left\{ \matrix{ 
\int dz \int d \vec r^{~\prime 2}_\bot  \vec r^{~\prime 2}_\bot 
K^2_0 (r^\prime_\bot Q) \int d \vec l^{~2}_\bot \vec l^{~2}_\bot 
\tilde{\sigma} (\vec l^{~2}_\bot, 
z(1-z),W^2),\cr
\frac{1}{4}\int dz \frac{1-2z(1-z)}{z(1-z)} \int d \vec r^{~\prime 2}_\bot 
\vec r^{~\prime 2}_\bot K^2_1 (r^\prime_\bot Q) 
\int d \vec l^{~2}_\bot \vec l^{~2}_\bot \tilde{\sigma}
(\vec l^{~2}_\bot, z(1-z), W^2).}\right.
\label{2.12}
\eqa
Making use of the mathematical identities \cite{GR},
\bqa
\int^\infty_0 dy y^3 K^2_0 (y) & = \frac{1}{3},\nonumber \\
\int^\infty_0 dy y^3 K^2_1 (y) & = \frac{2}{3},
\label{2.13}
\eqa
the photoabsorption cross section (\ref{2.12}), valid for
$Q^2  > \vec l^{~\prime 2}_{\bot Max} (W^2)$ from (\ref{2.10}) 
(and $x \cong Q^2/W^2 \ll 1$), reduces to the simple form
\be
\sigma_{\gamma^*_{L,T}p} (W^2, Q^2) = \alpha \sum_q Q^2_q \frac{1}{Q^2}
\left\{ \matrix{
\int dz \int d \vec l^{~2}_\bot \vec l ^{~2}_\bot \tilde{\sigma}
(\vec l^{~2}_\bot, z (1-z), W^2),\cr
2 \int dz \frac{1}{4} \frac{1-2z (1-z)}{z(1-z)} \int d \vec l^{~2}_\bot 
\vec l^{~2}_\bot \tilde{\sigma} (\vec l^{~2}_\bot, z(1-z),W^2). }\right.
\label{2.14}
\ee

According to our derivation, the large-$Q^2$ result (\ref{2.14}) 
is a consequence of the
transverse-position-space representation (\ref{2.1}) combined with color 
transparency (\ref{2.3}) that in turn rests on decent behavior of $\tilde \sigma (\vec l^{~2}_\bot,
z (1-z), W^2)$ as characterized by $\vec l^{~\prime 2}_{\bot~Max} (W^2)$.

For the ensuing discussion, it will be useful to represent the contribution of
the dipole cross section to the transverse cross section in (\ref{2.14}) in
terms of the contribution to the longitudinal one by introducing the factor
$\rho_W$, 
\bqa
& & \int dz \frac{1}{4} \frac{1-2z (1-z)}{z(1-z)} \int d \vec l_\bot^{~2} \vec
l_\bot^{~2} \tilde\sigma (\vec l_\bot^{~2} , z (1-z), W^2) \nonumber \\
& & = \rho_W \int dz \int d \vec l_\bot^{~2} \vec l_\bot^{~2} \tilde\sigma
(\vec l_\bot^{~2}, z (1-z), W^2). 
\label{2.14a}
\eqa
The cross section (\ref{2.14}) then becomes, 
\be
\sigma_{\gamma^*_{L,T}} (W^2, Q^2) = \alpha \sum Q^2_q \frac{1}{Q^2} \int dz
\int d \vec l_\bot^{~2} \vec l_\bot^{~2} \tilde\sigma (\vec l_\bot^{~2}, z
(1-z), W^2) \left\{ \matrix{ 1 , \cr 2 \rho_W , } \right.
\label{2.14b}
\ee
and the longitudinal-to-transverse ratio, $R (W^2, Q^2)$, at large $Q^2$ is
given by
\be
R (W^2, Q^2) \equiv \frac{\sigma_{\gamma^*_L p} (W^2, Q^2)}{\sigma_{\gamma^*_T
    p} (W^2, Q^2)} = \frac{1}{2\rho_W} .
\label{2.14c}
\ee
In (\ref{2.14a}) to (\ref{2.14c}), the index $W$ indicates a potential
dependence of $\rho_W$ on the energy $W$. Actually, we will find that $\rho_W$
is a $W$-independent constant, see Section 2.3.
The factor $1/2$ in (\ref{2.14c}) is due to the enhanced probability for
transverse photons to fluctuate into $q \bar q$ pairs relative to longitudinal
photons, compare (\ref{2.13}). 
The additional factor of $1 / \rho_W$ is associated with different
interactions of $q \bar q$ fluctuations originating from transverse,
$\gamma^*_T \rightarrow q \bar q$, and longitudinal, $\gamma^*_L \rightarrow q
\bar q$, photons, respectively.

By comparing the representation of the cross section in (\ref{2.14b}) with the one
in (\ref{2.11}), taking into account the $\vec r_\bot^{~2} \rightarrow 0$ form
of the dipole cross section in (\ref{2.3}), we obtain a substitution rule that connects
the longitudinal with the transverse photoabsorption cross section. Indeed,
substituting the replacement (using (\ref{2.3}))
\be
\sigma_{(q \bar q)p} (\vec r_\bot^{~2} , z(1-z), W^2) \rightarrow \sigma_{(q
  \bar q)p} ( \rho_W \vec r_\bot^{~2}, z(1-z), W^2)
\label{2.14d}
\ee
into the longitudinal cross section in (\ref{2.11}) in conjunction with 
\be
K^2_0 (r_\bot \sqrt{z(1-z)} Q) \rightarrow K^2_1 (r_\bot \sqrt{z(1-z)} Q)
\label{2.14e}
\ee
reproduces (\ref{2.14b}), which relates the transverse photoabsorption cross section to
the longitudinal one, 
\be
\sigma_{\gamma^*_L p} (W^2, Q^2) \rightarrow \sigma_{\gamma^*_T p} (W^2, Q^2).
\label{2.14f}
\ee
We thus have arrived at the conclusion that $q \bar q$ states originating from
transversely polarized photons, $\gamma^*_T \rightarrow q \bar q$, interact
with enhanced transverse size, 
\be
\vec r_\bot^{~2} \rightarrow \rho_W \vec r_\bot^{~2} , 
\label{2.14g}
\ee
relative to $q \bar q$ states stemming from $\gamma^*_L \rightarrow q \bar q$
transitions. 
Based on the interpretation of
$\rho_W$ in (\ref{2.14g}), in Section 2.3, we will show that the absolute
magnitude of $\rho_W$ is uniquely determined as $\rho_W = 4/3$.

It is frequently assumed that the dipole cross section in (\ref{2.1}) and
(\ref{2.2}),
i.e. $\tilde{\sigma} (\vec l^{~2}_\bot, z (1-z), W^2)$, does not depend on 
the configuration of the $q \bar q$ state, $z(1-z)$. According
to (\ref{2.14}), strict independence of $\tilde{\sigma} (\vec l^{~2}_\bot,
z (1-z), W^2)$ from $z(1-z)$ implies a logarithmic divergence in the
transverse photoabsorption cross section. The divergence is avoided by a
restriction on $0 \le z (1-z) < \frac{1}{4}$ given by
\be
z (1-z) > \epsilon .
\label{2.15}
\ee
This restriction corresponds to adopting an ansatz for 
$\tilde{\sigma} (\vec l^{~2}_\bot, z (1-z), W^2)$ of the form
\be
\tilde{\sigma}(\vec l^{~2}_\bot, z (1-z), W^2) \to \tilde{\sigma}
(\vec l^{~2}_\bot, W^2) \theta (z (1-z) - \epsilon),
\label{2.16}
\ee
as a ``minimal'' dependence\footnote{The factorization of the 
$z (1-z)$ dependence in (\ref{2.16}) strictly speaking amounts to
an assumption that does not necessarily follow from (\ref{2.14}).
Finiteness of (\ref{2.14}) can also be achieved by an appropriate 
correlation of the $z(1-z)$ and $\vec l_\bot$ dependences not of the form
(\ref{2.16}). Compare e.g. the specific model (\ref{3.3}) below.The ansatz
(\ref{2.16}) is explicitly realized by (\ref{3.17})}
of the dipole cross section on $z (1-z)$.

Taking into account the restriction (\ref{2.15}), the photoabsorption cross
section (\ref{2.14}) becomes\footnote{Here, with $\epsilon = {\rm const.}$, we
  exclude the more general case of $\epsilon = \epsilon (\vec l_\bot^{~2})$.}
\be
\sigma_{\gamma^*_{L,T}p} (W^2, Q^2) = \alpha \sum_q Q^2_q \frac{1}{Q^2}
\left\{ \matrix{
\int_{z(1-z) > \epsilon} dz \int d \vec l^{~2}_\bot \vec l ^{~2}_\bot 
\tilde{\sigma}
(\vec l^{~2}_\bot, W^2),\cr
2 \int_{z(1-z) > \epsilon} dz \frac{1}{4} \frac{1-2z (1-z)}{z(1-z)} 
\int d \vec l^{~2}_\bot 
\vec l^{~2}_\bot \tilde{\sigma} (\vec l^{~2}_\bot, W^2). }\right.
\label{2.17}
\ee
It may be rewritten as
\be
\sigma_{\gamma^*_{L,T}p} (W^2, Q^2) = \alpha \sum_q Q^2_q \frac{1}{Q^2}
\sqrt{1 - 4 \epsilon} \int d \vec l^{~2}_\bot \vec l^{~2}_\bot 
\tilde{\sigma} (\vec l^{~2}_\bot, W^2) \left\{ \matrix{
1, \cr
2 \rho (\epsilon), } \right.
\label{2.18}
\ee
i.e. $\rho_W$ in (\ref{2.14a}) becomes
\be
\rho_W = 
\rho (\epsilon) = \frac{\int_{z (1-z) > \epsilon} dz 
\frac{1 - 2z (1-z)}{z (1-z)}}{4 \int_{z (1-z) > \epsilon} dz} =
\frac{1}{4 \sqrt{1 - 4 \epsilon}} \int_{z(1-z) > \epsilon} dz 
\frac{1 - 2 z (1-z)}{z(1-z)}.
\label{2.19}
\ee
Explicitly, one finds
\be
\rho (\epsilon) = \frac{1}{2\sqrt{1 - 4 \epsilon}} 
\left( ln \frac{(1 + \sqrt{1 - 4 \epsilon})^2}{4 \epsilon} - 
\sqrt{1 - 4 \epsilon} \right) \simeq \frac{1}{2} \ln \frac{1}{\epsilon}.
\label{2.20}
\ee
We note that in Section 3 we will introduce the parameter $a$, related to
$\epsilon$ by $\epsilon = 1/6a$. The ratio $R$ of the longitudinal to the
transverse photoabsorption cross section from (\ref{2.14c}) according to 
(\ref{2.18}) is given by $1/2 \rho (\epsilon)$,
\be
R \equiv \frac{\sigma_{\gamma^*_L p} (W^2, Q^2)}{\sigma_{\gamma^*_T p}
(W^2, Q^2)} = \frac{1}{2 \rho (\epsilon)} .
\label{2.21}
\ee
The ratio $R$ in (\ref{2.21}) is independent of a particular parameterization of 
the $\vec l^{~2}_\bot$ dependence of the  
dipole cross section, that is for $\tilde{\sigma} (\vec l^{~2}_\bot, W^2)$
in (\ref{2.16}). 

With respect to subsequent discussions in Sections 2.2 and 2.3, we note the
origin of the $z(1-z)$-dependent factors in (\ref{2.14}) and (\ref{2.17}) from
the coupling of the $q \bar q$ states to the electromagnetic current. The
electromagnetic current determining the $\gamma^* (q \bar q)$ coupling of a
timelike photon of mass squared $M^2_{q \bar q} = \vec k_\bot^{~2} / z(1-z)$ is
given by \cite{Cvetic}
\be
\sum_{\lambda = - \lambda^\prime = \pm 1} \vert j_L^{\lambda, \lambda^\prime}
\vert^2 = 8 M^2_{q \bar q} z (1-z) = 8 \vec k^{~2}_\bot,
\label{2.22}
\ee
and
\bqa
\sum_{\lambda = - \lambda^\prime = \pm 1} \vert j_T^{\lambda, \lambda^\prime}
(+) \vert^2 & = & \sum_{\lambda = - \lambda^\prime = \pm 1} 
\vert j_T^{\lambda, \lambda^\prime} (-) \vert^2 = 2 M^2_{q \bar q}
(1-2z (1-z))= \nonumber \\ 
& = & 2 \vec k^2_\bot \frac{(1-2z (1-z))}{z (1-z)},
\label{2.23}
\eqa
for a longitudinal photon, $\gamma^*_L$, and a transverse one, $\gamma^*_T$,
respectively. Comparison of (\ref{2.22}) and (\ref{2.23}) with (\ref{2.14}) 
and (\ref{2.14a}) reveals that the size enhancement $\rho_W$
%(\ref{2.17}) 
is related to the difference of the longitudinal and transverse photon
couplings of dipole states carrying the transverse momentum 
$\vec l_\bot$ of the absorbed gluon. At large $Q^2$, the interaction of the 
photon according to (\ref{2.14}) reduces to interactions of 
fluctuations into $q \bar q$ dipole states carrying a quark transverse momentum
identical to the transverse momentum of  
the absorbed gluon, $\vec l_\bot$. 

According to (\ref{2.22}) and (\ref{2.23}), the normalized $z (1-z)$ distributions
$f_{L,T} (z(1-z))$ of a $q \bar q$ pair of fixed mass $M_{q \bar q}$
originating from a longitudinally and a transversely polarized photon are
given by \cite{Ku-Schi}
\be
f_L (z (1-z)) = 6 z(1-z),
\label{2.24}
\ee
and
\be
f_T (z (1-z)) = \frac{3}{2} (1 - 2 z (1-z)),
\label{2.25}
\ee
respectively. 

We end the present Section by stressing the simplicity of the
physical picture underlying the photoabsorption in DIS at low $x$ and 
sufficiently large $Q^2$. The photon fluctuates into a $q \bar q$ dipole
state. The $\gamma^*_{L,T} (q \bar q)$
transition strength is determined by the electromagnetic current
in (\ref{2.22}) and (\ref{2.23}). 
The $q \bar q$ dipole state entering (\ref{2.14}) and (\ref{2.17}) 
carries a quark (antiquark) 
transverse
momentum equal to the transverse momentum of the absorbed gluon, $\vec l_\bot$.
Summation over all fluctuations, the weight function $\tilde{\sigma} (\vec
l^{~2}_\bot, W^2)$ being characteristic for the 
transverse momentum distribution of the gluons in the nucleon, upon multiplication
by $1/Q^2$, determines the photoabsorption cross section. The representations,
(\ref{2.14}) and (\ref{2.17}), accordingly, explicitly 
demonstrate that the $q \bar q$ fluctuations
directly test the gluon distribution in the nucleon that is characterized by
$\tilde \sigma  (\vec l^{~2}_\bot, W^2)$.
The enhanced transverse photoabsorption cross section, due to $2\rho_W$ in
(\ref{2.14b}) and to $2\rho(\epsilon)$ in (\ref{2.18}),  
results from the enhanced 
transition of transverse photons into $q \bar q$ pairs, compare (\ref{2.13}) and (\ref{2.14}),
in conjunction with a $(q \bar q)p$ interaction of the $q \bar q$ pairs from
transverse photons with enhanced transverse size,  
compare (\ref{2.14d}) and (\ref{2.14g}).

\subsection{The photoabsorption cross section at large 
\boldmath $Q^2$,\unboldmath part II,
\boldmath $(q \bar q)^{J=1}_{L,T}$\unboldmath states.}

In this Section, we will represent the photoabsorption cross section in 
terms of scattering cross sections for dipole states $(q \bar q)^{J=1}_{L,T}$
with definite spin $J = 1$, and longitudinal as well as transverse
polarization, $L$ and $T$, respectively.

Upon introducing $\vec r^{~\prime}_\bot = \vec r_\bot \sqrt{z (1-z)}$
from (\ref{2.8}), the photoabsorption cross section (\ref{2.6}) becomes
\cite{MKDS}
\bqa
& & \sigma_{\gamma^*_{L,T}p} (W^2, Q^2) = \frac{3\alpha}{2 \pi^2} \sum_q
Q^2_q Q^2 \cdot 2 \cdot \label{2.29} \\
& & \hspace*{-1cm} \cdot  \left\{ \matrix{ 
\hspace{-1cm}  \int d^2 r^\prime_\bot K^2_0 (r^\prime_\bot Q) \int dz 2z (1-z) 
\sigma_{(q \bar q)p} \left( \frac{r^\prime_\bot}{\sqrt{z (1-z)}}, z (1-z), W^2
\right),\cr
\int d^2 r^\prime_\bot K^2_1 (r^\prime_\bot Q) \int dz \frac{1}{2} 
(z^2 + (1-z)^2) \sigma_{(q \bar q)p}  \left( 
\frac{r^\prime_\bot}{\sqrt{z (1-z)}}, z(1-z), W^2\right).} \right.
\nonumber
\eqa
The cross section in (\ref{2.29}) is written in such a manner that the 
appearance of the rotation functions, $d^1_{j j^\prime} (z)$, is 
explicitly displayed, i.e.
\bqa
& & \sigma_{\gamma^*_{L,T}p} (W^2, Q^2) = \frac{3\alpha}{2 \pi^2} \sum_q
Q^2_q Q^2 \cdot 2 \cdot \label{2.30}  \\
& & \hspace{-0.5cm}    \cdot \left\{ \matrix{ 
\int d^2 r^\prime_\bot K^2_0 (r^\prime_\bot Q) \int dz \left( d^1_{10} (z)
\right)^2    
\cdot \sigma_{(q \bar q)p} \left( \frac{r^\prime_\bot}{\sqrt{z (1-z)}}, z (1-z), W^2
\right),\cr
\hspace{-3cm}    \int d^2 r^\prime_\bot K^2_1 (r^\prime_\bot Q) \int dz \frac{1}{2} 
\left( (d^1_{1-1} (z))^2 + (d^1_{11} (z))^2 \right) \cdot   \cr 
\hspace{5cm}     \cdot \sigma_{(q \bar q)p}  \left( 
\frac{r^\prime_\bot}{\sqrt{z (1-z)}}, z(1-z), W^2\right).} \right.
\nonumber
\eqa
The rotation funtions originate from the $\gamma^*(q \bar q)$ couplings
via the electromagnetic currents in (\ref{2.22}) and (\ref{2.23}), rewritten as
\be
\sum_{\lambda = - \lambda = \pm 1} \vert j^{\lambda, \lambda^\prime}_L
\vert^2 = 4 M^2_{q \bar q} \left( d^1_{10} (z)\right)^2,
\label{2.31}
\ee
and
\be
\sum_{\lambda = - \lambda^\prime = \pm 1} \vert j^{\lambda, \lambda^\prime}_T
(+) \vert^2 = \sum_{\lambda = -\lambda = \pm 1} \vert 
j^{\lambda, \lambda^\prime}_T (-) \vert^2 = 4 M^2_{q \bar q} \frac{1}{2}
\left( (d^1_{1-1} (z))^2 + (d^1_{11} (z))^2 \right).
\label{2.32}
\ee
Integration over $dz$ in (\ref{2.31}) and (\ref{2.32}) defines the total longitudinal 
and transverse transition strengths for the $\gamma^*_L (q \bar q)$ and
$\gamma^*_T (q \bar q)$ transitions. 
Requiring factorization of these transition strengths in (\ref{2.30}), we represent
$\sigma_{\gamma^*_{L,T} p} (W^2, Q^2)$ in terms of the 
so-defined cross sections for scattering of
$(q \bar q)^{J=1}_{L,T}$ states on the proton, 
$\sigma_{(q \bar q)^{J=1}_{L,T}p}
(r^\prime_\bot, W^2)$,
\bqa
& & \sigma_{\gamma^*_{L,T}p} (W^2, Q^2) = \frac{3\alpha}{2 \pi^2} \sum_q
Q^2_q Q^2 \cdot \nonumber\\
& & \hspace*{-1cm} \cdot 2 \left\{ \matrix{ 
\int d^2 r^\prime_\bot K^2_0 (r^\prime_\bot Q) \int dz \left( d^1_{10} (z)
\right)^2  
\sigma_{(q \bar q)^{J=1}_L p} (r^\prime_\bot, W^2),\cr
\int d^2 r^\prime_\bot K^2_1 (r^\prime_\bot Q) \int dz \frac{1}{2} 
\left( (d^1_{1-1} (z))^2 + (d^1_{11} (z))^2 \right) 
\sigma_{(q \bar q)^{J=1}_T p}  
(r^\prime_\bot, W^2).} \right.
\label{2.33}
\eqa
Upon inserting the normalizations
\be
\int dz (d^1_{10} (z))^2 = \int dz (d^1_{1-1} (z))^2 = \int dz (d^1_{11}
(z))^2 = \frac{1}{3},
\label{2.34}
\ee
(\ref{2.33}) becomes
\be
\sigma_{\gamma^*_{L,T}p} (W^2, Q^2) = \frac{\alpha}{\pi} \sum_q Q^2_q Q^2
\int dr^{\prime 2}_\bot K^2_{0,1} (r^\prime_\bot Q) 
\sigma_{(q \bar q)^{J=1}_{L,T} p} (r^\prime_\bot, W^2).
\label{2.35}
\ee
By comparing (\ref{2.35}) with (\ref{2.30}), we find that the $J=1$ dipole cross-sections
introduced in (\ref{2.33}) are explicitly given by
\bqa
 & & \hspace{-0.5cm}  \sigma_{(q \bar q)^{J=1}_{L,T} p} (r^\prime_\bot, W^2) = 
\label{2.36} \\ 
& = & 3 \cdot  \left\{ \matrix{
\int dz  (d^1_{10} (z))^2 \sigma_{(q \bar q) p} \left( 
\frac{r^\prime_\bot}{\sqrt{z(1-z)}}, z (1-z), W^2 \right),\cr
\int dz \frac{1}{2} \left( (d^1_{1-1} (z))^2 + (d^1_{11} (z))^2 \right)
\sigma_{(q \bar q)} \left( \frac{r^\prime_\bot}{\sqrt{z(1-z)}},
z (1-z), W^2 \right).} \right. 
\nonumber
\eqa

We add the comment at this point that the $(q \bar q)^{J=1}_{L,T}p$ cross
sections in (\ref{2.33}) to (\ref{2.36}) may be identified 
as the $J=1$ parts of the
partial-wave expansions
\bqa
&& d^1_{10} (z) \sigma_{(q \bar q)p} \left( 
\frac{r^\prime_\bot}{\sqrt{z (1-z)}}, z (1-z), W^2 \right) = \nonumber \\
&& = d^1_{10} (z) \sigma_{(q \bar q)^{J=1}_L p} (r^\prime_\bot, W^2)
+ d^2_{10} (z) \sigma_{(q \bar q)^{J=1}_L} (r^\prime_\bot, W^2) + ...
\label{2.37}
\eqa
and 
\bqa
&& d^1_{1-1} (z) \sigma_{(q \bar q)p} \left( 
\frac{r^\prime_\bot}{\sqrt{z (1-z)}}, z (1-z), W^2 \right) = \nonumber \\
&& = d^1_{1-1} (z) \sigma_{(q \bar q)^{J=1}_{-1} p} (r^\prime_\bot, W^2)
+ d^2_{1-1} (z) \sigma_{(q \bar q)^{J=2}_{-1}} (r^\prime_\bot, W^2) + ...
\label{2.38}
\eqa
as well as
\bqa
&& d^1_{11} (z) \sigma_{(q \bar q)p} \left( 
\frac{r^\prime_\bot}{\sqrt{z (1-z)}}, z (1-z), W^2 \right) = \nonumber \\
&& = d^1_{11} (z) \sigma_{(q \bar q)^{J=1}_{+1} p} (r^\prime_\bot, W^2)
+ d^2_{11} (z) \sigma_{(q \bar q)^{J=2}_{+1}} (r^\prime_\bot, W^2) + ...
\label{2.39}
\eqa
These partial wave expansions explicitly demonstrate that the cross
section (\ref{2.36}) introduced by the factorization requirement in 
(\ref{2.33})
and (\ref{2.35}) stand for the cross sections for the scattering of
$(q \bar q)^{J=1}_{L,T}$ states on the proton. 

DIS at low $x \le Q^2/W^2 \ll 1$ and suffiently large $Q^2$ is recognized
as elastic diffractive forward scattering of $(q \bar q)^{J=1}_{L,T}$ fluctuations
of the photon on the proton, compare (\ref{2.35}). 

We return to the representation of the dipole cross section (\ref{2.2}) which
contains color transparency. Applying the projection (\ref{2.36}) to 
representation (\ref{2.2}), we obtain
\be
\sigma_{(q \bar q)^{J=1}_{L,T}p} (\vec r_\bot^{~\prime}, W^2) = \int 
d^2 \vec l^{~\prime}_\bot \bar \sigma_{(q \bar q)^{J=1}_{L,T}p} 
(\vec l^{~\prime 2}_\bot , W^2) (1 - e^{-i \vec l^{~\prime}_\bot \cdot
\vec r^{~\prime}_\bot}).
\label{2.40}
\ee
The relation between $\tilde{\sigma} (\vec l^{~\prime 2}_\bot z (1-z), 
z (1-z), W^2)$ 
in (\ref{2.2}), and $\bar \sigma_{(q \bar q)^{J=1}_{L,T}p} 
(\vec l^{~\prime 2}_\bot,
W^2)$ in (\ref{2.40}), is analogous to (\ref{2.36}), i.e.
\bqa
& &    \bar\sigma_{(q \bar q)^{J=1}_{L,T} p} (\vec l^{~\prime 2}_\bot, W^2) = 
\label{2.41} \\
& & = 3 \cdot
\left\{ \matrix{
\int dz  (d^1_{10} (z))^2 z (1-z) \tilde{\sigma} (\vec l^{~\prime 2}_\bot z 
(1-z), 
z (1-z), W^2), \cr
\int dz \frac{1}{2} \left( (d^1_{1-1} (z))^2 + (d^1_{11} (z))^2 \right)
z (1-z) \tilde{\sigma} (\vec l^{~\prime 2}_\bot z (1-z), z (1-z), W^2).
} \right. 
\nonumber
\eqa
Expanding (\ref{2.40}) for $\vec r^{~\prime 2}_\bot \to 0$, 
in analogy to (\ref{2.3}), we have
\be
\sigma_{(q \bar q)^{J=1}_{L,T} p} (\vec r^{~\prime 2}_\bot, W^2) = \frac{1}{4}
\pi \vec r^{~\prime 2}_\bot \int d \vec l^{~ \prime 2}_\bot \vec l^{~\prime 2}_\bot
\bar \sigma_{(q \bar q)^{J=1}_{L,T} p} (\vec l^{~\prime 2}_\bot, W^2),~
(\vec l^{~\prime 2}_{\bot~Max} (W^2) \vec r^{~ \prime 2}_\bot \ll 1).
\label{2.42}
\ee
Substituting (\ref{2.42}) into (\ref{2.35}) and integrating over 
$d \vec r^{~\prime 2}_\bot$ with
the help of (\ref{2.13}), we find the large-$Q^2$ representation
\be
\sigma_{\gamma^*_{L,T}p} (W^2, Q^2) = \alpha \sum_q Q^2_q \frac{1}{Q^2}
\frac{1}{6}
\left\{ \matrix{
\int d \vec l^{~\prime 2}_\bot \vec l^{~\prime 2}_\bot 
\bar\sigma_{(q \bar q)^{J=1}_L p} (\vec l^{~\prime 2}_\bot, W^2), \cr
2 \int d l^{~\prime 2}_\bot \vec l^{~\prime 2}_\bot 
\bar\sigma_{(q \bar q)^{J=1}_T p} (\vec l^{~\prime 2}_\bot, W^2)
} \right.
\label{2.43}
\ee
in terms of the $(q \bar q)^{J=1}_{L,T} p$ cross sections, 
$\bar \sigma_{(q \bar q)^{J=1}_{L,T}p} (\vec l^{~\prime 2}_\bot, W^2)$.
The representation (\ref{2.43}) is also obtained directly from (\ref{2.14}) by introducing
$\vec l^{~\prime 2}_\bot $ and inserting (\ref{2.41}).

The ratio of the integrals over the transverse and the longitudinal $(q \bar
q)^{J=1}p$ cross sections in (\ref{2.43})
must be identical to the factor $\rho_W$ already introduced in (\ref{2.14a}), 
\be
\int d\vec l^{~\prime 2}_\bot \vec l^{~\prime 2}_\bot \bar\sigma_{(q \bar
  q)^{J=1}_T p} (\vec l^{~\prime 2}_\bot , W^2) = \rho_W \int d \vec l^{~\prime
  2}_\bot \vec l^{~\prime 2}_\bot \bar\sigma_{(q \bar q)^{J=1}_L p} (\vec
l^{~\prime 2}_\bot , W^2).
\label{2.43A}
\ee

According to the proportionality (\ref{2.43A}), the dipole cross sections for
transversely and longitudinally polarized dipole states in (\ref{2.42}) become
related to each other via
\bqa
& & \bar\sigma_{(q \bar q)^{J=1}_T p} (\vec r^{~\prime 2}_\bot , W^2) =
\frac{1}{4} \pi \rho_W \vec r^{~\prime 2}_\bot \int d \vec l^{~\prime 2}_\bot
\vec l^{~\prime 2}_\bot \bar\sigma_{(q \bar q)^{J=1}_L p} (\vec l^{~\prime
  2}_\bot , W^2) \nonumber \\
& & = \bar\sigma_{(q \bar q)^{J=1}_L p} (\rho_W \vec r^{~\prime 2}_\bot , W^2),
~~~~~~(\vec l^{~\prime 2}_{\rm Max} (W^2) \vec r^{~\prime 2}_\bot \ll 1 ) . 
\label{2.43B}
\eqa 
According to (\ref{2.43B}), for $\vec r^{~\prime 2}$ sufficiently small,
the cross section for transversely polarized $(q \bar q)^{J=1}$ states on the
proton, $\bar\sigma_{(q \bar q)^{J=1}_T p} (\vec r^{~\prime 2}_\bot , W^2)$, is
obtained from the cross section for longitudinally polarized $(q \bar q)^{J=1}$
states, $\bar\sigma_{(q \bar q)^{J=1}_L p} (\vec r^{~\prime 2}_\bot
,\hfill\break  W^2)$, by
performing the substitution of $\vec r^{~\prime 2}_\bot$ by $\rho_W \vec
r^{~\prime 2}$, 
\be
\vec r^{~\prime 2}_\bot \rightarrow \rho_W \vec r^{~\prime 2}_\bot
\label{2.43C}
\ee
in $\bar\sigma_{(q \bar q)^{J=1}_L p} (\vec r^{~\prime 2}_\bot , W^2)$. 

Upon inserting the proportionality (\ref{2.43A}), the large-$Q^2$
photoabsorption cross section (\ref{2.43}) becomes
\be
\sigma_{\gamma^*_{L,T}p} (W^2, Q^2) = \alpha \sum_q Q^2_q \frac{1}{Q^2}
\frac{1}{6} \int d \vec l^{~\prime 2}_\bot \vec l^{~\prime 2}_\bot
\bar\sigma_{(q \bar q)^{J=1}_L p} (\vec l^{~\prime 2}_\bot , W^2) 
\left\{ \matrix{1 & \cr
                2 \rho_W & . }  \right. 
\label{2.43D}
\ee 

It is tempting to generalize the substitution law (\ref{2.43C}), $\vec
r^{~\prime}_\bot \rightarrow \sqrt{\rho_W} \vec r^{~\prime}_\bot$, from its
validity for $\vec r^{~\prime 2}_\bot \rightarrow 0$ to arbitrary values of
$\vec r^{~\prime}_\bot$ by rewriting (\ref{2.40}) as 
\be
\sigma_{(q \bar q)^{J=1}_{L,T}p} (\vec r^{~\prime 2}_\bot , W^2) = \int d^2 \vec
l^{~\prime 2}_\bot \bar\sigma_{(q \bar q)^{J=1}_L p} (\vec l^{~\prime 2}_\bot ,
W^2) \left\{ \matrix{ (1 - e^{-i \vec l^{~\prime}_\bot \cdot \vec r^{~\prime}_\bot}) ,
    & \cr
(1 - e^{-i \vec l_\bot^{~\prime}\cdot (\sqrt{\rho_W} \vec r^{~\prime}_\bot)} ) . & }
\right. 
\label{2.43E}
\ee
The representation (\ref{2.43E}), in the limit of $\vec r^{~\prime 2}_\bot
\rightarrow \infty$ implies a helicity-in\-depen\-dent color-dipole cross section
that is given by
\be
\sigma_{(q \bar q)^{J=1}_{L,T}p} (\vec r^{~\prime 2}_\bot \rightarrow \infty ,
  W^2) = \pi \int d \vec l^{~\prime 2}_\bot \bar\sigma_{(q \bar q)^{J=1}_L} (\vec
    l^{~\prime 2}_\bot , W^2) \equiv \sigma^{(\infty)} (W^2) . 
\label{2.43F}
\ee
The representation (\ref{2.43E}), accordingly, contains the dynamical
assumption (\ref{2.43F}). In this respect, (\ref{2.43F}) differs from the
representations (\ref{2.2}) and (\ref{2.40}) which are based on the gauge
invariance of the color-dipole interaction by itself. We will come back to
(\ref{2.43F}) in Section 2.5.

\subsection{The Ratio of \boldmath $R \equiv \sigma_{\gamma^*_L p} (W^2, Q^2)/
\sigma_{\gamma^*_T p} (W^2, Q^2).$\unboldmath }

The ratio of the longitudinal to the transverse photoabsorption cross section
at sufficiently large $Q^2$, according to (\ref{2.14b}) and (\ref{2.43D}), is determined by the
proportionality factor $1 / 2 \rho_W$. The factor $1/2$ stems from the
difference in the $\vec r_\bot^{~\prime}$ dependence of the photon wave
functions, compare (\ref{2.13}), for
longitudinally and transversely polarized photons. The factor $1 / \rho_W$,
according to (\ref{2.43A}) and (\ref{2.43B}), is associated with the enhancement of the
transverse dipole-proton cross section relative to the longitudinal one in the limit
of $\vec l^{~\prime 2}_{\bot{\rm Max}}(W^2) \vec r^{~\prime 2}_\bot \ll 1$.
According to (\ref{2.14g}), $\rho_W$ is identical to the factor that is 
responsible for the enhancement of the size, $\vec r_\bot^{~2}$, of $q \bar q$
states originating from $\gamma^*_T \rightarrow q \bar q$ transitions, relative
to the size of $q \bar q$ states from $\gamma^*_L \rightarrow q \bar q$
transitions.  

The enhancement of the transverse relative to the longitudinal $q \bar q$-dipole-proton
cross section is recognized as a consequence of the enhanced transverse size of
transversely relative to longitudinally polarized dipole states.
Longitudinally and transversely polarized $(q \bar q)^{J=1}$ states,
$(q \bar q)^{J=1}_L$ and $(q \bar q)^{J=1}_T$, determining the cross sections in
(\ref{2.43}), differ in the
transverse-momentum distribution of the quark (antiquark). According to
(\ref{2.22}) to (\ref{2.25}), as a consequence of the $\gamma^*_{L,T}
\rightarrow (q \bar q)_{L,T}^{J=1}$ transitions, the average value of the square of
the transverse momentum, $\vec l^{~2}_\bot = z (1-z) \vec l^{~\prime 2}_\bot$, of
a quark (antiquark) in the $(q \bar q)^{J=1}_{L,T}$ state is given by
\be
\langle \vec l_\bot^{~2} \rangle^{\vec l^{~ \prime 2}_\bot = const}_{L,T}
= \vec l^{~\prime 2}_\bot 
\left\{ \matrix{
6 \int dz z^2 (1-z)^2 = \frac{4}{20} \vec l^{~\prime 2}_\bot, \cr
\frac{3}{2} \int dz~ z(1-z) (1 - 2z (1-z)) = \frac{3}{20} \vec l^{~\prime 2}_\bot .
} \right.
\label{2.43G}
\ee
The $q \bar q$ states of fixed mass $\vec l^{~\prime 2}_\bot$ from longitudinal
photons predominantly originate with $z(1-z)\not= 0$, in contrast to the $q
\bar q$ states from transverse photons which originate predominantly from
$z(1-z)\cong 0$, compare (\ref{2.22}) and (\ref{2.23}). 
The average transverse momentum for a $(q \bar q)^{J=1}_L$
state originating from the $\gamma^*_L \rightarrow (q \bar q)^{J=1}_L$
transition, according to (\ref{2.43G}), is enhanced by the factor
$4/3$\footnote{The left-hand and right-hand sides in (\ref{2.43H}) belong to
  the same value of $\vec l_\bot^{~\prime 2} = const.$, but the ratio, $4/3$,
is independent of the specific value chosen for $\vec l_\bot^{~\prime 2}$.}, 
\be
\langle \vec l_\bot^{~2} \rangle^{\vec l_\bot^{~\prime 2} = {\rm const}}_{(q \bar
  q)^{J=1}_L } = \frac{4}{3}
\langle \vec l_\bot^{~2} \rangle^{\vec l^{~\prime 2}_\bot = {\rm const}}_{(q \bar
 q)^{J=1}_T }.
\label{2.43H}
\ee
Longitudinally polarized photons produce $(q \bar q)^{J=1}$ pairs with
(relatively)  \hfill\break 
``large'' internal quark transverse momentum, while transversely
polarized photons lead to $(q \bar q)^{J=1}$ states of ``small'' internal quark
transverse momentum. 

By invoking the uncertainty principle, $(q \bar q)^{J=1}_L$ states originating
from longitudinally polarized photons accordingly have ``small'' transverse
size, while $(q \bar q)^{J=1}_T$ states from transversely polarized photons have relatively
``large'' transverse size. The enhancement factor, when passing from
``small-size'' longitudinally polarized $(q \bar q)^{J=1}_L$ states to
``large-size'' transversly polarized $(q \bar q)^{J=1}_T$ states, from
(\ref{2.43H}) is accordingly given by $4/3$ 
i.e. the factor $\rho_W$ in $\vec r_\bot^{~\prime 2} \rightarrow \rho_W \vec
r_\bot^{~ \prime 2}$ in (\ref{2.43C})    
and (\ref{2.14g})\footnote{Note that by comparing (\ref{2.14b}) and 
  (\ref{2.43}), one finds $\int dz \int d \vec l_\bot^{~2} \vec l_\bot^{~2}
  \tilde\sigma (\vec l_\bot^{~2} , z (1 - z), W^2) =   
\frac{1}{6} \int d \vec l_\bot^{~\prime 2}
    \vec l_\bot^{~\prime 2} \bar\sigma_{(q \bar q)^{J=1}_L p} 
(\vec l_\bot^{~\prime 2}, W^2)$.  
The right-hand side in the longitudinal photoabsorption cross section
(\ref{2.12}) may be rewritten as 
$$
\sigma_{\gamma^*_L p} (W^2, Q^2) = \frac{3\alpha}{2} \sum_q Q^2_q \cdot Q^2
\int d \vec r_\bot^{~\prime 2} \vec r^{~\prime 2}_\bot K^2_0 (r^\prime_\bot Q) \frac{1}{6} \int d \vec
l_\bot^{~\prime 2} \vec l_\bot^{~\prime 2} \bar\sigma_{(q \bar q)^{J=1}_L}
(\vec l_\bot^{~\prime 2} , W^2) ,
$$
thus explicitly connecting the dipole size $\vec r^{~\prime}_\bot$ with the $(q \bar q)^{J=1}_L$ state
of fixed mass $\vec l_\bot^{~\prime 2}$, as required for the above argument.}
is equal to $4/3$, \cite{Ku-Schi}
\be
\rho_W \equiv \rho = \frac{4}{3}. 
\label{2.43I}
\ee
The factor $\rho_W = \rho$ is independent of the energy, $W$, since the Lorentz
boost from e.g. the $(q \bar q)^{J=1}$ rest frame to the $\gamma^* p$ frame
does not affect the ratio of the transverse momenta, $\vec l_\bot$, in the $(q
\bar q)^{J=1}_T$ and the $(q \bar q)^{J=1}_L$ state.  

The ratio $R$ for sufficiently large $Q^2$ is given by
\be
R = \frac{\sigma_{\gamma^*_L p} (W^2, Q^2)}{\sigma_{\gamma^*_T p} (W^2, Q^2)}
= \frac{1}{2 \rho} = 
\left\{ \matrix{
0.5~~~{\rm for}~~ \rho = 1, \cr
\frac{3}{8} = 0.375~~~{\rm for}~~ \rho = \frac{4}{3}.
} \right.
\label{2.43J}
\ee
In (\ref{2.43J}), for comparison, in addition to the case of transverse-size
enhancement of $\rho = 4/3$, we have also indicated the case of $\rho = 1$
obtained from helicity independence, i.e. by replacing the transverse-size
enhancement by the simplifying ad hoc assumption of
equality of the $(q \bar q)^{J=1} p$ cross sections for longitudinal and 
transverse $(q \bar
q)^{J=1}$ states. The transverse-size enhancement is responsible for the
deviation of $R$ from $R=0.5$. 

In the case of the ansatz (\ref{2.16}), from (\ref{2.20}), with
$\rho(\epsilon)=4/3$, one finds 
\be
\epsilon \cong 0.0303.
\label{2.43K}
\ee

Our examination of the longitudinal-to-transverse ratio $R$ at large
$Q^2$ may be summarized as follows. The ratio is first of all
determined by a factor $1/2$, originating from the ratio of the
probabilities to find a $q \bar q$ with size
parameter squared, $\vec r^{~\prime 2}_\bot = \vec r^{~2}_\bot
z (1-z)$, in a longitudinally and a transversely polarized photon; 
compare (\ref{2.12}) to (\ref{2.14}), and (\ref{2.43}). 
The second factor, $1/\rho$ in (\ref{2.43J}), results from the 
different dependence on the configuration variable $z(1-z)$ of $q \bar q$
states from 
longitudinally and transversely polarized photons implying interactions of $(q
\bar q)$ states with different
average transverse momenta squared, $\vec l^{~2}_\bot$, 
of the quark (antiquark) in the $(q \bar
q)^{J=1}_{L,T}$ states, compare (\ref{2.43H}).  
Invoking the
uncertainty relation with respect to the scattering of these $(q \bar
q)^{J=1}_{L,T}$ states on the proton, one arrives at the fixed value of $\rho
= 4/3$ in (\ref{2.43J}) for the transverse-size enhancement that enters
(\ref{2.43B}) and determines the value of $R$ in (\ref{2.43J}).

In terms of the proton structure functions, $F_L (x,Q^2)$ and
$F_2 (x, Q^2)$, the result (\ref{2.43J}) for $R$ at large $Q^2$ becomes
\be
F_L (x,Q^2) = \frac{1}{1+2 \rho} F_2 (x,Q^2) = 
\left\{ \matrix{
0.33 F_2 (x,Q^2),~~~(\rho = 1),\cr
0.27 F_2 (x,Q^2),~~~(\rho = \frac{4}{3}).
} \right.
\label{2.51}
\ee
The prediction (\ref{2.51}) of $F_L = 0.27 F_2$ is consistent with the
experimental results from the H1 and ZEUS collaboration. Compare figs. 2
and 3.
\begin{figure}[h!]
\centerline{\epsfig{file=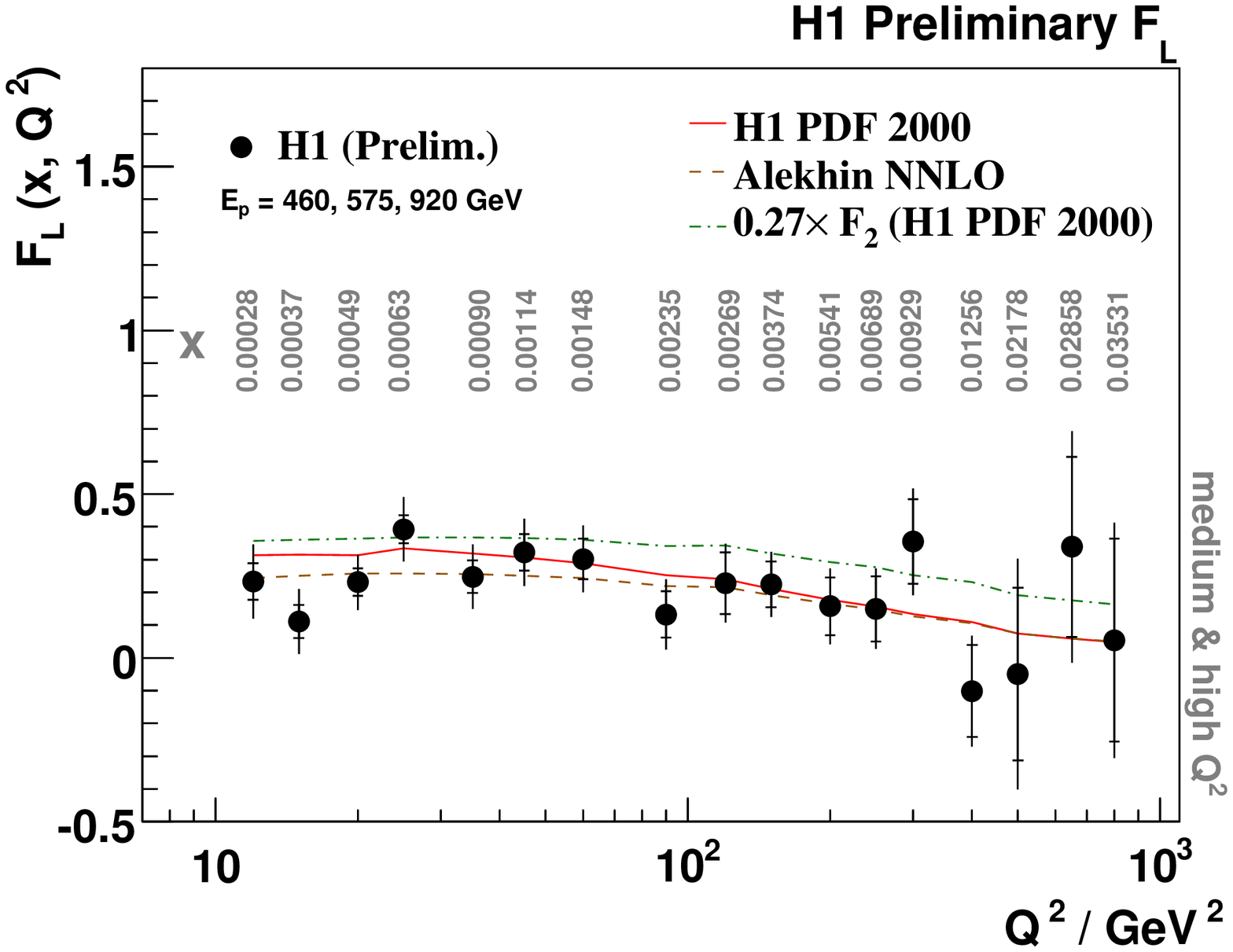,width=8cm} }
\caption{{\footnotesize The H1 experimental results for the longitudinal proton structure function,
$F_L (x, Q^2)$, compared with the prediction of $F_L (x, Q^2) = 0.27 \times F_2
(x, Q^2)$ from transverse-size enhancement of transversely relative to
longitudinally polarized $(q \bar q)^{J=1}$ states.}} 
\end{figure}
%Fig.2
\begin{figure}[h!]
\centerline{\epsfig{file=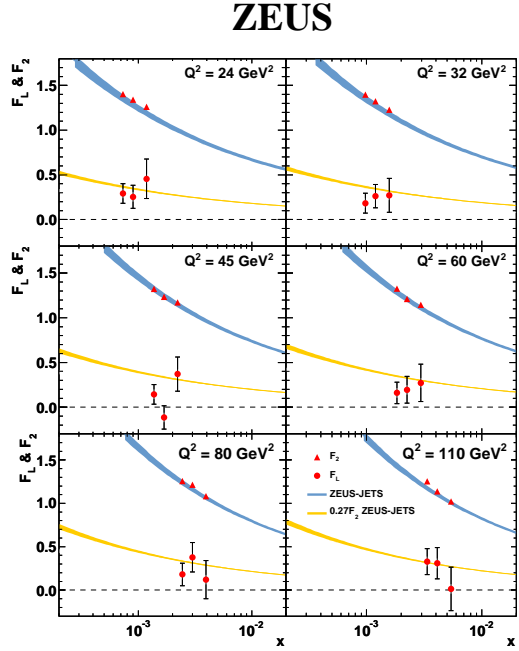,width=8cm}}
\caption{{\footnotesize The ZEUS experimental results
for the longitudinal proton structure function,
$F_L (x, Q^2)$, compared with the prediction of $F_L (x, Q^2) = 0.27 \times F_2
(x, Q^2)$ from transverse-size enhancement of transversely relative to
longitudinally polarized $(q \bar q)^{J=1}$ states.}}
\end{figure}
%Fig.3

\subsection{Discussion on the Representations of the CDP in Sections
2.1 and 2.2}.

The CDP of DIS at low x is based on a life-time argument concerning massive
hadronic fluctuations of the photon. The argument is identical to the one
put forward in the space-time interpretation \cite{Gribov, Niels} 
of generalized vector 
dominance in the early 1970ies. The life-time in the rest frame of the
nucleon of a hadronic fluctuation of mass $M_{q \bar q}$, given by the
covariant expression \cite{Golec}
\be
\frac{1}{\Delta E} = \frac{1}{x + \frac{M^2_{q \bar q}}{W^2}} \frac{1}{M_p}
\gg \frac{1}{M_p},
\label{2.52}
\ee
becomes large in comparison with the inverse of the proton mass, $M_p$,
provided $x \cong Q^2/W^2 \ll 1$ and the c.m. energy, $W$, is sufficiently
large. The $\gamma^* p$ interaction with the nucleon at low x, accordingly, proceeds via
the interaction of hadronic $q \bar q$ fluctuations of timelike
four-momentum squared identical to $M^2_{(q \bar q)}$.
More definitely, the integration over the dipole cross section 
$\sigma_{(q \bar q)p} (r_\bot, z(1-z), W^2)$ in transverse position space
in (\ref{2.1}) describes the interaction of a continuum of massive $q \bar q$
states.
The dipole cross section depends on $W^2$  
\footnote{Compare also
ref. \cite{Ewerz}.}, just as any other
purely hadronic interaction cross section. In particular, the dipole cross section 
does not depend on the virtuality, $Q^2$, of the photon, and consequently,
it does not depend on $x$. 

The dipole cross section in (\ref{2.1}) does not
refer to a definite spin $J$ of the massive $q \bar q$ continuum states. The
interaction with the nucleon, nevertheless, proceeds via the spin 
$J=1$ projection of the dipole cross section $\sigma_{(q \bar q)p}
(r_\bot, z (1-z), W^2)$, compare the discussion in Section 2.2, in particular
the relations (\ref{2.35}) and (\ref{2.36}).

The $W$ dependence of the dipole cross section explicitly, via 
$\tilde \sigma (\vec l^{~2}_\bot, z(1-z), W^2)$ in (\ref{2.14b}) and
$\bar \sigma_{(q \bar q)^{J=1}_L p} (\vec l^{~\prime 2}_\bot, W^2)$
in (\ref{2.43D}) with (\ref{2.43I}), enters the large-$Q^2$ approximation of the photoabsorption
cross section. Inserting (\ref{2.14b}) and (\ref{2.43D}) into the proton structure
function,
\bqa
F_2 (x,Q^2) & = & \frac{Q^4(1-x)}{4 \pi^2 \alpha (Q^2 + (2M_px)^2)} 
\left(\sigma_{\gamma^*_L p}
(W^2, Q^2) + \sigma_{\gamma^*_T p} (W^2, Q^2)\right)\cong \nonumber \\
& \cong & \frac{Q^2}{4 \pi^2 \alpha} \left(\sigma_{\gamma^*_L p} (W^2, Q^2)
+ \sigma_{\gamma^*_T p} (W^2, Q^2) \right)
\label{2.53}
\eqa
one finds
\be
F_2 (x, Q^2) = \frac{\sum_q Q^2_q}{4 \pi^2} \int dz 
\int d \vec l^{~2}_\bot
\vec l^{~2}_\bot \tilde \sigma (\vec l^{~2}_\bot, z(1-z), W^2) (1 + 2 \rho),
\label{2.54}
\ee
and
\be
F_2 (x, Q^2) = \frac{\sum_q Q^2_q}{4 \pi^2} \frac{1}{6} \int d 
\vec l^{~\prime 2}_\bot \vec l^{~\prime 2}_\bot 
\bar \sigma_{(q \bar q)^{J=1}_L p} (\vec l^{~\prime 2}_\bot, W^2)
\left( 1+2 \rho \right),
\label{2.55}
\ee
where $\rho$ for $Q^2$ sufficiently large is given in (\ref{2.43I}).
According to the right-hand sides in (\ref{2.54}) and (\ref{2.55}),
the structure function only depends on $W^2 = Q^2/x$ in the
color-transparency region of sufficiently large $Q^2$ and sufficiently
small $x < 0.1$.

The $W^2$ dependence in (\ref{2.54}) and (\ref{2.55}) can be empirically tested by
plotting the experimental data for the proton structure function $F_2 (x \cong
Q^2/W^2, Q^2)$ as a function of $1/W^2$. 

\begin{figure}[h!]
\centerline{\epsfig{file=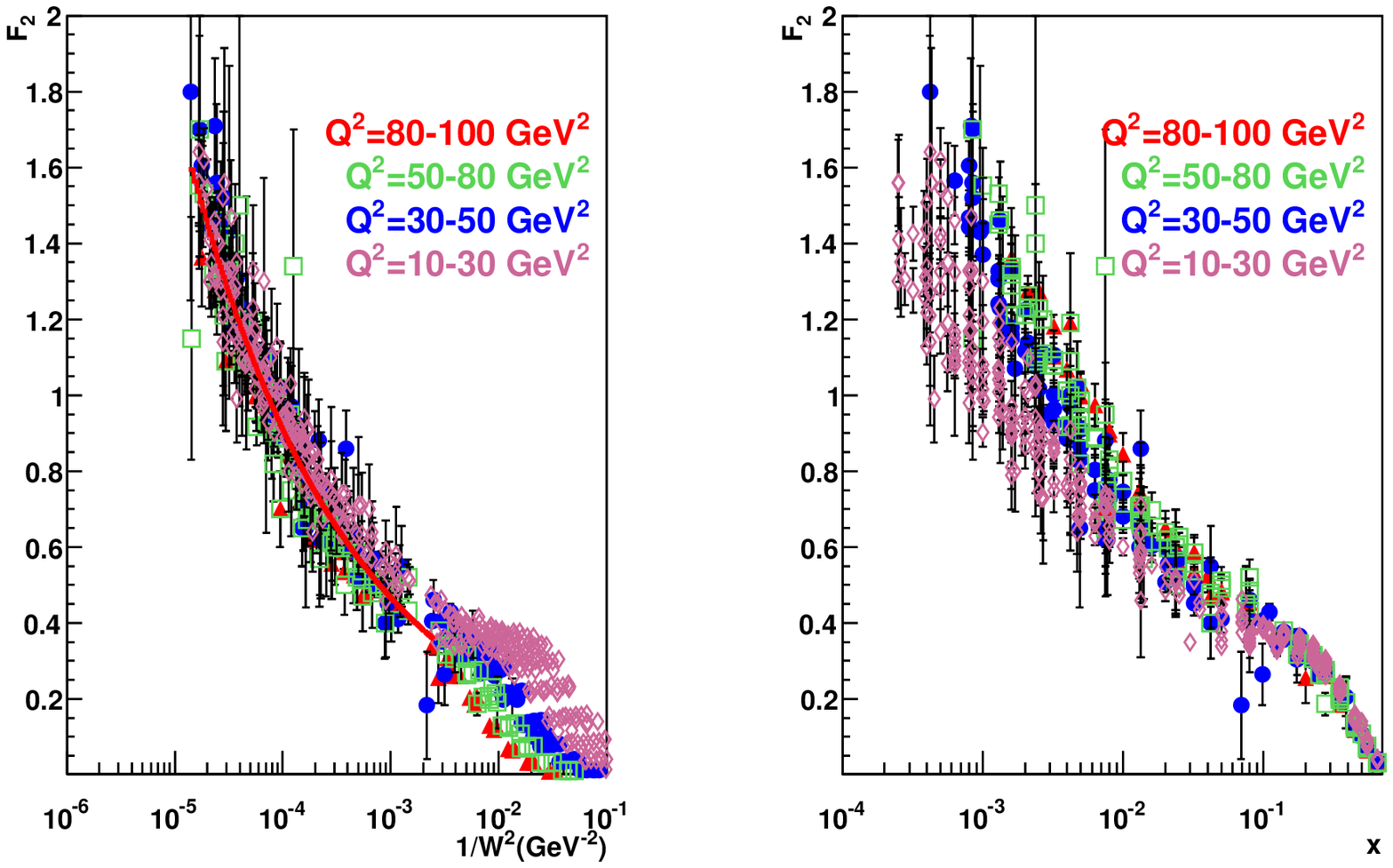,width=11cm} } 
~\hspace{3cm} Fig.4a   \hspace{4cm} Fig.4b
\caption{{\footnotesize In fig.4a we show the experimental data for $F_2(x\cong Q^2/W^2, Q^2)$ as a
function of $1/W^2$, and in fig.4b, for comparison, as a function of $x$. The
theoretical prediction 
based on (\ref{(q)}) and (\ref{(r)}) is also shown in fig.4a.}}
\end{figure}
%Fig.4a und b 
In fig. 4a, we show\footnote{Figure 4 was kindly prepared by Prabhdeep Kaur
  (compare thesis, in preparation).} 
the experimental data from HERA for $Q^2$ in the large range of 
$10 GeV^2 \le Q^2 \le 100 GeV^2$ as a function of $1/W^2$. 
In the relevant range of $x\cong Q^2/W^2 < 0.1$, approximately corresponding to
$1/W^2 \le 10^{-3}$, the
experimental data show indeed a tendency to lie on a single line, quite in
contrast to the range of $1/W^2 \ge 10^{-3}$. The opposite tendency of the
experimental data, approximate clustering around a single line for $x \ge 0.1$,
but stronger deviations from a single line at $x \le 0.1$ is seen in fig. 4b,
where the same experimental data for $F_2 (x, Q^2)$ are plotted in the usual
manner as a function of Bjorken $x$. The replacement of $W^2$ by $W^2 \simeq Q^2/x$,
when passing from fig. 4a to fig. 4b now yields the well-known increased
violation of Bjorken scaling in the diffraction region of $x < 0.1$. 
Compare Section 2.7 for a discussion of the theoretical prediction shown in fig.4a.

We summarize: DIS at low $x$ proceeds via the imaginary part of the
forward scattering amplitude of a continuum of massive 
$(q \bar q)^{J=1}_{L,T}$ states. The interaction of the $q \bar q$
color dipole with the gluon field of the proton, by gauge invariance,
fulfills (\ref{2.2}) and (\ref{2.40}), implying color transparency, (\ref{2.3}) and 
(\ref{2.42}). For
$Q^2$ sufficiently large (with $x \cong Q^2/W^2 \ll 0.1$ sufficiently small), 
the structure function $F_2 (x, Q^2)$ only
depends on the single variable $W$. No details of perturbative
QCD beyond the gauge-invariant color-dipole interaction are needed to
deduce the CDP of Sections 2.1 and 2.2, and this (approximate) dependence of
$F_2 (x, Q^2)$ on the single variable $W^2$ for $Q^2$ sufficiently large. 
In particular, no reference
to details of the perturbative gluon density of the proton is needed. In this connection,
also compare the derivation of the CDP in ref. \cite{Cvetic} as well as   
the formally much more
complete and elaborate derivation in ref. \cite{Ewerz}.

By starting from the $\vec l_\bot$-factorization approach,
under certain assumptions, one may introduce a
CDP-like representation \cite{Nikolaev,Golec-Biernat}
for the photoabsorption cross section containing $x$
instead of $W^2$ in the dipole cross section in (\ref{2.1}). 
Such a representation does not factorize the $Q^2$ dependence inherently
connected with the photon wave function and the $W$ dependence that governs the
$(q \bar q)p$ dipole interaction. As a consequence, the CDP-like representation
is ill-suited to represent the transition from large $Q^2$ to small $Q^2$
including the solely $W$-dependent 
cross section of $Q^2 = 0$ photoproduction. Examining and understanding this
transition to low $x \cong Q^2 / W^2$ photoabsorption, however, is the main aim
and also the essential achievement of the CDP-representation of DIS at low $x$. 

\subsection{Low-x Scaling}

A model-independent analysis of the experimental data on DIS from HERA has
revealed \cite{DIFF2000, SCHI}  that the photoabsorption cross section, 
$\sigma_{\gamma^*p} (W^2, Q^2)$, at low $x$ is a function of the low-x
scaling variable\footnote{Scaling in terms of a different, x-dependent 
instead of W-dependent, scaling variable was found in ref. \cite{GB}}
\be
\eta (W^2, Q^2) = \frac{Q^2 + m^2_0}{\Lambda^2_{sat} (W^2)},
\label{2.62}
\ee
i.e. a function of the single variable $\eta (W^2, Q^2)$,
\be
\sigma_{\gamma^*p} (W^2, Q^2) = \sigma_{\gamma^*p} 
\left(\eta (W^2, Q^2)\right).
\label{2.63}
\ee
In (\ref{2.62}) and (\ref{2.63}), the ``saturation scale'', 
$\Lambda^2_{sat} (W^2)$, empirically increases as $\Lambda^2_{sat} (W^2)
\sim (W^2)^{C_2}$, with $C_2 \cong 0.27$ and $m^2_0 \cong 0.15 GeV^2$
\cite{DIFF2000,SCHI}.
The empirical analysis of the experimental data showed that 
$\sigma_{\gamma^*p} \left( \eta (W^2, Q^2)\right)$ 
for large $\eta (W^2, Q^2) \gg 1$ is inversely proportional to
$\eta (W^2, Q^2)$,
\be
\sigma_{\gamma^*p} (W^2, Q^2) \sim \sigma^{(\infty)} (W^2) 
\frac{1}{\eta (W^2, Q^2)},
\label{2.64}
\ee
while for small values of $\eta (W^2, Q^2) \ll 1$, the dependence on
$\eta (W^2, Q^2)$ is logarithmic,
\be
\sigma_{\gamma^*p} (W^2, Q^2) \sim \sigma^{(\infty)} (W^2) \ln 
\frac{1}{\eta (W^2, Q^2)},~~(\eta (W^2, Q^2) \ll 1).
\label{2.65}
\ee
In (\ref{2.64}) and (\ref{2.65}) the cross section $\sigma^{(\infty)}(W^2)$
empirically was found to be of
hadronic size and approximately constant, $\sigma^{(\infty)}(W^2) \simeq
const.$, as a function of the energy $W$.

In the present Section 2.5, we will show that not only the existence of
the scaling behavior (\ref{2.63}), but also the observed functional dependence of
the cross section, as $1/\eta (W^2, Q^2)$ for large $\eta (W^2, Q^2)$, and
as $\ln (1/\eta (W^2, Q^2))$ for small $\eta (W^2, Q^2)$, in (\ref{2.64}) and
(\ref{2.65}), respectively, is a general and direct consequence of the color-dipole
nature of the interaction of the hadronic fluctuations of the photon with
the color field in the nucleon. No specific parameterization of the 
color-dipole-proton cross section, $\sigma_{(q \bar q)p} (r_\bot, z (1-z),
W^2)$, must be introduced to deduce the empirically observed 
functional dependence
in (\ref{2.64}) and (\ref{2.65}).

The ensuing analysis will be based on the representation of the
photoabsorption cross section in Section 2.2 in terms of the scattering
of $(q \bar q)^{J=1}_{L,T}$ states on the proton. Compare (\ref{2.35}) in 
particular, as well as the longitudinal and transverse dipole cross
sections given by (\ref{2.40}). The representation (\ref{2.40}) of the dipole
cross section, as a consequence of (\ref{2.2}), is solely based on the 
gauge-invariant coupling of the color-dipole state to the gluon field
in the nucleon.

Upon angular integration, (\ref{2.40}) becomes 
\bqa
\sigma_{(q \bar q)^{J=1}_{L,T}p} (r^\prime_\bot, W^2) && =
\pi \int d \vec l^{~\prime 2}_\bot \bar \sigma_{(q \bar q)^{J=1}_{L,T} p}
(\vec l^{~\prime 2}_\bot , W^2). \nonumber \\
&& \cdot \left( 1 - \frac{\int d \vec l^{~\prime 2}_\bot 
\bar \sigma_{(q \bar q)^{J=1}_{L,T} p} (\vec l^{~\prime 2}_\bot, W^2) J_0
(l^\prime_\bot r^\prime_\bot)}{\int d \vec l^{~\prime 2}_\bot
\bar \sigma_{(q \bar q)^{J=1}_{L,T} p} (\vec l^{~\prime 2}_\bot, W^2)}
\right), 
\label{2.66}
\eqa
where $r^\prime_\bot \equiv \sqrt{\vec r^{~\prime 2}_\bot}$ and 
$J_0 (l^\prime_\bot r^\prime_\bot)$ denotes the Bessel function of
order zero. We assume that the integrals in (\ref{2.66}) do exist and are
determined by the integrands in a restricted range of 
$\vec l^{~\prime 2}_\bot < \vec l^{~\prime 2}_{\bot~Max} \equiv 
l^{\prime 2}_{\bot~Max} (W^2)$, where
$\bar \sigma_{(q \bar q)^{J=1}_{L,T} p} (\vec l^{~\prime 2}_\bot, W^2)$
is appreciably different from zero. The resulting dipole cross section (\ref{2.66}),
for any fixed value of $r^\prime_\bot$,  
strongly depends on the variation of the phase, $l^\prime_\bot r^\prime_\bot$,
in (\ref{2.40}) and (\ref{2.66}) as a function of $l^\prime_\bot < l^\prime_{\bot~Max}
(W^2)$. 

Indeed, if for a given value of $r^\prime_\bot$ the phase
$l^\prime_\bot r^\prime_\bot$ in the relevant range of $l^\prime_\bot <
l^\prime_{\bot~Max} (W^2)$
is always smaller than unity, i.e.
\be
0 < l^\prime_\bot r^\prime_\bot < l^\prime_{\bot~Max} (W^2) r^\prime_\bot
\ll 1,
\label{2.67}
\ee
the second term in the bracket of (\ref{2.66}) essentially cancels the first one,
since
\be
J_0 (l^\prime_\bot r^\prime_\bot) \cong 1 - \frac{1}{4} (l^\prime_\bot
r^\prime_\bot)^2 + \frac{1}{4^3} (l^\prime_\bot r^\prime_\bot)^4 + \cdots .
\label{2.68}
\ee
Substitution of (\ref{2.68}) into (\ref{2.66}) implies the proportionality of the
dipole cross section to $r^{\prime 2}_\bot$ already given in (\ref{2.42}). 
Combining (\ref{2.42}) with (\ref{2.43B}) and (\ref{2.43I}), we find
\bqa
& & \sigma_{(q \bar q)^{J=1}_{L,T} p} (r^{\prime 2}_\bot, W^2) =
\label{2.69} \\
& &= \frac{1}{4} \pi r^{\prime 2}_\bot \int d \vec l^{~\prime 2}_\bot
\vec l^{~\prime 2}_\bot \bar \sigma_{(q \bar q)^{J=1}_L p} 
(\vec l^{~\prime 2}_\bot, W^2) \left\{ \matrix{
1,\cr
\rho ,} \right.  \left(r^{\prime 2}_\bot \ll
\frac{1}{l^{\prime 2}_{\bot~Max} (W^2)}\right).
\nonumber
\eqa

In the limiting case of
\be
l^\prime_{\bot Max} (W^2) r^\prime_\bot \gg 1,
\label{2.70}
\ee
alternative to (\ref{2.67}), the rapid oscillation of the Bessel
function under variation of $0 < l^\prime_\bot < l^\prime_{\bot~Max} (W)$
at fixed $r^\prime_\bot$ implies a vanishing contribution of the second
term in (\ref{2.66}). The dipole cross section (\ref{2.66}) in this limit
is not proportional to the dipole size $\vec r_\bot^{~\prime 2}$, but, in
distinction from (\ref{2.69}), becomes identical to the $\vec r^{~\prime
  2}_\bot$-independent limit $\sigma^{(\infty)}_{L,T} (W^2)$ of normal hadronic
size,
\bqa
\sigma_{(q \bar q)^{J=1}_{L,T} p} (r_\bot^{~\prime 2}, W^2) && \cong \pi
\int d \vec l^{~\prime 2}_\bot \bar \sigma_{(q \bar q)^{J=1}_{L,T} p} 
(\vec l^{~\prime 2}_\bot , W^2) 
\equiv \sigma^{(\infty)}_{L,T} (W^2),
\nonumber \\ 
& & \left(r^{\prime 2}_\bot \gg
\frac{1}{l^{\prime 2}_{\bot~Max} (W^2)} \right).
\label{2.71}
\eqa

We note that the $r_\bot^\prime$-independent limit on the right-hand side in
(\ref{2.71}) obtained at any fixed value of $r_\bot^\prime$ for $l^{\prime
  2}_{\bot \max} (W^2) \rightarrow \infty$ coincides with the limit of
$r^\prime_\bot \rightarrow \infty$ at fixed energy, $W$, or fixed $\vec
l^{~\prime 2}_{\bot \max}(W^2)$. A small $q \bar q$ dipole at infinite energy
yields the same cross section as a sufficiently large dipole at finite energy
$W$. 

The gauge-invariant color-dipole interaction with the gluon thus implies
the emergence of two scales, the helicity-dependent integral\footnote{For
  generality, we keep the distinction between $\sigma^{(\infty)}_L (W^2)$ and
  $\sigma^{(\infty)}_T (W^2)$, even though the essential conclusions of this
  Section do not depend on whether this distinction is kept or replaced by the
  equality (\ref{2.43F}).} over  \\ 
$\bar \sigma_{(q \bar q)^{J=1}_{L,T} p} (\vec l^{~\prime 2}_\bot, W^2)$ in (\ref{2.71})
and the first moment of $\bar \sigma_{(q \bar q)^{J=1}_L} (l^{\prime 2}_\bot,
W^2)$ in (\ref{2.69}), which determine the dipole cross section for relatively large
$r^{\prime 2}_\bot$ and relatively small $r^{\prime 2}_\bot$, respectively.
Whether (\ref{2.71}) or (\ref{2.69}) is relevant for a chosen value of $r^\prime_\bot$
depends on the value of $l^{\prime 2}_{\bot~Max} (W^2)$ that in
turn depends on the $W$-dependence of the $(q \bar q)^{J=1}_{L,T}p$ dipole
cross section,
$\bar \sigma_{(q \bar q)^{J=1}_{L,T}p} 
(\vec l^{~\prime 2}_\bot ,W^2)$.

It is appropriate to introduce and use the normalized distribution in 
$\vec l^{~\prime 2}_\bot$,
\bqa
\Lambda^2_{sat} (W^2) & \equiv &  \frac{\int d \vec l^{~\prime 2}_\bot
\vec l^{~\prime 2}_\bot \bar \sigma_{(q \bar q)^{J=1}_L p} 
(\vec l^{~\prime 2}_\bot , W^2)}{\int d \vec l^{~\prime 2}_\bot
\bar \sigma_{(q \bar q)^{J=1}_L p} (\vec l^{~\prime 2}_\bot, W^2)}= 
\nonumber \\
& = & 
\frac{1}{\sigma^{(\infty)}_L(W^2)} \pi \cdot  \int d \vec l^{~\prime 2}_\bot
\vec l^{~\prime 2}_\bot \bar \sigma_{(q \bar q)^{J=1}_L p} 
(\vec l^{~\prime 2}_\bot , W^2),
\label{2.72}
\eqa
as the second scale besides $\sigma^{(\infty)}_{L,T} (W^2)$ from 
(\ref{2.71})\footnote{
The scale $\Lambda^2_{sat} (W^2)$ in (\ref{2.72}) is to be identified with the
parameter $\Lambda^2_{sat} (W^2)$ in (\ref{2.62}) that was introduced in the
fit \cite{DIFF2000, SCHI} to the experimental data.}.
The $r^{\prime 2}_\bot \to 0$ limit in (\ref{2.69}) then becomes
\be
\sigma_{(q \bar q)^{J=1}_{L,T} p} (r^{\prime 2}_\bot, W^2) = \frac{1}{4}
r^{\prime 2}_\bot \sigma^{(\infty)}_L (W^2) \Lambda^2_{sat} (W^2)
\left\{ \matrix{
1, \cr
\rho,
} \right.~\left(r^{\prime 2}_\bot \ll \frac{1}
{l^{\prime 2}_{\bot~Max} (W^2)} \right).
\label{2.73}
\ee
The cross section $\sigma^{(\infty)}_{L,T} (W^2)$, as a consequence of the 
color-dipole interaction in (\ref{2.2}) and (\ref{2.40}),
according to (\ref{2.71}) and
(\ref{2.73}), is of relevance for both, the $r^{\prime 2}_\bot \to \infty$ as
well as the $r^{\prime 2}_\bot \to 0$ behavior of the dipole cross section.

Before returning to the photoabsorption cross section, we add a further
comment on the dipole cross section (\ref{2.66}) and its important limits in
(\ref{2.69}) (or, equivalently, in (\ref{2.73})) and in (\ref{2.71}). 
The dependence of the dipole cross section (\ref{2.66}) on
$r^\prime_\bot$ is determined by the destructive interference originating from  
the (negative) second term in the bracket in (\ref{2.66}). At any fixed
value of $r^\prime_\bot$, for sufficiently high energy, i.e. 
with increasingly greater values of $\vec l^{~\prime 2}_{\bot~Max} (W^2)$, 
the vanishing of this term, due
to strong oscillations of the integrand leads to the $\vec
r^{~\prime}_\bot$-independent limit of a cross section of hadronic size in 
(\ref{2.71}). With increasing energy a transition occurs from the region of
color transparency (\ref{2.69}), where the cross section is proportional to the
dipole size, $\vec r^{~\prime 2}_\bot$, to the saturation regime (\ref{2.71})
characterized by a cross section that is independent of the
dipole size, $\vec r^{~\prime 2}_\bot$; the interaction of a colorless $q \bar
q$ dipole is in the saturation regime replaced by the interaction of a colored
quark and a colored antiquark thus producing a cross section of hadronic
size. Both, color transparency, as well as the transition to the hadronlike
saturation behavior, are recognized as a genuine consequence of the
gauge-invariant color-dipole interaction (\ref{2.1}). 
It is a misconception to associate the saturation regime
with an increased density in a small-size region of the proton: in the
high-energy limit of (\ref{2.71}) the cross section is not proportional to the
dipole size, and therefore it cannot be interpreted as the product of a (small)
dipole size with a high-gluon-density region.  

We turn to the photoabsorption cross section in (\ref{2.29}). The integration over
$d^2 \vec r^{~\prime}_\bot$ in (\ref{2.29}) at fixed $Q^2$ is dominated by
\be
\vec r^{~\prime 2}_\bot \equiv r^{\prime 2}_\bot \le \frac{1}{Q^2}.
\label{2.74}
\ee
Compare (\ref{2.7}) and (\ref{2.8}). The resulting photoabsorption cross section for fixed
$Q^2$ then depends on whether the limiting case of either (\ref{2.69}) (or equivalently
(\ref{2.73})) or of (\ref{2.71}) is relevant for $r^{\prime 2}_\bot \le 1/Q^2$.

For the case of
\be
r^{\prime 2}_\bot < \frac{1}{Q^2} \ll \frac{1}{l^{\prime 2}_{\bot~Max}
(W^2)},
\label{2.75}
\ee
the $r^{\prime 2}_\bot \to 0$ expression in (\ref{2.73}) is relevant. This region of
relatively large $Q^2$ was treated in Section 2.2. Compare (\ref{2.42}) and (\ref{2.43}).
Introducing $\Lambda^2_{sat} (W^2)$ from (\ref{2.72}) and (\ref{2.73}) on the right-hand
side of (\ref{2.43D}), with $\rho_W = \rho$ from (\ref{2.43I}),
we find
\be
\sigma_{\gamma^*_{L,T}p} (W^2, Q^2) = \frac{\alpha}{\pi} \sum_q Q^2_q
\frac{1}{6} \sigma^{(\infty)}_L (W^2) \frac{\Lambda^2_{{\rm sat}} (W^2)}{Q^2} 
\left\{ \matrix{
1, \cr
2 \rho.
} \right.
\label{2.76}
\ee
The total photoabsorption cross section is given by
\bqa
\sigma_{\gamma^* p} (W^2, Q^2) & = & \sigma_{\gamma^*_L p} (W^2, Q^2) +
\sigma_{\gamma^*_T p} (W^2, Q^2) \nonumber \\
& = & \frac{\alpha}{\pi} \sum_q Q^2_q (1 + 2 \rho) \frac{1}{6}
\sigma^{(\infty)}_L (W^2) \frac{\Lambda^2_{sat} (W^2)}{Q^2}.
\label{2.77}
\eqa
Unitarity requires the hadronic dipole cross section, $\sigma^\infty_{L,T} (W^2)$,
from (\ref{2.71}) to only weakly\footnote{Actually, a logarithmic increase of
  $\sigma^{(\infty)}_{L,T} (W^2)$ is allowed.} depend on $W^2$,
\be
\sigma^{(\infty)}_{L,T} (W^2) \cong~{\rm const.}
\label{2.78}
\ee
Moreover, motivated by quark confinement and/or 
quark-hadron duality \cite{ST}, the divergence of $r^{\prime 2}_\bot \to \infty$
for $Q^2 \to 0$ in (\ref{2.74}) must be replaced by
\be
r^{\prime 2}_\bot < \frac{1}{Q^2 + m^2_0},
\label{2.79}
\ee
where $m^2_0$ actually depends on the quark flavor. 
For light quarks, $m^2_0 \lsim m^2_\rho$, where $m_\rho$ is the
$\rho^0$ meson mass, is relevant. Replacing\footnote{Actually, realistic values
  of $\Lambda^2_{sat}(W^2)$ fulfill the hierarchy of $\Lambda^2_{sat}(W^2) \gg
  m^2_0$, such that in the relevant range of $\eta (W^2, Q^2) = (Q^2 + m^2_0) /
  \Lambda^2_{sat} (W^2) \approx Q^2 / \Lambda^2_{sat} (W^2)$. The replacement
  of $Q^2 \rightarrow Q^2 + m^2_0$ in the case of (\ref{2.80}) is of formal nature.} 
$Q^2 \to Q^2 + m^2_0$ in (\ref{2.77}),
and identifying the resulting (inverse) ratio with the empirical parameter
$\eta (W^2, Q^2)$ in (\ref{2.62}), we have
\be
\sigma_{\gamma^* p} (W^2, Q^2) = \frac{\alpha}{\pi} \sum_q Q^2_q (1 + 2 \rho)
\frac{1}{6} \sigma^{(\infty)}_L (W^2) \frac{1}{\eta (W^2, Q^2)},~~~
(\eta (W^2, Q^2) \gg 1),
\label{2.80}
\ee
where $\eta (W^2, Q^2) \gg 1$ as a consequence of (\ref{2.75}) and
(\ref{2.72}).
With (\ref{2.78}), this is the empirically established scaling behavior (\ref{2.64}).

We turn to the case of
\be
\frac{1}{l^{\prime 2}_{\bot~Max} (W^2)} \ll \frac{1}{Q^2},
\label{2.81}
\ee
alternative to (\ref{2.75}), and relevant in particular for large values of
the energy $W$ and relatively small values of $Q^2$. In this case of
(\ref{2.81}), within the integration domain of $r^{\prime 2}_\bot < 1/Q^2$ from
 (\ref{2.74}), we have to discriminate two different regions. For
\be
r^{\prime 2}_\bot < \frac{1}{\vec l^{~\prime 2}_{\bot~Max} (W^2)}
\ll \frac{1}{Q^2},
\label{2.82}
\ee
we have color transparency (\ref{2.73}). In distinction from (\ref{2.75}), color
transparency only holds in a small restricted domain of the full
integration interval $r^{\prime 2}_\bot < 1/Q^2$. For the remaining
integration domain,
\be
\frac{1}{\vec l^{~\prime 2}_{\bot~Max} (W^2)} < r^{\prime 2}_\bot
< \frac{1}{Q^2},
\label{2.83}
\ee
the $r^\prime_\bot$-independent dipole cross section (\ref{2.71}) becomes relevant.

It is useful to split the integration domain into the sum of two different
ones. Noting that according to the definition (\ref{2.72}),
\be
\Lambda^2_{sat} (W^2) \lsim l^{\prime 2}_{\bot~Max} (W^2),
\label{2.84}
\ee
we use $\Lambda^2_{sat} (W^2)$ as the splitting parameter of the integral
over $dr^{\prime 2}_{\bot}$. The photoabsorption cross section (\ref{2.35}) then
becomes
\bqa
\sigma_{\gamma^*_{L,T}p} (W^2, Q^2) & = & 
\frac{\alpha}{\pi} \sum_q Q^2_q Q^2
\left( \int^{\frac{1}{\Lambda^2_{sat}(W^2)}}_0 d r^{\prime 2}_\bot +
\int^{\frac{1}{Q^2}}_{\frac{1}{\Lambda^2_{sat}(W^2)}} d r^{\prime 2}_\bot
\right) \cdot
\nonumber \\
& & \cdot K^2_{0,1} (r^\prime_\bot Q) 
\sigma_{(q \bar q)^{J=1}_{L,T}p} (r^{\prime 2}_\bot , W^2).
\label{2.85}
\eqa

The main contribution to the photoabsorption cross section is due to the
second term on the right-hand side in (\ref{2.85}). The first term will 
subsequently be shown to be negligible compared with the second one.
Only taking into account the second term, upon introducing the 
$r^\prime_\bot$-independent dipole
cross section from (\ref{2.71}), we find
\be
\sigma_{\gamma^*_{L,T}p} (W^2, Q^2) = \frac{2 \alpha}{\pi} \sum
Q^2_q Q^2       
 \left\{ \matrix{
\sigma^{(\infty)}_L (W^2) 
\int^{\frac{1}{Q}}_{\frac{1}{\Lambda_{sat}(W^2)}} dr^\prime_\bot 
r^\prime_\bot K^2_0 (r^\prime_\bot Q),\cr
\sigma^{(\infty)}_T (W^2)
\int^{\frac{1}{Q}}_{\frac{1}{\Lambda_{sat} (W^2)}} d r^\prime_\bot
r^\prime_\bot K^2_1 (r^\prime_\bot Q).
} \right.
\label{2.86}
\ee

The cross section in the high-energy limit, (\ref{2.86}), as a consequence 
of the factorization of the dipole cross section (\ref{2.71}), is directly
given by an integral over the photon wave function, compare e.g. (\ref{2.86})
with the general expression in (\ref{2.35}).

In the integration domain (\ref{2.83}) of $r^\prime_\bot Q < 1$, relevant in
(\ref{2.85}) and (\ref{2.86}), upon introducing
$y = r^\prime_\bot Q$, we can approximate $K^2_{0,1} (r^\prime_\bot Q)
= K^2_{0,1} (y)$ by
\bqa
K^2_0 (y) &\simeq & \ln^2 y, \nonumber \\
K^2_1 (y) &\simeq & \frac{1}{y^2}, ~~~~~~(y < 1).
\label{2.87}
\eqa
We find
\be
\sigma_{\gamma^*_{L,T}p} (W^2, Q^2) = \frac{2 \alpha}{\pi} \sum Q^2_q
 \left\{ \matrix{
\frac{1}{4} \sigma^{(\infty)}_L (W^2),~~~~~~~~~~~~~~ \cr
\frac{1}{2} \sigma_T^{(\infty)} (W^2)\ln \frac{\Lambda^2_{sat} (W^2)}{Q^2} . }
\right. 
\label{2.88}
\ee
The longitudinal cross section becomes small in this limit of very high energy
$W$ and comparatively small values of $Q^2$. According to (\ref{2.88}), the
longitudinal cross section may be neglected, and the total cross section is
given by 
\be
\sigma_{\gamma^*p} (W^2, Q^2) = \frac{\alpha}{\pi} \sum Q^2_q 
\sigma^{(\infty)}_T (W^2) \ln \frac{\Lambda^2_{sat} (W^2)}{Q^2}.
\label{2.89}
\ee
With the replacement of $Q^2 \to Q^2 + m^2_0$, compare (\ref{2.79}), and upon
introducing $\eta (W^2, Q^2)$ from (\ref{2.62}), we indeed have derived the
empirically observed logarithmic dependence,
\be
\sigma_{\gamma^* p} (W^2, Q^2) = \frac{\alpha}{\pi} \sum Q^2_q
\sigma^{(\infty)}_T (W^2) 
\ln \frac{1}{\eta (W^2, Q^2)}~,~ 
(\eta (W^2, Q^2) \ll 1) ,
\label{2.90}
\ee
where $\sigma_T^{(\infty)} (W^2) \cong ~{\rm const}$, compare (\ref{2.78}).

Combining (\ref{2.76}) and (\ref{2.88}), the ratio $R$ of the longitudinal and
the transverse parts of the photoabsorption cross section is given by
\be
R (W^2, Q^2) = \frac{1}{2} \left\{ \matrix{
\frac{\sigma_L^{(\infty)}(W^2)}{\sigma_T^{(\infty)}(W^2)}
\frac{1}{\ln \frac{1}{\eta (W^2, Q^2)}},   (\eta (W^2, Q^2) \ll 1),\cr
\frac{1}{\rho}, ~~~~~~~~~~~~~~~~~~~ (\eta (W^2, Q^2) \gg 1).} \right.
\label{2.90a}
\ee
In the limit of $\eta (W^2, Q^2) \ll 1$, i.e. for $W^2 \to \infty$ at
fixed $Q^2$, the longitudinal part of the photoabsorption cross section
becomes vanishingly small compared with the transverse part. In the limit of
$\eta (W^2, Q^2) \gg 1$, we have $\rho = 4/3$ from transverse-size enhancement,
while $\rho = 1$ under the ad hoc assumption of helicity independence. 

We finally have to convince ourselves that the first term in (\ref{2.85}) can
be neglected relative to the second one. Inserting (\ref{2.73}) into
(\ref{2.35}), the contribution of the first term becomes
\be
\sigma^{(I)}_{\gamma^*_{L,T} p} (W^2, Q^2) = \frac{\alpha}{\pi} \sum_q
Q^2_q \frac{1}{2} \sigma_L^{(\infty)} (W^2)
\frac{\Lambda^2_{sat} (W^2)}{Q^2} \left\{ \matrix{
\int^{\frac{Q}{\Lambda_{sat} (W^2)}}_0 dy y^3 K^2_0 (y), \cr
\rho \int^{\frac{Q}{\Lambda_{sat} (W^2)}}_0 dy y^3 K^2_1 (y). 
} \right.
\label{2.91}
\ee
Evaluation of the integrals upon inserting (\ref{2.87}) yields
\bqa
\sigma^{(I)}_{\gamma^* p} (W^2, Q^2) & = & \frac{\alpha \sum_q Q^2_q}{\pi}
\frac{1}{4} \sigma_L^{(\infty)} (W^2) \left(\frac{1}{2} 
\frac{Q^2}{\Lambda^2_{sat}(W^2)} \ln^2 \frac{Q^2}{\Lambda^2_{sat} (W^2)}
+ \rho \right) \nonumber \\
& \cong & \frac{\alpha \sum Q^2_q}{\pi} \frac{1}{4} \sigma_L^{(\infty)} 
(W^2) \rho ,
~~~~~(\Lambda^2_{sat} (W^2) \gg Q^2).
\label{2.92}
\eqa
Since (\ref{2.89}) is enhanced by $\ln (\Lambda^2_{sat} (W^2)/Q^2)$, we can 
neglect (\ref{2.92}) for sufficiently large $\Lambda^2_{sat} (W^2)$.

The resulting cross sections (\ref{2.80}) and (\ref{2.90})  establish the empirically
observed low-x scaling behavior as a consequence of the interaction
of the $q \bar q$-fluctuations of the (real or virtual) photon as
color-dipole states. Low-$x$ scaling is recognized as a genuine
consequence of the CDP in the formulation given in (\ref{2.35}) and (\ref{2.40})
that is based on (\ref{2.1}) and (\ref{2.2}). ``Saturation'' i.e. the slow
logarithmic increase as $\ln \Lambda^2_{sat} (W^2)$ in (\ref{2.90}),
is not based on a specific model assumption. It occurs as 
a consequence of the transition of the $(q \bar q) p$ interaction from the
color-transparency region to the hadronic one. This transition occurs for any
given $Q^2$, or any fixed dipole size, provided 
the energy is sufficiently high such that the $q \bar q$ state does not
interact as a colorless dipole, but rather as a system of two colored quarks.

\subsection{The photoproduction limit for \boldmath $W^2 \to \infty$
\unboldmath
at fixed
\boldmath $Q^2 > 0$.\unboldmath}

In Section 2.5, we found that the CDP from (\ref{2.1}) and (\ref{2.2}) 
implies that
the photoabsorption cross section at low $x \cong Q^2/W^2 \ll 1$
depends on the single scaling variable $\eta (W^2, Q^2)$ from (\ref{2.62}) and
(\ref{2.72}). Moreover, the dependence of $\sigma_{\gamma^*_{L,T} p} (\eta
(W^2, Q^2))$, for small and large values of $\eta (W^2, Q^2)$ was
found to be uniquely determined without adopting a specific 
parameterization for the dipole cross section, compare (\ref{2.90}) and (\ref{2.80}),
\bqa
\sigma_{\gamma^*p} (W^2, Q^2) & = & \sigma_{\gamma^*p} (\eta
(W^2, Q^2)) = \label{2.93} \\
& = & \frac{\alpha}{\pi} \sum_q Q^2_q 
\left\{ \matrix{
\sigma^{(\infty)}_T (W^2)
\ln \frac{1}{\eta (W^2, Q^2)},~~~(\eta (W^2, Q^2) \ll 1), \cr
\frac{1}{6} ( 1 +2 \rho ) \sigma_L^{(\infty)} (W^2)  \frac{1}{\eta(W^2, Q^2)},
~~~ (\eta (W^2,Q^2) \gg 1).} \right.
\nonumber
\eqa
where unitarity restricts $\sigma_{L,T}^{(\infty)} (W^2)$ to being
at most weakly dependent on $W$. 
In this Section 2.6, we present a more detailed discussion of
the important $\eta (W^2, Q^2) \to 0$ limit of $W^2 \to \infty$ at fixed
values of $Q^2$.

We explicitly assume $\Lambda^2_{sat} (W^2)$ to increase with the
energy, $W$. There are convincing theoretical arguments for this 
assumption, independent of the analysis of the experimental data that
was referred to in the discussion related to (\ref{2.62}) to (\ref{2.65}).

Note that the absorption of a gluon of transverse momentum $\vec l_\bot$
by a $q \bar q$ fluctuation leads to ``diagonal'' as well as
``off-diagonal'' transitions with respect to the mass, $M_{q \bar q}$,
of the $q \bar q$ fluctuations,
\bqa
(q \bar q)_{M_{q \bar q}} \to \left\{ \matrix{
(q \bar q)_{M_{q \bar q}},~~~~~~~{\rm ``diagonal")}, \cr
(q \bar q)_{M^\prime_{q \bar q} \not= M_{q \bar q}},~~~~~~~~
{\rm (``off-diagonal")}.
} \right.
\label{2.94}
\eqa
The mass difference in the second line of (\ref{2.94}) is proportional to
$\vec l^{~\prime 2}_\bot = \vec l^{~2}_\bot / z (1-z)$, 
or to $\Lambda^2_{sat} (W^2)$ from (\ref{2.72}), on the average,
\be
\Delta M^2_{q \bar q} \equiv M^{\prime 2}_{q \bar q} - 
M^2_{q \bar q} \sim \Lambda^2_{sat} (W^2).
\label{2.95}
\ee
This connection excludes $\Lambda^2_{sat} (W^2) = const.$ unless one
is willing to postulate the mass difference between incoming and outgoing
$q \bar q$ states in hadronic diffraction to be equal to a fixed 
value that is $W$-independent, even for $W \to \infty$. Constancy, 
$\Lambda^2_{sat} (W^2) = {\rm const.}$, would imply a
$W$-dependence of the photoabsorption cross section (\ref{2.93}) that is
exclusively determined by the factorized cross section $\sigma_{L,T}^{(\infty)}
(W^2)$ from (\ref{2.71}), entirely independent of the details of the dynamics of the gluon
field in the proton related to $\Lambda^2_{sat} (W^2)$ from (\ref{2.72}). One
accordingly can safely dismiss the assumption of $\Lambda^2_{sat} (W^2)
= const$ on theoretical grounds, independently of its 
inconsistency with the experimental data, compare (\ref{2.62}) to (\ref{2.65}).
A further argument on the increase of $\Lambda^2_{sat} (W^2)$ with the
energy may be based on the 
consistency of the CDP with a description of the proton structure function in
terms of sea quark and gluon distributions and their evolution with $Q^2$. This
will be discussed below, compare Section 2.7.

Considering the limit of $\eta (W^2, Q^2) \to 0$, or $W^2 \to \infty$ at
fixed $Q^2$, we introduce the ratio of the virtual to the real 
photoabsorption cross section, and from (\ref{2.93}) we find \cite{SCHI}
\bqa
\lim_{W^2 \to \infty \atop {Q^2~{\rm fixed}}} 
\frac{\sigma_{\gamma^*p} (\eta (W^2, Q^2))}{\sigma_{\gamma^*p} (\eta
(W^2, Q^2 = 0))} 
&& = \lim_{W^2 \to \infty \atop {Q^2~{\rm fixed}}} 
\frac{\ln \left( \frac{\Lambda^2_{sat}(W^2)}{m^2_0} 
\frac{m^2_0}{(Q^2 + m^2_0)} \right)}{\ln 
\frac{\Lambda^2_{sat} (W^2)}{m^2_0}} = \nonumber \\
&& = 1 + \lim_{W^2 \to \infty \atop {Q^2~{\rm fixed}}} 
\frac{\ln \frac{m^2_0}{Q^2 + m^2_0}}{\ln 
\frac{\Lambda^2_{sat} (W^2)}{m^2_0}} = 1.
\label{2.96}
\eqa
At sufficiently large $W$, at any fixed value of $Q^2$, the $\gamma^*p$
cross section approaches the $Q^2$-independent $(Q^2 = 0)$ photoproduction
limit. We stress again that this result (\ref{2.96}) is independent of any
particular parameterization of the dipole cross section. It is solely
based on the CDP (\ref{2.1}) with the general form of the dipole cross section
(\ref{2.2}) required by the gauge-invariant two-gluon coupling of the $q \bar q$
fluctuation in the forward-Compton-scattering amplitude. 

The $(Q^2, W^2)$ plane corresponding to (\ref{2.93}) and (\ref{2.96}) is
simple. It consists of only two regions separated by the line $\eta (W^2), Q^2)
\cong 1$, compare fig.5. Below this line i.e. for $\eta (W^2, Q^2) \gg 1$, we have color
transparency with $\sigma_{\gamma^* p} (W^2, Q^2) \sim \Lambda^2_{\rm sat}
(W^2) / Q^2$, while for $\eta (W^2, Q^2) \ll 1$, we have hadronlike saturation
behavior.  
\begin{figure}
\centerline{\epsfig{file=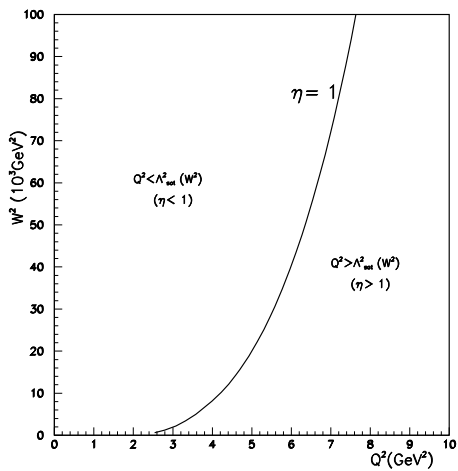,width=7cm}}
\caption{{\footnotesize The $(Q^2, W^2)$ plane showing the line $\eta (W^2, Q^2) = 1$ separating the
large-$Q^2$ and the small-$Q^2$ region.}}
\end{figure}
%Fig.5
Without explicit parameterization of $\Lambda^2_{sat} (W^2)$, 
the relation (\ref{2.96}) does not determine the energy scale, at which the limit
of photoproduction is reached in (\ref{2.96}). The limit (\ref{2.96}) was first given
\cite{SCHI} under the assumption of a specific ansatz for the dipole cross section in
(\ref{2.2}),
\be
\sigma_{(q \bar q)p} (r_\bot, z (1-z), W^2) = \sigma^{(\infty)} (W^2) 
(1 - J_0 (r_\bot z (1-z) \Lambda_{sat} (W^2))
\label{2.97}
\ee
that was used in a successful fit \cite{DIFF2000, SCHI} to the 
experimental data from HERA. By extrapolating
the fit to the experimental data based on (\ref{2.97}) to $W^2 \to \infty$ at
fixed $Q^2$, one finds the limiting behavior (\ref{2.96}). Inserting the fit
result \cite{SCHI}\footnote{The original fit \cite{SCHI} with $\Lambda^2 (W^2)
  = C_1 (W^2 + W^2_0)^{C_2}$ and $W^2_0 = 1081 \pm 124 GeV^2$ can in good
approximation be replaced by (\ref{2.98}).}
\bqa
\Lambda^2_{sat} (W^2) & = & (0.34 \pm 0.06) \left( \frac{W^2}{1 GeV^2} \right)^{C_2}
GeV^2 ,
\nonumber \\
C_2 & = & 0.27 \pm 0.01 , \label{2.98} \\
m^2_0 & = & 0.15 GeV^2 \pm 0.04 GeV^2 , \nonumber
\eqa
into (\ref{2.96}) allows one to examine the approach to the photoabsorption limit
in (\ref{2.96}). As expected from the logarithmic behavior in (\ref{2.96}), exceedingly
high energies are needed to approach this limit. Compare Table 1 for
a specific example.

\begin{table}[h]
\begin{center}
\begin{tabular}{|c|c|c|} \hline
$Q^2 [GeV^2]$ & $W^2 [GeV^2]$ & 
$\frac{\sigma_{\gamma^*p} (\eta (W^2, Q^2))}{\sigma_{\gamma p} (W^2)}$ \\
\hline \hline
1.5 & $2.5 \times 10^7$ & 0.5 \\ \hline
    & $1.26 \times 10^{11}$ & 0.63 \\ \hline
\end{tabular}
\caption{The approach to $\sigma_{\gamma^*p} (\eta (W^2, Q^2))$ to the
photoproduction limit, $\sigma_{\gamma p} (W^2) = \sigma_{\gamma^*p} 
(\eta (W^2, Q^2 = 0))$, for $W^2\rightarrow\infty$ at fixed $Q^2 > 0$. }
\end{center}
\end{table}

More recently, fits to the low-x DIS data based on various parameterizations
of the photoabsorption cross section of the general form 
\be
\sigma_{\gamma^*p} (W^2, Q^2) \sim l^{\lambda_{eff}(Q^2)}
\label{2.99}
\ee
were examined by Caldwell \cite{CAL}, in particular in view of an extrapolation to the
above limit of large $W$ at fixed $Q^2$. The ansatz (\ref{2.99}), with
\be
l = \frac{1}{2 M_px} \cong \frac{1}{2 M_p} \frac{W^2}{Q^2},
\label{2.100}
\ee
was motivated by the lifetime, or coherence length, of a hadronic 
fluctuation according to (\ref{2.52}).

The particular fit based on
\be
\sigma_{\gamma^*p} (W^2, Q^2) = \sigma_0 (Q^2) l^{\lambda_{eff}(Q^2)},
\label{2.101}
\ee
and individually carried out for a series of values of $Q^2$ in the
interval $0.15 \lsim Q^2 \lsim 400 GeV^2$, led to an intersection of the
straight lines in the representation of the log of 
$\sigma_{\gamma^*p} (W^2, Q^2)$ against the log of the coherence length
$l$. The intersection, interpreted as indication for the approach to 
a $Q^2$-independent
limit at large $W^2$, occured at
\be
W^2 \cong 10^9 Q^2,
\label{2.102}
\ee
to be compared with the (not yet fully asymptotic) results for $W^2$ from our
approach in Table 1. It is of interest that the large-$W$ extrapolation
of a fit to the experimental data based on the simple 
intuitively well-motivated, but still fairly ad-hoc
ansatz (\ref{2.101}) implies a saturation effect similar to the one predicted 
from the CDP, the validity of which stands on firm theoretical grounds.
Not every ansatz for a successful fit in terms of the 
variable $l$ in (\ref{2.100}), however, as pointed out in ref. \cite{CAL}, implies
an approach to a $Q^2$-independent saturation limit. Precise empirical evidence for
the limiting behavior (\ref{2.96}) presumably requires experiments at energies 
substantially above the ones explored at HERA.

\subsection{The CDP, the Gluon Distribution Function and 
Evolution} 

The CDP of DIS corresponds\footnote{With respect to the present Section 2.7, 
compare also ref.\cite{23a}}
to the low-$x$ approximation of the pQCD-improved
parton model in which the
interaction of the (virtual) photon occurs by interaction with the
quark-antiquark sea in the proton via $\gamma^*$ gluon $\rightarrow q \bar q$
fusion, compare fig.6.
\begin{figure}[h!]
\centerline{\epsfig{file=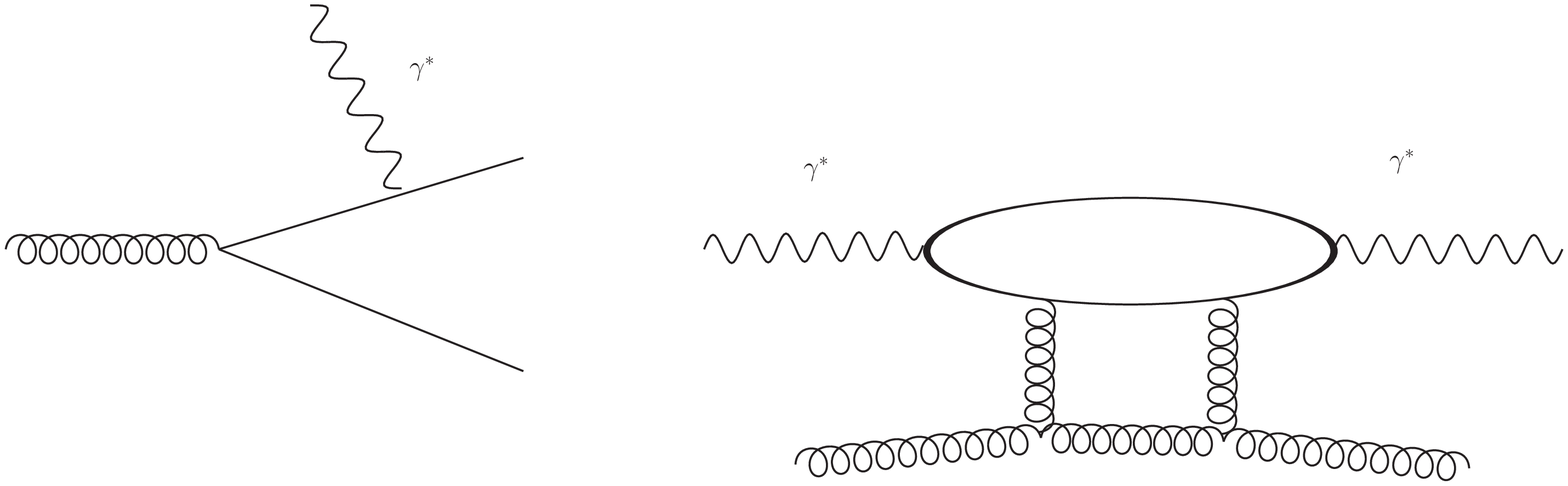,width=7cm}}
\par
\hspace{4cm} (a)  \hspace{3cm} (b)   
\caption{~ \hfill\break  {\footnotesize
(a) Photon-gluon fusion. \hfill\break
(b) Higher order contributions to photon-gluon $\rightarrow q \bar q$ fusion
resolving the lower blob in fig.1. The lower part of the diagram must be
extended to become a gluon ladder.}}
\end{figure}
%Fig.6a und b
The longitudinal structure function, $F_L (x,Q^2)$, in this low-$x$ or CDP
approximation of pQCD solely depends on the gluon density, $g (x, Q^2)$, \cite{Martin}
\be
F_L(x,Q^2) = \frac{\alpha_s (Q^2)}{3\pi} \sum_q Q^2_q \cdot 6 I_g (x, Q^2)
\label{(a)}
\ee
with 
\be
I_g (x, Q^2) \equiv \int^1_x \frac{dy}{y} \left( \frac{x}{y}\right)^2 \left( 1
  - \frac{x}{y} \right) y g (y , Q^2) .
\label{(b)}
\ee
where $G (y, Q^2) \equiv y g (y, Q^2)$. For a wide range of different gluon
distributions, independently of their specific form, the integration in
(\ref{(b)}) yields a result that is proportional to the gluon density at a
rescaled value $x / \xi_L$  \cite{Martin} i.e. 
\be 
F_L ( \xi_L x, Q^2) = \frac{\alpha_s (Q^2)}{3\pi} \sum_q Q^2_q G (x, Q^2).
\label{(c)}
\ee
The rescaling factor $\xi_L$ in (\ref{(c)}) has the preferred value of $\xi_L
\cong 0.40$ \cite{Martin}. The interaction of the longitudinally polarized photon with the
quark (antiquark) originating from gluon $\rightarrow q \bar q$ splitting, via
$F_L (\xi_L x, Q^2)$, in good approximation thus fully determines the $x$ and
$Q^2$ dependence of the gluon distribution function. 

We turn to the structure function $F_2 (x, Q^2)$. In the DIS scheme of pQCD, at
low $x$ and sufficiently large $Q^2$, $F_2 (x, Q^2)$ is proportional to the
singlet or sea quark distribution $\sum(x, Q^2)$, 
\be
\sum (x, Q^2) = \sum^{n_f}_{q=1} (q_q (x) + \bar q_q (x)).
\label{(d)}
\ee
For four flavors of quarks, $n_f = 4$, and flavor-blind quark distributions, 
the structure function is given by 
\be
F_2 (x, Q^2) = x\sum (x, Q^2) \frac{1}{4} \sum_q Q^2_q = \frac{5}{18} x \sum (x
, Q^2) .
\label{(e)}
\ee
In the CDP approximation, $\gamma^*$ gluon $\rightarrow q \bar q$ fusion, the
evolution of $F_2 (x , Q^2)$ with $Q^2$ is determined by the gluon distribution
according to \cite{Lipatov}
\be
\frac{\partial F_2 (x, Q^2)}{\partial \ln Q^2} = \frac{\alpha_s (Q^2)}{\pi}
\sum_q Q^2_q \int^1_x dz P_{q g} (z) G \left( \frac{x}{z} , Q^2 \right) , 
\label{e-a}
\ee
where in leading order of pQCD
\be
P_{q g} (z) = P_{q g}^{(0)} = \frac{1}{2} ( z^2 + (1 - z)^2) .
\label{e-b}
\ee
The evolution equation (\ref{e-a}), again for a wide range of choices for the
gluon distribution, may be represented by the proportionality \cite{Prytz}
\be
\frac{\partial F_2 (\xi_2 x , Q^2)}{\partial \ln Q^2} = \frac{\alpha_s
  (Q^2)}{3\pi} \sum_q Q^2_q G  (x ,Q^2) .
\label{(f)}
\ee
The rescaling factor in this case is given by $\xi_2 \cong 0.50$ \cite{Prytz}. 

The validity of (\ref{(c)}) and (\ref{(f)}) and the values of the rescaling
factors $(\xi_L , \xi_2) = (0.40 , 0.50)$ will be reexamined below by evaluating
the relations (\ref{(a)}) for $F_L (x , Q^2)$ and (\ref{e-a}) for $F_2 (x,
Q^2)$ for the specific gluon distribution to be obtained by requiring
consistency with the CDP approach.   

We introduce the ratio of $F_2 (x, Q^2)$ and $F_L (x , Q^2)$ by employing
the form of this ratio in (\ref{2.51} ), 
but allowing for a potential dependence of $\rho \equiv \rho (x, Q^2)$ on the
kinematic variables $x$ and $Q^2$, 
\be
F_L (x,Q^2) = \frac{1}{2 \rho + 1} F_2 (x , Q^2).
\label{(g)}
\ee
Replacing the right-hand side of (\ref{(f)}) by $F_L (\xi_L x , Q^2)$ from
(\ref{(c)}), and subsequently replacing $F_L (\xi_L x, Q^2)$ by $F_2 (\xi_L x,
Q^2)$ 
according to the
defining relation (\ref{(g)}), the evolution equation (\ref{(f)}) becomes 
\be
(2 \rho + 1) \frac{\partial}{\partial \ln Q^2} F_2 \left( \frac{\xi_2}{\xi_L}
 x, Q^2 \right) = F_2 (x, Q^2),
\label{(h)}
\ee
or, in terms of the flavor singlet distribution (\ref{(d)}) according to (\ref{(e)}), 
\be
(2 \rho + 1) \frac{\partial}{\partial \ln Q^2} \frac{\xi_2}{\xi_L} \sum \left( \frac{\xi_2}{\xi_L}
x , Q^2\right) = \sum (x, Q^2) . 
\label{(i)}
\ee

By alternatively replacing $F_L (x, Q^2)$ in (\ref{(g)})
by the gluon distribution from (\ref{(c)}), upon inserting the resulting
expression for $F_2 (x, Q^2)$ into the evolution equation (\ref{(f)}), we find
an evolution equation for the gluon density that reads 
\be
\frac{\partial}{\partial \ln Q^2} (2\rho + 1) \alpha_s (Q^2) G
\left( \frac{\xi_2}{\xi_L} x, Q^2 \right) = \alpha_s (Q^2) G (x , Q^2) . 
\label{(j)}
\ee
Comparing (\ref{(j)}) with (\ref{(i)}), we conclude: if and only if 
\be
(2\rho + 1) = {\rm const.} , 
\label{(k)}
\ee
the evolution of the gluon density multiplied by $\alpha_s (Q^2)$  
in (\ref{(j)}) coincides with the quark-singlet
evolution according to (\ref{(h)}) and (\ref{(i)}). 

Identical evolution of the $q \bar q$ sea originating from $\gamma^*$ gluon $\rightarrow q
\bar q$ fusion (fig.6a) and the gluon distribution multiplied by $\alpha_s (Q^2)$
appears as natural consequence of the fact that the $q \bar q$ state seen by
the photon originates from the gluon: 
the evolution of the sea distribution, measured by the interaction with the
photon, directly yields the evolution of the gluon distribution. 

In the CDP, according to Section 2.4, specifically according to (\ref{2.55}), 
and supported by the experimental results in fig.4, the structure function $F_2 (x, Q^2)$
for $x < 0.1$ and $Q^2$ sufficiently large, depends on the single variable $W^2$, 

\be
F_2 (x, Q^2) = F_2 (W^2 = \frac{Q^2}{x}) . 
\label{(l)} 
\ee
Independently of the specific form of the functional dependence of $F_2 (x,
Q^2)$ on $W^2$, according to (\ref{(l)}), the $Q^2$ dependence and the $x$
dependence of $F_2 (x, Q^2)$ are intimately related to each other. This is a
consequence of the $W$ dependence of the dipole cross section in (\ref{2.1}),
compare 
(\ref{2.76}) and (\ref{2.53}). 
In terms of the energy variable $W$, the evolution equation (\ref{(h)}) becomes 
\be
(2 \rho_W + 1) \frac{\partial}{\partial \ln W^2} F_2 \left( \frac{\xi_L}{\xi_2}
  W^2 \right) = F_2 (W^2). 
\label{(m)}
 \ee
Since according to (\ref{2.1}) the longitudinal as well as the transverse
photoabsorption cross section depend on $W^2$, also the potential dependence of
$\rho$ on $x$ and $Q^2$ is restricted to $W$, and in (\ref{(m)}) , this is indicated by
$\rho_W$. 

We assume a power-law dependence for $F_2 (W^2)$ on $W^2$, 
\be
F_2 (W^2) \sim (W^2)^{C_2} = \left( \frac{Q^2}{x} \right)^{C_2} . 
\label{(n)}
\ee
 We note that the dependence of $F_2 (x, Q^2) = F_2 (W^2)$ in (\ref{(n)}) on a fixed
 (i.e. $Q^2$-independent) constant power $C_2$ of $1/x$ coincides with the
 $x\rightarrow 0$ so-called ``hard'' Pomeron solution \cite{Adel} of pQCD that
 rests on a $(1/x)^\lambda$ input assumption for the flavor-singlet quark as
 well as the gluon distribution ($\lambda = {\rm const})$. A fixed power of $1/x$, as
 $(1/x)^{\epsilon_0}$, also appears in the Regge approach to DIS based on a
 linear combination of a ``soft'' and a ``hard'' Pomeron, with the fit
 parameter of the hard Pomeron contribution being given by $\epsilon_0 \cong
 0.43$ \cite{Dom}.

It is a unique feature of the CDP, however, that the $1/x$ dependence and the
$Q^2$ dependence of $F_2 (x, Q^2)$ (for $x < 0.1$ and $Q^2$ sufficiently large,
$Q^2 \ge 10 {\rm GeV}^2$) are determined by one and
the same constant power $C_2$, compare (\ref{(n)}). 

Inserting the power-law (\ref{(n)}) into the evolution equation (\ref{(m)}), we
find the constraint 
\be
(2\rho_W + 1) C_2 \left( \frac{\xi_L}{\xi_2} \right)^{C_2} = 1 . 
\label{(o)}
\ee
Consistency of the power law (\ref{(n)}) for the $W$ dependence with the
flavor-singlet evolution (\ref{(m)}) thus implies the remarkable constraint
(\ref{(o)}) that connects the exponent $C_2$ of the $1/x$ dependence with the
longitudinal-to-transverse ratio of the photoabsorption cross sections, $2
\rho_W$, or, equivalently with the ratio of $F_2 (x, Q^2)$ and $F_L (x, Q^2)$
in (\ref{(g)}). 
Constancy of $C_2$ implies constancy of $\rho_W = \rho = {\rm const}$, and vice
versa. 

In the CDP, from (\ref{2.43I}),
$\rho$ has the constant and fixed value of $\rho = 4/3$. With this CDP value
of $\rho = 4/3$, we find from (\ref{(o)}) (compare also
\cite{23a})\footnote{Note that (\ref{(p)}) differs from the result in
  \cite{KU-SCHI} by taking into account the rescaling factor $\xi_L = 0.4$ as
  well as $\rho = 4/3$.}
\be
C_2 = \frac{1}{2\rho + 1} \left( \frac{\xi_2}{\xi_L} \right)^{C_2} = 
0.29
\label{(p)}
\ee 
where the preferred value of $\xi_2 / \xi_L = 0.5 / 0.4 = 1.25$ was
inserted. We note that the (ad hoc) variation of this value in the interval $1
\le \xi_2 / \xi_L \le 1.5$ around the above value of $\xi_2 / \xi_L = 1.25$
yields $0.27 \le C_2 \le 1.31$. The result $C_2 = 0.29$ accordingly is fairly
insensitive under variation of the rescaling factors $\xi_2$ and $\xi_L$. 

We specify (\ref{(n)}) 
by adopting the theoretical result for the exponent $C_2 = 0.29$ from (\ref{(p)})
and by
introducing a proportionality constant, $f_2$, 
\be
F_2 (W^2) = f_2 \cdot \left( \frac{W^2}{1{\rm GeV}^2} \right)^{C_2 = 0.29} . 
\label{(q)}
\ee
Via an eye-ball fit to the experimental data for $F_2 (W^2)$ as a function
of $1/W^2$ in fig. 4a, we find 
\be
f_2 = 0.063 . 
\label{(r)}
\ee
The theoretical prediction (\ref{(q)}) with (\ref{(r)}) is shown in fig.4a. 
A detailed comparison with the experimental data, separately
for distinct values of $Q^2$ in the relevant range of $10 {\rm GeV}^2 \le Q^2
\le 100 {\rm GeV}^2$ shows agreement with the single-free-parameter fit
(\ref{(q)}) to the structure function $F_2 (W^2)$ in (\ref{(q)}) for
$10{\rm GeV}^2 \le Q^2 \le 100{\rm GeV}^2$. Compare the discussion in Section 5, in particular
figs.16 and 17. 

According to (\ref{(e)}), the flavor-singlet quark or sea distribution is proportional to the structure
function $F_2 (W^2)$, 
\be
x \sum (x, Q^2) = \frac{18}{5} f_2 \cdot \left( \frac{W^2}{1 {\rm
      GeV}^2}\right)^{C_2 =0.29}.
\label{(s)}
\ee 
Employing the proportionality (\ref{(c)}) of the gluon distribution to the
longitudinal structure function $F_L (\xi_L x, Q^2)$, and expressing $F_L
(\xi_L x, Q^2)$ in terms of $F_2 (\xi_L x, Q^2)$ according to (\ref{(g)}), we
find that also the gluon distribution can be directly deduced from the
experimental data for the structure function $F_2 (x, Q^2) = F_2 (W^2 = Q^2 /
x)$, 
\bqa
\alpha_s (Q^2) G (x, Q^2) & = & \frac{3\pi}{\sum_q Q^2_q} F_L (\xi_L x , Q^2)
\nonumber \\
 & = & \frac{3\pi}{\sum_q Q^2_q} \frac{1}{(2\rho + 1)} F_2 (\xi_L x, Q^2) 
\label{(t)} \\
& = & \frac{3\pi}{\sum_q Q^2_q (2\rho + 1)} \frac{f_2}{\xi_L^{C_2 = 0.29}} \left(
  \frac{W^2}{1 {\rm GeV}^2} \right)^{C_2 = 0.29} , \nonumber
\eqa
where (\ref{(q)}) was inserted in the last step. 

This is the appropriate point to add a remark, as previously announced, on the
validity of the representations (\ref{(c)}) and (\ref{(f)}) in terms of the
rescaling factors $(\xi_L, \xi_2)$.
It will turn out that indeed without loss of generality (\ref{(a)}) and
(\ref{e-a}) for our gluon distribution may be replaced by (\ref{(c)}) and
(\ref{(f)}).  

Inserting the gluon distribution (\ref{(t)}) into the representations of $F_L
(x, Q^2)$ and $F_2 (x, Q^2)$ in (\ref{(a)})  and (\ref{e-a}), one may explicitly
test the validity of the proportionalities to the gluon distribution in
(\ref{(c)}) and (\ref{(f)}) that originate from (\ref{(a)}) and (\ref{e-a}). 
One finds that the above choice of the rescaling
factors, $(\xi_L, \xi_2)= (0.4, 0.5)$, yields a small discrepancy between the 
evaluation of the integrals over the gluon distribution and the representation
in terms of the rescaling factors that amounts to about 4\% and 6.5 \% for
$F_L (x , Q^2)$ and $F_2 (x, Q^2)$, respectively. The discrepancy is reduced to
less than 0.5\% , for the choice of $(\xi_L , \xi_2) = (0.45, 0.40)$. This
implies a change of $C_2 = 0.29$ to $C_2 = 0.26$ in (\ref{(p)}), close to the
value of $C_2 = 0.27 \pm 0.01$ found in the fit in refs.\cite{DIFF2000,SCHI}.
For the
comparison with the experimental data, the difference between $C_2 = 0.26$ and
$C_2 = 0.29$ is not very important. We use
$C_2 = 0.29$ in fig.4a and in the more extensive comparison with the
experimental data in figs.16 and 17 in Section 5. 

\begin{figure}
\epsfig{file=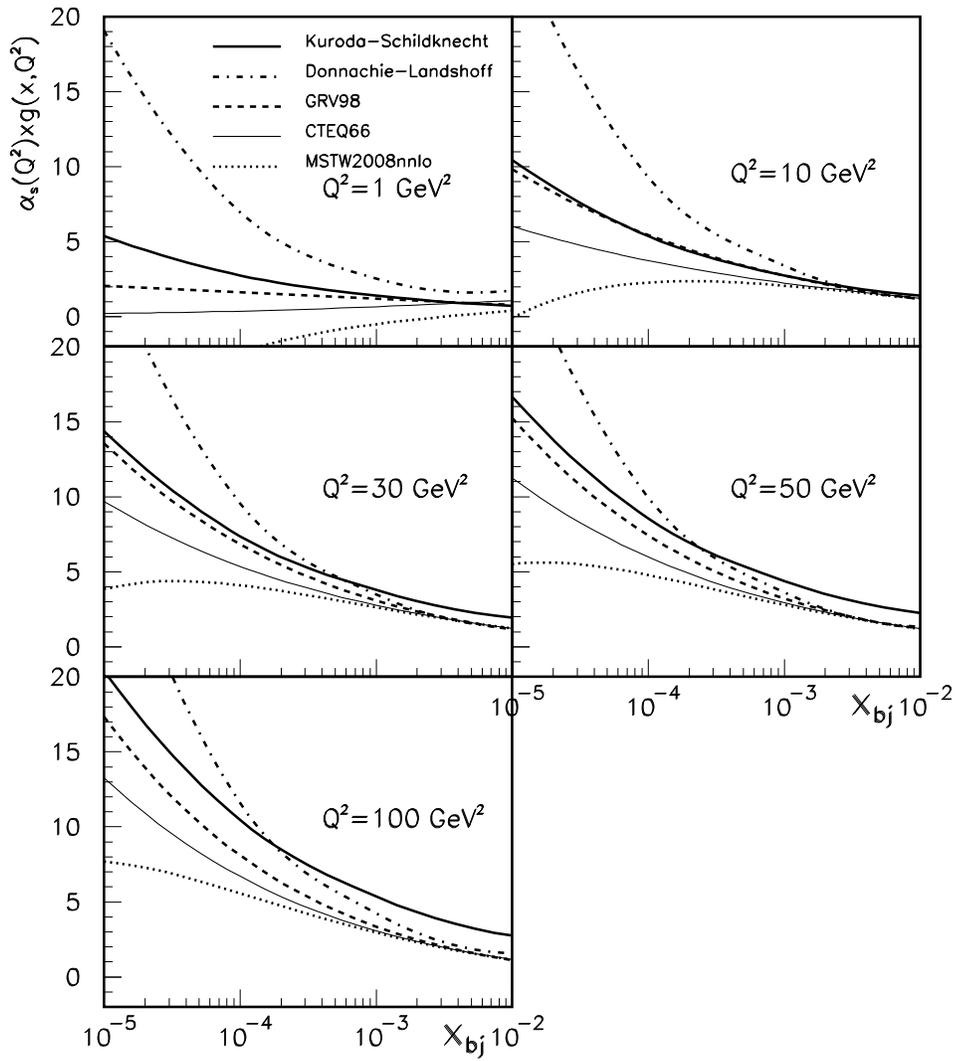,width=15.5cm}   
%\vspace{1.5cm}
\caption{\footnotesize
The gluon-distribution function from (\ref{(t)}) compared with the gluon
distributions from the hard-Pomeron part of a Regge fit \cite{Dom} to $F_2 (x,
Q^2)$,  and from the $F_2 (x, Q^2)$ fits GRV \cite{GRV},
CTEQ \cite{28a} and MSTW\cite{28b}.}
\end{figure}
%Fig.7a und b
In fig.7, we compare our gluon distribution from (\ref{(t)}) with various
gluon distributions obtained in fits to the experimental results for $F_2 (x, Q^2)$. 
Compare refs.\cite{Dom,GRV} and the Durham Data Base \cite{Dur}\footnote{The
  gluon-distribution functions in fig.7 marked GRV, MSTW and CTEQ were
  extracted from the Durham Data Base \cite{Dur}}. 
The gluon distributions from the various fits
were multiplied by $\alpha_s (Q^2)$, where 
\be
\alpha_s^{NLO} (Q^2) = \frac{1}{bt} \left[ 1 - \frac{b^\prime \ln t}{bt}\right]   . 
\label{(u)}
\ee
with
\be
b = \frac{33 - 2 n_f}{12\pi} , ~~~b^\prime = \frac{153 - 19 n_f}{2\pi (33 -
  2n_f)} , 
\label{(v)}
\ee
and $t = \ln (Q^2 / \Lambda^2_{QCD} ), n_f = 4$ and $\Lambda_{QCD} = 340 {\rm
  MeV}$ corresponding to $\alpha_s (M^2_Z) = 0.113$.

According to fig.7, there is a considerable spread between the gluon-distribution
functions extracted from experimental data of the structure function $F_2 (x,
Q^2)$ by different collaborations. The gluon-distribution function
corresponding to the hard Pomeron of the Regge fit \cite{Dom} in general lies
above our result. The results from the so-called global analysis by the
CTEQ \cite{28a} and MSTW \cite{28b} collaborations are lower than ours. The
fact that our results are fairly close to the results from GRV \cite{KU-SCHI} seems no
accident and deserves further examination. 

Our relation (\ref{(t)}) obtained as a consequence of the low-$x$ pQCD
approximations (\ref{(a)}) and (\ref{e-a}) and the $W$ dependence of $F_{L,2}
(x, Q^2) = F_{L,2} (W^2)$ from the CDP is transparent and simple as far as the
underlying assumptions are concerned. 
The extracted gluon distribution only depends
on the single normalization parameter $f_2$ that was adjusted to
the experimental data. 
The gluon distribution can directly be read off from the experimental data for
$F_2 (W^2 = Q^2 / x)$ shown in fig.4 by multiplication of these data with the
constant given in (\ref{(t)}). 

We end this Section with the following summarizing comments: 
\begin{itemize}
\item[i)] 
The starting point for our extraction of the gluon distribution is the low-$x$
approximation of the pQCD-improved parton model that relates the gluon
distribution to the longitudinal structure function, $F_L (x, Q^2)$, compare
(\ref{(a)}). This relation is supplemented by the $W$-dependence of the
structure functions $F_L (W^2 = Q^2 / x)$ and $F_2 (W^2 = Q^2 / x)$ and their
proportionality via the constant factor of $1/(2\rho + 1)$, both the $W$
dependence and the proportionality being extracted from the CDP and being
supported empirically. Finally, a power-law dependence, $F_2 \sim (W^2)^{C_2} =
(Q^2 / x)^{C_2}$ is inserted, with $C_2 = 0.29$ predicted from sea-quark
evolution. The extraction of the gluon distribution depends on only one fitted
normalization constant, $f_2$. 

\item[ii)]
The gluon distribution resulting from (\ref{(t)}) lies within the range of
gluon distributions available in the literature. We note that our extraction of
the gluon distribution from the data on $F_2 (x, Q^2) = F_2 (W^2 = Q^2 / x)$ is
not based on a resolution of the $ggpp$ vertex, the lower blob in fig.1. The
consistency of our gluon distribution with the ones in the literature indicates
that the gluon distribution does not as sensitively depend on details of the
structure of the $ggpp$ vertex as usually expected, assumed or elaborated
upon. Compare the BFKL approach \cite{Kur} to DIS at low $x$, as well as the double
asymptotic scaling (DAS) solution \cite{25a, 25b, Cavola} of DGLAP evolution \cite{Lipatov}    
based on replacing the unresolved lower part of the diagram in fig.1 by the lower part of
the diagram in fig.6b that has to be extended by a gluon ladder. 
We conjecture that our gluon distribution nevertheless, in the sense of a
numerical approximation, is consistent with the DGLAP gluon evolution equation at low
$x$ that supplements the evolution of the flavor-singlet quark distribution
solely employed in our analysis. 

\item[iii)]
As mentioned, our $1/x$ dependence (\ref{(t)}), $(2\rho + 1) \alpha_s (Q^2) xg (x, Q^2)
\sim x \sum (x , Q^2) \sim (1 / x)^{C_2}$ with fixed exponent $C_2$, is closely
related to DGLAP evolution with the input constraint of a hard Pomeron \cite{Adel}. 
We differ from ref.\cite{Adel}
insofar, as we have the necessary constraint of  $(2 \rho + 1) = {\rm const}$,
compare (\ref{(o)}), while the analysis of
ref.\cite{Adel} led to $(2\rho + 1) \alpha_s (Q^2) = {\rm const.}$. 

\item[iv)]
Our $(1 / x)^{C_2}$ dependence is analogous to the $(1 / x)$ dependence of the
hard Pomeron component of the Regge approach \cite{Dom}. However, we predict $C_2 = 0.29$
from sea-quark evolution, the value being consistent with experiment, while the analogous
parameter $\epsilon_0\cong 0.43$ in the Regge approach is a pure fit-parameter. Moreover,
the CDP contains a smooth transition to low $Q^2$, including $Q^2= 0$, rather
than relying on the addition of a soft Pomeron. In the language of Pomeron
exchange, the CDP only knows of a single Pomeron which is relevant for both small
and large values of $Q^2$.

\end{itemize}

\section{Models for the Dipole Cross Section}

\renewcommand{\theequation}{\arabic{section}.\arabic{equation}}
\setcounter{equation}{0}

In Section 2, without adopting a specific parameterization for the dipole 
cross section, we found the proportionalities (\ref{2.93}) of the total
photoabsorption cross section to $\ln (1/\eta (W^2, Q^2))$, for 
$\eta (W^2, Q^2) \ll 1$, and to $1/\eta (W^2, Q^2)$ for $\eta (W^2, Q^2) \gg
1$. Any specific parameterization of the dipole cross section
has to interpolate between these two limits.

In Section 3.1, we will remind ourselves of a previously employed ansatz for the
dipole cross section that implies $R(W^2, Q^2) = 1/2$ at large $Q^2$
for the ratio of $R (W^2,Q^2) = \sigma_{\gamma^*_Lp} (W^2, Q^2)/
\sigma_{\gamma^*_Tp} (W^2, Q^2)$. In Section 3.2, we introduce a more
general ansatz that allows for the transverse-size reduction and associated
enhancement of the transverse relative to the longitudinal photoabsorption
cross section from Section 2.3.

\subsection{A Dipole Cross Section Implying \boldmath$R = 0.5$\unboldmath}

The ansatz for the dipole cross section in (\ref{2.1}), previously employed in 
a successful fit to the experimental data on the total cross section,
$\sigma_{\gamma^* p} (W^2, Q^2)$, is given by \cite{DIFF2000}
\be
\sigma_{(q \bar q)p} (\vec r_\bot, z (1-z), W^2) = \sigma^{(\infty)} (W^2)
\left( 1 - J_0 \left( r_\bot \sqrt{z (1-z)} \Lambda_{sat} (W^2) \right)
\right),
\label{3.1}
\ee
where $\sigma^{(\infty)} (W^2)$ is of hadronic size and weakly
dependent on $W$, while $\Lambda^2_{sat} (W^2)$ increases as a small
power of $W^2$. Since the cross section (\ref{3.1}) depends on the product
$\vec r^{~\prime}_\bot = \vec r_\bot \sqrt{z (1-z)}$, the longitudinal
and transverse $J=1$ projections in (\ref{2.36}) become identical,\footnote{For
clarity, in terms of $(q \bar q)^{J=1}$ helicities, $\bar \sigma_{(q \bar q)^{J=1}_L} \equiv
\bar \sigma_{(q \bar q)^{J=1}_0}$ and $\bar \sigma_{(q \bar q)^{J=1}_T}
\equiv \frac{1}{2} \left( \bar \sigma_{(q \bar q)^{J=1}_+}  +
\bar \sigma_{(q \bar q)^{J=1}_-} \right) = \bar \sigma_{(q \bar
q)^{J=1}_+}$}
\bqa
&&\sigma_{(q \bar q)^{J=1}_L p} \left( r^\prime_\bot \Lambda_{sat} (W^2)\right)
= \sigma_{(q \bar q)^{J=1}_T p} \left( r^\prime_\bot \Lambda_{sat} (W^2) \right)
\nonumber \\
&& = \sigma^{(\infty)} (W^2) \left( 1 - J_0 (r^\prime_\bot \Lambda_{sat}
(W^2) \right) \label{3.2} \\
&& = \sigma^{(\infty)} (W^2) \left\{ \matrix{
\frac{1}{4} \vec r^{~\prime 2}_\bot \Lambda^2_{sat} (W^2),~~~{\rm for} 
~\vec r^{~\prime 2}_\bot \Lambda^2_{sat} (W^2) \to 0,\cr
1,~~~~~~~~~~~~~~~~~~~~~{\rm for}~ \vec r^{~\prime 2}_\bot \Lambda^2_{sat} (W^2)
\to \infty. } \right.\nonumber
\eqa

With respect to momentum space, the ansatz (\ref{3.1}), according to
(\ref{2.2}), corresponds to
\be
\tilde \sigma (\vec l^{~2}_\bot, z (1-z), W^2) = 
\frac{\sigma^{(\infty)} (W^2)}{\pi}
\delta \left(\vec l^{~2}_\bot - z (1-z) \Lambda^2_{sat} (W^2) \right).
\label{3.3}
\ee
Its $J=1$ projections, according to (\ref{2.41}), are given by
\bqa
&& \bar \sigma_{(q \bar q)^{J=1}_L p} (\vec l^{~\prime 2}_\bot, W^2) =
\bar \sigma_{(q \bar q)^{J=1}_T p} (\vec l^{~\prime 2}_\bot , W^2) = \nonumber \\
&& = \frac{\sigma^{(\infty)} (W^2)}{\pi} \delta \left(\vec l^{~\prime 2}_\bot
- \Lambda^2_{sat} (W^2) \right).
\label{3.4}
\eqa
Substitution of (\ref{3.3}) and (\ref{3.4}) into (\ref{2.2}) and (\ref{2.40}), respectively, takes
us back to (\ref{3.1}) and (\ref{3.2}).

We remark that helicity independence, the equality of the cross sections for
scattering of the  $J=1$ projections for
longitudinally and transversely polarized $(q \bar q)^{J=1}$ states in 
(\ref{3.2}) and (\ref{3.4}), is a general consequence of the dependence of the
ansatz (\ref{3.1}) on the variable $r_\bot^\prime = r_\bot \sqrt{z(1-z)}$. 
Any dipole cross section in (\ref{2.1}) fulfilling 
\bqa
\sigma_{(q\bar q)p} (\vec r_\bot , z(1-z), W^2) = \sigma_{(q \bar q)p} \left(
  \vec r_\bot \sqrt{z(1-z)}, W^2\right) ,
\label{3.5}
\eqa
together with color transparency (\ref{2.2}), implies helicity independence and
$R(W^2, Q^2) = 1/2$ at large $Q^2$. 
Indeed, consistency of (\ref{3.5}) with (\ref{2.2}), 
\bqa
& & \sigma_{(q \bar q)p} (r_\bot \sqrt{z(1-z)}, W^2) = \int d^2 \vec
l_\bot^{~\prime} z(1 -z)\tilde\sigma (\vec l_\bot^{~\prime 2} z (1-z), z(1-z),
W^2) \cdot \nonumber \\ 
& & \cdot \left( 1 - e^{-i \vec l_\bot^{~\prime} \cdot \vec
    r_\bot^{~\prime}}\right) ,
\label{3.6}
\eqa
requires $z(1-z) \tilde\sigma (\vec l_\bot^{~\prime 2} z (1-z), z(1-z), W^2)$
to be independent of $z(1-z)$. Under this constraint, (\ref{2.41}) implies
helicity independence and $R(W^2, Q^2) = 1/2$ according to (\ref{2.43}). 

The ansatz (\ref{3.1}) for the dipole cross section has to be supplemented by a
constraint on the masses of the contributing $q \bar q$ fluctuations that is
best incorporated by returning from transverse position space to momentum space. The
constraint reads 
\be
m^2_0 \le M^2_{q \bar q}, M^{\prime 2}_{q \bar q} \le m^2_1 (W^2), 
\label{3.7}
\ee
where the notation, i.e. $M^2_{q \bar q}, M^{\prime 2}_{q \bar q}$, 
for the masses of the $(q \bar q)$ dipole states, indicates that incoming and
outgoing $q \bar q$ masses in the forward Compton amplitude of fig.1 do not
necessarily agree with each other. The lower bound, $m_0^2$, depends on the
flavor of the actively contributing quarks. For up and down quarks the value of
$m_0$ must be somewhat below the $\rho^0$ mass. The upper bound, $m^2_1 (W^2)$,
depends on the available energy. In most applications of the CDP, the
approximation of $m^2_1 (W^2)\rightarrow \infty$ is employed that restricts the
kinematic range 2.of applicability of the CDP. For the present discussion we put
$m^2_1 (W^2) \rightarrow \infty$. We will come back to a finite value of $m^2_1
(W^2)$ in Section 4. 

According to dimensional analysis, with $m^2_1 (W^2)\rightarrow\infty$, the
photoabsorption cross section resulting from (\ref{3.1}) in addition to the
dependence on $\eta (W^2, Q^2) = (Q^2 + m^2_0) / \Lambda^2_{\rm sat} (W^2)$
will depend on $m^2_0 / \Lambda^2_{\rm sat} (W^2)$. For the realistic case of
$m^2_0 / \Lambda^2_{\rm sat} (W^2) \ll 1$, the total photoabsorption cross
section $\sigma_{\gamma^*p} (W^2, Q^2) = \sigma_{\gamma^*_T p} (W^2, Q^2) +
\sigma_{\gamma^*_L p} (W^2, Q^2)$ takes the remarkably simple explicit form \cite{DIFF2000}
\bqa
\sigma_{\gamma^* p} (W^2, Q^2)& = & 
\sigma_{\gamma^* p} (\eta (W^2, Q^2)) + O \left( \frac{m^2_0}{\Lambda^2_{\rm
      sat} (W^2)} \right) =  \nonumber \\ 
& &= \frac{\alpha R_{e^+ e^-}}{3\pi} \sigma^{(\infty)} (W^2) I_0 (\eta) + O \left(
  \frac{m^2_0}{\Lambda^2_{\rm sat} (W^2)} \right) , 
\label{3.8}
\eqa
where
\bqa
I_0 (\eta (W^2, Q^2)) & = & 
\frac{1}{\sqrt{1 + 4\eta (W^2, Q^2)}} \ln \frac{\sqrt{1 + 4 \eta (W^2, Q^2)}
  +1}{\sqrt{1+4\eta(W^2, Q^2)}-1} \cong  \label{3.9}  \\
& \cong& \left\{  \matrix{  \ln \frac{1}{\eta(W^2, Q^2)} + O (\eta \ln \eta ), 
~~{\rm for}~ \eta (W^2, Q^2) \rightarrow \frac{m^2_0}{\Lambda^2_{\rm sat}
  (W^2)}, \cr
\frac{1}{2\eta (W^2, Q^2)} + O \left( \frac{1}{\eta^2}\right) , ~ {\rm
  for}~ \eta (W^2 , Q^2) \rightarrow \infty ,  } \right. \nonumber
\eqa
and 
\be
R_{e^+ e^-} = 3 \sum_q Q^2_q .
\label{3.10}
\ee
As expected, since (\ref{3.1}) fulfills color transparency, compare
(\ref{3.2}), the result
(\ref{3.8}) with (\ref{3.9}) and $\sigma^{(\infty)}(W^2)\cong {\rm const}$ 
constitutes an example for the general result
in (\ref{2.80}) and 
(\ref{2.90}) from Section 2.5. 

For further reference, we give the explicit parameterization of the
ansatz (\ref{3.1}) and the values of the parameters obtained in the fit to the
experimental data. The ``saturation scale'', $\Lambda^2_{sat} (W^2)$ is
given by \cite{DIFF2000,SCHI}
\be
\Lambda^2_{sat} (W^2) = B \left( \frac{W^2}{W^2_0} + 1 \right)^{C_2},
\label{3.11}
\ee
with
\bqa
B &=& 2.24 \pm 0.43~GeV^2 , \nonumber \\
W^2_0 &=& 1081 \pm 124~GeV^2 ,  \\
C_2 &=& 0.27 \pm 0.01 . \nonumber 
\label{3.12}
\eqa
In good approximation, (\ref{3.11}) becomes
\be
\Lambda^2_{sat} (W^2) = C_1 \left( \frac{W^2}{1~GeV^2} \right)^{C_2},
\label{3.13}
\ee
with 
\be
C_1 = 0.34 \pm 0.06 {\rm GeV}^2 
\label{3.14}
\ee
i.e. $\Lambda^2_{{\rm sat}}(W^2)$ is in good approximation determined by only
two parameters, the normalization scale $C_1$ and the exponent $C_2$. 

The hadronic cross section, $\sigma^{(\infty)} (W^2)$, was obtained \cite{DIFF2000} by requiring
consistency with the Regge fit to the measured $Q^2 = 0$ photoproduction cross section.
It determines the product of $R_{e^+e^-} \sigma^{(\infty)} (W^2)$,
where $R_{e^+e^-} = 3 \sum_q Q^2_q$. With three active flavors,
$R_{e^+e^-} = 2$, and 
\be
\sigma^{(\infty)} (W^2) \cong 30 mb = 77.04~GeV^{-2}.
\label{3.15}
\ee
The value of the lower bound, $m^2_0$, in (\ref{3.7}) is given by 
\be
m^2_0 = 0.15 \pm 0.04 GeV^2 . 
\label{3.16}
\ee

\subsection{The Ansatz for the Dipole Cross Section implying
\boldmath $R = 1/2\rho (\epsilon)$\unboldmath}

Returning to the discussion in Section 2, compare in particular (\ref{2.16}),
we generalize (\ref{3.3}) to become
\footnote{The quantities $\bar \sigma^{(\infty)}
(W^2)$ and $\bar \Lambda^{~2}_{sat} (W^2)$ are proportional to
$\sigma^{(\infty)} (W^2)$ and $\Lambda^2_{sat} (W^2)$ introduced by the
defining relations (\ref{2.71}) and (\ref{2.72}). The constant proportionality factors will be
given below.}
\be
\tilde \sigma (\vec l^{~2}_\bot, z(1-z), W^2) =   
\frac{\bar\sigma^{(\infty)} (W^2)}{\pi} \delta \left( \vec l^{~2}_\bot -
\frac{1}{6} \bar \Lambda^2_{sat} (W^2) \right)  \Theta (z (1-z) - \epsilon).       
\label{3.17}
\ee
With respect to transverse position space, according to (\ref{2.2}), we obtain
from (\ref{3.17}), 
\bqa
& & \hspace{-0.8cm}  \sigma_{(q \bar q) p} (r_\bot, z(1-z), W^2) = \bar\sigma^{(\infty)} (W^2) \left( 1
  - J_0 (r_\bot \frac{\bar\Lambda_{\rm sat}(W^2)}{\sqrt 6} ) \right) \Theta (z
(1-z) - \epsilon ) \nonumber \\
& & \hspace{-0.8cm}  \cong \bar\sigma^{(\infty)} (W^2) \Theta (z (1-z)-\epsilon) 
\left\{ \matrix{ \frac{1}{4} \frac{\bar\Lambda^{~2}_{\rm sat} (W^2)}{6} \vec
    r_\bot^{~2}, & {\rm for}~ \vec r_\bot^{~2} \rightarrow 0, \cr
1 , & {\rm for}~ \vec r_\bot^{~2} \rightarrow \infty . }  \right.
\label{3.18} 
\eqa

The $\delta$-function in (\ref{3.17}), via 
$\bar \Lambda^{~2}_{sat} (W^2)$, specifies the $W$-dependence of the integral
$\int d \vec l^{~2}_\bot \vec l^{~2}_\bot~ \tilde \sigma (\vec l^{~2}_\bot,
W^2)$ that, according to (\ref{2.18}), determines the photoabsorption cross
section at large $Q^2$. The $\Theta$-function in (\ref{3.17}), compare
(\ref{2.16}), provides the necessary
$W$-dependent cut on $\vec l^{~\prime 2}_\bot = \vec l^{~2}_\bot / z(1-z)$. It forbids
$q \bar q$ fluctuations of infinitely large mass to occur
as a result of gluon absorption at finite energy, $W$.
The $J=1$ projections of the ansatz (\ref{3.17}), by substitution of (\ref{3.17}) into
(\ref{2.41}), are found to be given by  
\bqa
&&\bar \sigma_{(q \bar q)^{J=1}_{L,T}p} \left( \vec l^{~\prime 2}_\bot, 
\bar \Lambda^{~2}_{sat} (W^2) \right)  = f_{L,T} \left(\vec l^{~\prime 2}_\bot,
\bar \Lambda^{~2}_{sat} (W^2) \right)\cdot \nonumber \\ 
&&~~~~~~~~\cdot \Theta \left( \vec l^{~\prime 2}_\bot - \frac{2}{3} 
\bar \Lambda^{~2}_{sat}
(W^2)\right) \Theta \left( a \bar \Lambda^{~2}_{sat} 
(W^2) - \vec l^{~\prime 2}_\bot\right),
\label{3.19}
\eqa
where
\be
f_L \left( \vec l^{~\prime 2}_\bot, \bar \Lambda^{~2}_{sat} (W^2) \right) =
\frac{\bar\sigma^{(\infty)} (W^2)}{3 \pi} 
\frac{\bar \Lambda^{~4}_{sat}(W^2)}{\vec l^{~\prime 6}_\bot}
\frac{1}{\sqrt{1 - \frac{2 \bar \Lambda^{~2}_{sat} 
(W^2)}{3 \vec l^{~\prime 2}_\bot}}},
\label{3.20}
\ee
and
\be
f_T \left( \vec l^{~\prime 2}_\bot, \bar \Lambda^{~2}_{sat} (W^2) \right) =
\frac{3 \vec l^{~ \prime 2}_\bot}{2 \bar \Lambda^{~2}_{sat} (W^2)}
\left( 1 - \frac{1}{3} 
\frac{\bar \Lambda^{~2}_{sat} (W^2)}{\vec l^{~\prime 2}_\bot} \right)
f_L \left(\vec l^{~\prime 2}_\bot, \bar \Lambda^{~2}_{sat} (W^2) \right).
\label{3.21}
\ee
The constant $a$ in (\ref{3.19}) is related to $\epsilon$ in (\ref{3.18}) by
$\epsilon = 1/6 a$, where $a \gg 1$. Comparison of (\ref{3.19}) with (\ref{3.4})
reveals that the peak as a function of $\vec l_\bot^{~\prime 2}$ at $\vec
l_\bot^{~\prime 2} = \Lambda^2_{\rm sat} (W^2)$ in (\ref{3.4}) is replaced by a
broad distribution in the interval $(2/3) \bar\Lambda^2_{\rm sat} (W^2) \le \vec
l_\bot^{~\prime 2} \le a \bar\Lambda^2_{\rm sat}(W^2)$.
For $\vec l^{~\prime 2}_\bot > \bar \Lambda^{~2}_{sat} (W^2)$, the
transverse part of the dipole cross section in (\ref{3.21}) becomes
enhanced by a factor of $\vec l_\bot^{~\prime 2} / \Lambda^2_{sat} (W^2)$ 
relative to the longitudinal one.

Inserting the $J=1$ dipole cross section (\ref{3.19}), with (\ref{3.20}) 
and (\ref{3.21}),
into the large-$Q^2$ form of the photoabsorption cross section in (\ref{2.43}),
we find (with $Q^2 \gg \bar \Lambda^2_{sat} (W^2)$)
\be
\sigma_{\gamma^*_{L,T}p} (W^2, Q^2) = \frac{\alpha}{\pi} \sum_q
Q^2_q \frac{1}{Q^2} \frac{1}{6} \bar \sigma^{~(\infty)} (W^2) 
\bar \Lambda^{~2}_{sat} (W^2) \sqrt{1 - \frac{2}{3a}} \left\{
\matrix{ 1,\cr
2 \rho \left( \epsilon = \frac{1}{6a} \right), } \right.
\label{3.22}
\ee
where $2 \rho \left( \epsilon = \frac{1}{6a} \right)$ coincides with the
result given in (\ref{2.20}). Here, we assumed $m^2_1(W^2)\rightarrow
\infty$. The generalization to finite values of $m^2_1 (W^2)$ will be given in
Section 4, compare (\ref{4.28}).

The photoabsorption cross section (\ref{3.22}) may be expressed in terms of
the cross section $\sigma_L^{(\infty)} (W^2)$ and the scale $\Lambda^2_{sat}
(W^2)$ introduced in Section (\ref{2.5}) in terms of integrals over the
longitudinal part of the $J=1$ dipole cross section. Compare (\ref{2.71}) and
(\ref{2.72}). Evaluating (\ref{2.71}) and (\ref{2.72}) for 
the ansatz (\ref{3.19}), we find
\bqa
\sigma_L^{(\infty)} (W^2) && = \bar \sigma^{~(\infty)} (W^2) 
\left( 1 + \frac{1}{3a} \right) \sqrt{1 - \frac{2}{3a}} \\
&& \cong \bar \sigma^{~(\infty)} (W^2) \left( 1 - \frac{1}{9a^2} \right),
~~~(a > 1) \nonumber
\label{3.23}
\eqa
and
\be
\Lambda^2_{sat} (W^2) = \bar \Lambda^2_{sat} (W^2) \frac{1}{1+ \frac{1}{3a}}.
\label{3.24}
\ee
The photoabsorption cross section (\ref{3.22}) may accordingly be written in terms
of $\sigma^\infty_L (W^2)$ and $\Lambda^2_{sat} (W^2)$ to become
\bqa
\sigma_{\gamma^*_{L,T}p} (W^2, Q^2)& =& \frac{\alpha}{\pi} \sum_q
Q^2_q \frac{1}{6} \sigma_L^{(\infty)} (W^2) 
\frac{\Lambda^2_{sat} (W^2)}{Q^2}
\left\{ \matrix{ 1 \cr
2 \rho \left( \epsilon = \frac{1}{6a} \right) } \right. , 
\nonumber \\
& & ( Q^2 \gg \Lambda^2_{sat} (W^2)) . 
\label{3.25} 
\eqa
The result (\ref{3.25}) correctly coincides with the general result (\ref{2.76}).

A comparison of (\ref{3.25}) with (\ref{3.8}) and the $\eta(W^2,
Q^2)\rightarrow\infty$ limit in (\ref{3.9}) shows that the large-$Q^2$ cross
section (\ref{3.25}) formally corresponds to the polarization-dependent
replacement in (\ref{3.1}) of 
\be
\Lambda^2_{sat} (W^2) \rightarrow \left\{ \matrix{ 
\Lambda^2_{sat, L} (W^2) = \Lambda^2_{sat} (W^2) , \cr
\Lambda^2_{sat , T} (W^2) = \rho (\epsilon ) \Lambda^2_{sat} (W^2) . } \right.
\label{3.26}
\ee
combined with the substitution
\be
\sigma^{(\infty)} (W^2) \rightarrow
\sigma_L^{(\infty)} (W^2)
\label{3.27}
\ee 
The justification of the resulting cross section (\ref{3.25}) 
rests on the ansatz (\ref{3.18}), since the dipole cross section in
(\ref{2.1}), and accordingly in 
({\ref{3.1}) must be independent of the polarization indices $T$ and $L$ of
  $q \bar q$ dipole fluctuations. The replacement (\ref{3.26}) with
  (\ref{3.27}) is nevertheless illuminating for an intuitive understanding of
  the transition from (\ref{3.1}) to the ansatz (\ref{3.17}).

\section{The Evaluation of the Photoabsorption \hfill\break 
Cross Section, Analytic Results. }
\renewcommand{\theequation}{\arabic{section}.\arabic{equation}}
\setcounter{equation}{0}

For the evaluation of the ansatz for the photoabsorption cross section
presented in Section 3, we return to momentum
space. Inserting the representation for the longitudinal and the
transverse part of the $J=1$ dipole cross section (\ref{2.40}) into
(\ref{2.35}), and employing the momentum-space representation of the
modified Bessel functions $K_{0,1} (r^\prime_\bot Q)$, one finds
(compare Appendix A)
\bqa
&& \hspace{-0.2cm}  \sigma_{\gamma^*_L p} (W^2, Q^2) = \frac{\alpha R_{e^+e^-}}{3 \pi}
Q^2 \int d \vec l^{~\prime 2}_\bot \bar \sigma_{(q \bar q)^{J=1}_L p}
(\vec l^{~\prime 2}_\bot, W^2) \cdot \nonumber \\
&& \hspace{-0.2cm}  \cdot \int d M^2 \int d M^{\prime 2} w (M^2, M^{\prime 2}, 
\vec l^{~\prime 2}_\bot) \left( \frac{1}{(Q^2 + M^2)^2} -
\frac{1}{(Q^2 + M^2) (Q^2 + M^{\prime 2})} \right) \nonumber \\
&& ~
\label{4.1}
\eqa
and
\bqa
&& \hspace{-0.5cm} \sigma_{\gamma^*_T p} (W^2, Q^2) = \frac{\alpha R_{e^+e^-}}{3 \pi}
\int d \vec l^{~\prime 2}_\bot \bar \sigma_{(q \bar q)^{J=1}_T p}
(\vec l^{~\prime 2}_\bot, W^2) \cdot \nonumber \\
&& \hspace{-0.5cm} \cdot \int d M^2 \int d M^{\prime 2} w (M^2, M^{\prime 2}, 
\vec l^{~\prime 2}_\bot) \left( \frac{M^2}{(Q^2 + M^2)^2} - \frac{1}{2}
\frac{M^2 + M^{\prime 2}-\vec l^{~\prime 2}_\bot)}{(Q^2 + M^2) 
(Q^2 + M^{\prime 2})} \right). \nonumber \\
&& ~
\label{4.2}
\eqa
In the transition from (\ref{2.35}) to (\ref{4.1}) and (\ref{4.2}), we
introduced the $q \bar q$ masses,
\be
M^2 = \frac{\vec k^{~2}_\bot}{z(1-z)} \equiv \vec k^{~\prime 2}_\bot,
\label{4.3}
\ee
in terms of the quark transverse momentum, $\vec k_\bot$, and 
\be
M^{\prime 2} = \frac{(\vec k_\bot + \vec l_\bot)^2}{z (1-z)},
\label{4.4}
\ee
in terms
of the transverse momentum of the quark upon absorption of the gluon.

In (\ref{4.1}) and (\ref{4.2}), $R_{e^+e^-} = 3 \sum_q Q^2_q$, where
the sum runs over the actively contributing quarks. The Jacobian
$w(M^2, M^{\prime 2}, \vec l^{~\prime 2}_\bot)$ in (\ref{4.1}) and
(\ref{4.2}) is given by \cite{Cvetic}
\be
w (M^2, M^{\prime 2}, \vec l^{~\prime 2}_\bot) =
\frac{1}{2 MM^\prime \sqrt{1 - \cos^2 \phi}} =
\frac{1}{2 M \sqrt{\vec l^{~\prime 2}_\bot} \sqrt{(1-\cos^2 \vartheta)}},
\label{4.5}
\ee
where $\phi$ denotes the angle between $\vec k_\bot$ and $(\vec k_\bot
+ \vec l_\bot)$, and $\vartheta$ denotes the angle between $\vec k_\bot$
and $\vec l_\bot$. Since
\be
\cos^2 \phi = \frac{1}{4 M^2 M^{\prime 2}} \left( M^2 + M^{\prime 2} -
\vec l_\bot^{~\prime 2} \right)
\label{4.6}
\ee
is symmetric under exchange of $M^2$ and $M^{\prime 2}$, also
$w (M^2, M^{\prime 2}, \vec l^{~\prime 2}_\bot)$ in (\ref{4.5}) is
symmetric under exchange of $M^2$ and $M^{\prime 2}$. The integrands in
(\ref{4.1}) and (\ref{4.2}) may be cast into a form that is fully 
symmetric under exchange of $M^2$ and $M^{\prime 2}$, thus explicitly
displaying the symmetry of the virtual forward-Compton-scattering amplitude
from fig. 1.
It describes the process $\gamma^* p \to \gamma^*p$ in terms of the
``diagonal'' transitions $M_{(q \bar q)} \to M_{(q \bar q)}$ and
$M^\prime_{(q \bar q)} \to M^\prime_{(q \bar q)}$ and the ``off-diagonal''
ones $M_{(q \bar q)} \leftrightarrow M^\prime_{(q \bar q)}$, in
a symmetric manner.

The integrations in (\ref{4.1}) and (\ref{4.2}) have to fulfill the
restrictions 
\be
m^2_0 \le M^2, M^{\prime 2} \le m^2_1 (W^2).
\label{4.7}
\ee
The lower bound, $m^2_0$, in (\ref{4.7}) corresponds to vanishing
$\gamma^* \to q \bar q$ transitions, as soon as $\vec k_\bot^{~2}$ (and $(\vec
k_\bot + \vec l_\bot )^2$ )  
become sufficiently small. A vanishing value of $\vec k^{~2}_\bot$ would
imply contributions to the Compton-forward-scattering amplitude of
states of unbounded transverse size that do not occur 
as a consequence of quark confinement. Via quark-hadron
duality in $e^+e^-$ annihilation, the value of $m_0$ must be somewhat 
below the $\rho^0$ mass\footnote{A refined treatment has to discriminate
  between the masses of the different quark flavors, and, in particular, has to
  introduce a larger lower limit for the charm contribution to the cross
  section.}. 
The upper limit, $m^2_1 (W^2)$, in (\ref{4.7})
follows from the restriction on the lifetime, (\ref{2.52}), of a
hadronic $q \bar q$ fluctuation that requires $M^2$ and $M^{\prime 2}$
to be strongly bounded for any finite value of the energy, $W$.
Quantitatively, for a typical HERA energy of $W = 225 GeV$, the crude
estimate of $M^2_{q \bar q}/W^2 = 0.01$ requires $m_1 (W) = 22.5 GeV$.
This value is approximately consistent with the mass range of the
diffractive continuum that is directly related to the scattering of
$q \bar q$ fluctuations relevant for the total photoabsorption cross
section. Obviously, the mass bound, $m^2_1 = m^2_1 (W^2)$, increases 
with increasing energy.

For the evaluation of (\ref{4.1}) and (\ref{4.2}) with the restriction 
of (\ref{4.7}) on $M^2$ and $M^{\prime 2}$, it is convenient to replace
the integration over $dM^{\prime 2}$ by an integration over 
$d \vartheta$. Noting that
\be
M^{\prime 2} (M^2, \vec l^{~\prime 2}_\bot, \cos \vartheta) =
M^2 + \vec l^{~\prime 2}_\bot + 2 M \sqrt{\vec l_\bot^{~\prime 2}} 
\cos \vartheta,
\label{4.8}
\ee
and
\be
\frac{\partial M^{\prime 2} (M^2, \vec l^{~\prime 2}_\bot, \cos \vartheta)}
{\partial \vartheta} = - \frac{1}{w (M^2, M^{\prime 2}, \vec l^{~\prime 2}_\bot)},
\label{4.9}
\ee
upon incorporating the restrictions in (\ref{4.7}), the integrations in
(\ref{4.1}) and (\ref{4.2}) simplify to become
\bqa
&& \int d M^2 \int d M^{\prime 2} w (M^2, M^{\prime 2}, 
\vec l^{~\prime 2}_\bot) = \nonumber \\
&& = \int^{m^2_1 (W^2)}_{m^2_0} d M^2 \int^\pi_0 d \vartheta -
\int^{(\sqrt{\vec l^{~\prime 2}_\bot} + m_0)^2}_{(\sqrt{\vec l^{~\prime
2}_\bot} - m_0)^2} d M^2 \int^\pi_{\vartheta_0 (M^2, \vec l^{~\prime 2}_\bot)}
d \vartheta  \nonumber \\
&& - \int^{m^2_1 (W^2)}_{(m_1 (W^2) - 
\sqrt{\vec l^{~\prime 2}_\bot})^2} dM^2 \int^{\vartheta_1 
(M^2, \vec l^{~\prime 2}_\bot)}_0 d \vartheta  . \nonumber\\
&~
\label{4.10}
\eqa2.
The first term in (\ref{4.10}) takes care of the bound on $M^2$ in
(\ref{4.7}), ignoring, however, the restriction on $\vartheta$ induced 
by the bound
on $M^{\prime 2}$. The second and the third term in (\ref{4.10}) correct
for this ignored restriction on $M^{\prime 2}$. The bounds on the
angles, $\vartheta_0 (M^2, \vec l^{~\prime 2}_\bot)$ and
$\vartheta_1 (M^2, \vec l^{~\prime 2}_\bot)$ in (\ref{4.10}),
are obtained from the lower and the upper bound on $M^{\prime 2}
(M^2, \vec l^{~\prime 2}_\bot,  \cos \vartheta)$ implied by
(\ref{4.8}) and are given by
\be
\cos \vartheta_{0,1} (M^2, \vec l^{~\prime 2}_\bot) =
\frac{m^2_{0,1} - M^2 - \vec l^{~\prime 2}_\bot}
{2 M \sqrt{\vec l^{~\prime 2}_\bot}}.
\label{4.11}
\ee
Here $m^2_1$ stands for $m^2_1 \equiv m^2_1 (W^2)$.
In terms of the $d M^2 d \vartheta$ integration (\ref{4.10}),
the photoabsorption cross sections in (\ref{4.1}) and (\ref{4.2}) become
\bqa
&& \sigma_{\gamma^*_Lp} (W^2, Q^2)  = \frac{\alpha R_{e^+e^-}}{3 \pi}
\int d \vec l^{~\prime 2}_\bot \bar \sigma_{(q \bar q)^{J=1}_L p}
(\vec l^{~\prime 2}_\bot, W^2) \cdot \label{4.12} \\
&& \cdot \int d M^2 \int d \vartheta \left( \frac{Q^2}{(Q^2+M^2)^2}-
\frac{Q^2}{(Q^2 + M^2) (Q^2 + M^{\prime 2} (M^2, \vec l^{~\prime 2}_\bot,
\cos \vartheta))} \right)  \nonumber
\eqa
and
\bqa
& & \sigma_{\gamma^*_Tp} (W^2, Q^2) = \frac{\alpha R_{e^+e^-}}{3 \pi}
\int d \vec l^{~\prime 2}_\bot \bar \sigma_{(q \bar q)^{J=1}_T p}
(\vec l^{~\prime 2}_\bot , W^2) \cdot  \label{4.13}   \\
&& \cdot \int d M^2 \int d \vartheta \left( \frac{M^2}{(Q^2 + M^2)^2} -
\frac{M^2 + M^{\prime 2} (M^2, \vec l^{~\prime 2}_\bot, \cos \vartheta)
- \vec l^{~\prime 2}_\bot}{2 (Q^2 + M^2) (Q^2 + M^{\prime 2} (M^2,
\vec l^{~\prime 2}_\bot, \cos \vartheta))} \right). \nonumber 
\eqa
The integrations in (\ref{4.12}) and (\ref{4.13}), according to 
(\ref{4.10}), lead to
a sum of three terms,
\be
\sigma_{\gamma^*_{L,T}p} (W^2, Q^2) = \sigma^{dom}_{\gamma^*_{L,T}p}
(W^2, Q^2) + \Delta \sigma^{(m^2_0)}_{\gamma^*_{L,T}p} (W^2, Q^2) +
\Delta \sigma^{(m^2_1 (W^2))}_{\gamma^*_{L,T}p} (W^2, Q^2).
\label{4.14}
\ee
The first term will be dominant. The correction due to the lower bound
$m^2_0$ will turn out to be small, of order 1 \%. The third term in 
(\ref{4.14}) will be found to yield a somewhat larger contribution, of order
10 \%, dependent on the values of the kinematical variables.

For the dominant term, the integration of (\ref{4.12}) and (\ref{4.13})
with the integration domain given by the first term in (\ref{4.10}), can
be carried out analytically. We concentrate on the dominant term, and 
for the correction terms refer to Appendix B.

Upon integration over $d \vartheta$ of (\ref{4.12}) and (\ref{4.13}),
the dominant contributions to the photoabsorption cross section become
\cite{MKDS}
\bqa
&& \sigma^{dom}_{\gamma^*_L p} (W^2, Q^2) = \frac{\alpha R_{e^+e^-}}{3}
\int d \vec l^{~\prime 2}_\bot \bar \sigma_{(q \bar q)^{J=1}_L p}
(\vec l^{~\prime 2}_\bot, W^2) \cdot  \label{4.15} \\
&& \cdot \int^{m^2_1 (W^2)}_{m^2_0} d M^2 \left( 
\frac{Q^2}{(Q^2 + M^2)^2} - \frac{Q^2}{(Q^2 + M^2) \sqrt{X (M^2, 
\vec l^{~\prime 2}_\bot, Q^2)}} \right),\nonumber 
\eqa
and
\bqa
&&\sigma^{dom}_{\gamma^*_T p} (W^2, Q^2) = \frac{\alpha R_{e^+e^-}}{3}
\int d \vec l^{~\prime 2}_\bot \bar \sigma_{(q \bar q)^{J=1}_T p} 
(\vec l^{~\prime 2}_\bot, W^2) \cdot \nonumber \\
&& \cdot \int^{m^2_1 (W^2)}_{m^2_0} \frac{d M^2}{2} 
\left( \frac{1}{(Q^2 + M^2)}
- \frac{2 Q^2}{(Q^2 + M^2)^2} - \frac{1}{\sqrt{X (M^2, \vec l^{~\prime 2}_\bot,
Q^2)}} + \right. \nonumber \\
&& \left. \frac{2 Q^2 + \vec l^{~\prime 2}_\bot}{(Q^2 + M^2) 
\sqrt{X (M^2, \vec l^{~\prime 2}_\bot, Q^2)}} \right), 
\label{4.16}
\eqa
where
\be
X (M^2, \vec l^{~\prime 2}_\bot, Q^2) \equiv (M^2 - \vec l^{~\prime 2}_\bot
+ Q^2)^2 + 4 Q^2 \vec l^{~\prime 2}_\bot.
\label{4.17}
\ee
Carrying out the integration over $dM^2$ in (\ref{4.15}) and
(\ref{4.16}), we finally obtain
\bqa
& & \sigma^{dom}_{\gamma^*_{L,T}p} (W^2, Q^2) = \frac{\alpha R_{e^+e^-}}{3}
\int d \vec l^{~\prime 2}_\bot \bar \sigma_{(q \bar q)^{J=1}_{L,T} p}
(\vec l^{~\prime 2}_\bot, W^2) \cdot \nonumber \\
&& \cdot \left( I_{L,T} (\vec l^{~\prime 2}_\bot, m^2_1 (W^2), Q^2) -
I_{L,T} (\vec l^{~\prime 2}_\bot, m^2_0, Q^2) \right), \nonumber \\
&& ~
\label{4.18}
\eqa
where $I_{L,T} (\vec l^{~\prime 2}_\bot, M^2, Q^2)$ denotes the 
indefinite integrals over $dM^2$ in (\ref{4.15}) and (\ref{4.16}). They
are given by
\bqa
&& \hspace{-0.7cm}  I_L (\vec l^{~\prime 2}_\bot, M^2, Q^2) = \frac{- Q^2}{Q^2
  + M^2} +  \frac{Q^2}{\sqrt{\vec l^{~\prime 2}_\bot 
(\vec l^{~\prime 2}_\bot + 4 Q^2)}} \cdot
\label{4.19} \\
&&\hspace{-0.7cm}  \cdot \ln \frac{\sqrt{\vec l^{~\prime 2}_\bot
(\vec l^{~\prime 2}_\bot + 4 Q^2)} \sqrt{X (M^2, \vec l^{~\prime 2}_\bot,
Q^2)} + \vec l^{~\prime 2}_\bot (3Q^2 - M^2 + \vec l^{~\prime 2}_\bot)}
{Q^2+M^2} \nonumber
\eqa
and
\bqa
&& \hspace{-0.7cm}  I_T (\vec l^{~\prime 2}_\bot, M^2, Q^2) = \label{4.20} \\ 
& & \hspace{-0.7cm} = \frac{Q^2}{Q^2+M^2} +
\frac{1}{2} \ln \frac{Q^2 + M^2}{\sqrt{X (M^2, \vec l^{~\prime 2}_\bot, Q^2)}
+ M^2 - \vec l^{~\prime 2}_\bot + Q^2} 
- \frac{2 Q^2 + \vec l^{~\prime 2}_\bot}{2 \sqrt{ \vec l^{~\prime 2}_\bot
(\vec l^{~\prime 2}_\bot + 4 Q^2)}}  \cdot \nonumber \\ 
& & \hspace{-0.7cm}  \cdot \ln \frac{\sqrt{\vec l^{~\prime 2}_\bot
(\vec l^{~\prime 2}_\bot + 4 Q^2)} \sqrt{X (M^2, \vec l^{~\prime 2}_\bot, 
Q^2)} + \vec l^{~\prime 2}_\bot (3Q^2 - M^2 + \vec l^{~\prime 2}_\bot)}
{Q^2 + M^2}. \nonumber 
\eqa
The representation (\ref{4.18}) of the (dominant part of the) photoabsorption
cross section does not depend on a specific ansatz for the dipole cross
section. The representation (\ref{4.18}) only relies on the general form
of the CDP given by (\ref{2.1}) with (\ref{2.2}) and by (\ref{2.35}) with
(\ref{2.40}) that follow from (\ref{2.1}) and (\ref{2.2}). In other words,
(\ref{4.18}) only rests on the low-x kinematics and the formation of 
$q \bar q$ color-dipole fluctuations that interact as color dipoles with 
the gluon field in the nucleon. In most applications of the CDP one considers
the limit of $m^2_1 (W^2) \to \infty$ that restricts the kinematic range
of validity of the CDP. In this limit of $\Delta 
\sigma_{\gamma^*_{L,T}p}^{(m^2_1
(W^2))} (W^2, Q^2) = 0$, the photoabsorption cross section is well 
represented by the dominant term (\ref{4.18}) evaluated for $m^2_1 (W^2)
\rightarrow \infty$, since $\Delta 
\sigma_{\gamma^*_{L,T}p}^{(m^2_0)}
(W^2,Q^2)$ can be neglected.

The evaluation of (\ref{4.18}) for the case of the ansatz (\ref{3.4}) of
the dipole cross section with helicity independence is straightforward.
For the sum of the longitudinal and the transverse cross section, for
$m_1^2(W^2)\rightarrow\infty$, the result is
given in (\ref{3.8}) with (\ref{3.9}). 

For the evaluation of the more general ansatz (\ref{3.19}), 
it will be convenient to replace the integration variable 
$\vec l^{~\prime 2}_\bot$ by
\be
y = \frac{2}{3} \frac{\bar\Lambda^2_{sat} (W^2)}{\vec l^{~\prime 2}}.
\label{4.21}
\ee
The $J=1$ dipole cross sections (\ref{3.20}) and (\ref{3.21}) then become
\be
f_L (y, \bar \Lambda^2_{sat} (W^2)) = \frac{9}{8}
\frac{\bar \sigma^{(\infty)} (W^2)}{\pi \bar \Lambda^2_{sat} (W^2)}
\frac{y^3}{\sqrt{1-y}},2.
\label{4.22}
\ee
and
\be
f_T(y, \bar \Lambda^2_{sat} (W^2)) = \frac{(1- \frac{1}{2}y)}{y} f_L (y,
\bar \Lambda^2_{sat} (W^2)).
\label{4.23}
\ee
Explicitly, the photoabsorption cross section (\ref{4.18}) for the
ansatz (\ref{3.19}) is then given by
\bqa
&& \sigma^{dom}_{\gamma^*_L p} (W^2, Q^2) = \frac{\alpha R_{e^+e^-}}{4} 
\frac{\bar \sigma^\infty (W^2)}{\pi} \int^1_{2/3 a} dy \frac{y}{\sqrt{1-y}}
\cdot \label{4.24} \\
&& \cdot \left( I_L \left( \frac{2}{3} \frac{\bar \Lambda^2_{sat} (W^2)}{y},
m^2_1 (W^2), Q^2 \right) - I_L \left( \frac{2}{3} \frac{\bar \Lambda^2_{sat}
(W^2)}{y}, m^2_0, Q^2 \right) \right)  \nonumber 
\eqa
and
\bqa
&&  \sigma^{dom}_{\gamma^*_T p} (W^2, Q^2) = \frac{\alpha R_{e^+e^-}}{4} 
\frac{\bar \sigma^{(\infty)} (W^2)}{\pi} \int^1_{2/3 a} dy
\frac{1-y/2}{\sqrt{1-y}} \cdot  \label{4.25}\\
&& \cdot \left( I_T \left( \frac{2}{3} \frac{\bar \Lambda^2_{sat} (W^2)}{y},
m^2_1 (W^2), Q^2 \right) - I_T \left( \frac{2}{3} 
\frac{\bar \Lambda^2_{\rm sat} (W^2)}{y}, m^2_0, Q^2 \right) \right). \nonumber
\eqa
We note that the replacements 
\bqa
&& \bar \sigma^{(\infty)} (W^2) \to \sigma^{(\infty)} (W^2), \nonumber \\
&& \frac{2}{3} \frac{\bar \Lambda^2_{sat} (W^2)}{y} \to \Lambda^2_{sat} (W^2),
\label{4.26}
\eqa
and the formal replacements
\bqa
&& \int^1_{2/3a} dy \frac{y}{\sqrt{1-y}} \to \frac{4}{3}~, \nonumber \\
&& \int^1_{2/3a} dy \frac{1 - \frac{y}{2}}{\sqrt{1-y}} \to \frac{4}{3}
\label{4.27}
\eqa
in (\ref{4.24}) and (\ref{4.25}) take us back to the 
photoabsorption cross section for the dipole cross section (\ref{3.4})
with helicity independence that is obtained by substitution of (\ref{3.4}) into (\ref{4.18}). 

The correction terms $\Delta\sigma^{(m^2_0)}_{\gamma^*_{L,T}p}(W^2, Q^2) $ and 
$\Delta\sigma^{(m^2_1)}_{\gamma^*_{L,T}p}(W^2, Q^2) $ from (\ref{4.14}) that
are to be added to the dominant parts of the cross sections (\ref{4.24}) and
(\ref{4.25}) are explicitly given in Appendix B, compare (\ref{B9}) and
(\ref{B10}). 

The evaluation of the cross sections in (\ref{4.24}) and (\ref{4.25}) together
with the correction terms (\ref{B9}) and (\ref{B10}) in general requires
numerical integration.\footnote{A computer program can be provided on request.} 

A simple analytic approximation of the cross sections
can be derived, however, for the limit of $Q^2 \gg \bar\Lambda^2_{sat}
(W^2)= \Lambda^2_{\rm sat} (W^2) \cdot (1 + 1/3 a) \cong \Lambda^2_{\rm sat}
(W^2)$, or $\eta (W^2, Q^2) \gg 1$. Ignoring the negligible contribution from
$\Delta^{(m^2_0)}_{\gamma^*_{L,T}p} (W^2, Q^2)$, the analytic approximation for
the sum of $\sigma^{dom}_{\gamma^*_{L,T}p}(W^2, Q^2)$ and
$\Delta\sigma^{(m^2_1)}_{\gamma^*_{L,T}p}(W^2, Q^2)$ is given by
\bqa
&  & \sigma_{\gamma^*_{L,T}p} (W^2, Q^2) = \sigma_{\gamma^*_{L,T}p} 
(\eta (W^2, Q^2), \xi) = \label{4.28} \\
& & = \frac{\alpha R_{e^+ e^-}}{18}
\frac{\sigma_L^{(\infty)}(W^2)}{\pi}\frac{1}{\eta (W^2, Q^2)} 
\left\{ \matrix{G_L (\eta (W^2, Q^2), \xi) , ~~~~~~~~~~~~~~& \cr
      2\rho (\varepsilon = \frac{1}{6a}) G_T (\eta (W^2, Q^2), \xi) ,&  \cr } \right.  \nonumber
\eqa
where
\bqa
G_L (\eta (W^2, Q^2) , \xi) & = & G_L \left( \frac{\xi}{\eta(W^2, Q^2)}\right) 
\label{4.29}\\
& = & \frac{\left(\frac{\xi}{\eta}\right)^3 + 3
  \left(\frac{\xi}{\eta}\right)^2}{\left( 1 + \frac{\xi}{\eta}\right)^3} =
  \left\{  \matrix{1, & {\rm for} \frac{\xi}{\eta} \rightarrow \infty , \cr
0.98, & {\rm for} \frac{\xi}{\eta} = 10 , \cr
0.5 , & {\rm for} \frac{\xi}{\eta}=1 ,  } \right. \nonumber
\eqa
and 
\bqa
G_T (\eta (W^2, Q^2) , \xi) & = & G_T \left( \frac{\xi}{\eta(W^2, Q^2)}\right) 
\label{4.30}\\
& = & \frac{2 \left( \frac{\xi}{\eta}\right)^3 + 3 \left(
    \frac{\xi}{\eta}\right)^2 + 3 \left( \frac{\xi}{\eta}\right)}{2 \left( 1 +
    \frac{\xi}{\eta} \right)^3} = 
\left\{  \matrix{1, & {\rm for} \frac{\xi}{\eta} \rightarrow \infty , \cr
0.88, & {\rm for} \frac{\xi}{\eta} = 10 , \cr
0.5 , & {\rm for} \frac{\xi}{\eta}=1 ,  } \right. \nonumber
\eqa
and $\rho (\epsilon = \frac{1}{6a})$ is given by (\ref{2.20}). Compare Appendix
    C for the derivation of (\ref{4.28}) to (\ref{4.30}).
In (\ref{4.28}) to (\ref{4.30}), $\eta \equiv \eta (W^2, Q^2) = (Q^2 + m^2_0)
    /\Lambda^2_{sat} (W^2)$ denotes the low-$x$ scaling variable defined by
    (\ref{2.62}), and the parameter $\xi$ specifies $m^2_1 (W^2)$ via  
\be
m^2_1 (W^2) = \xi \Lambda^2_{sat} (W^2) = \frac{\xi}{\eta (W^2, Q^2)} (Q^2 + m^2_0). 
\label{4.31}
\ee
where the approximation of $m^2_0 \cong 0$ is valid, since we are concerned
with $Q^2 \gg \Lambda^2_{sat} (W^2) \gg m^2_0$.
With (\ref{4.28}), we have obtained the generalization of (\ref{3.25}) to the
case of a finite upper bound, $m^2_1 (W^2)$, for the masses of the $q \bar q$
fluctuations. The limit of $\xi / \eta \rightarrow \infty$, or $\xi \rightarrow
\infty$ at fixed $\eta (W^2, Q^2)$, yields the frequently employed
approximation of the CDP that ignores the upper bound on the masses of the
contributing $q \bar q$ fluctuations. 
Since $\xi$ must be finite, compare (\ref{3.7}) and (\ref{4.31}), this
approximation of the CDP breaks down as soon as $\eta (W^2, Q^2)$ becomes
sufficiently large. 

According to (4.28}), the ratio of the longitudinal to the transverse
photoabsorption cross section for $Q^2 \gg \Lambda^2_{sat} (W^2)$ is given by
\be
R (W^2, Q^2) = \frac{\sigma_{\gamma^*_L p} (\eta (W^2, Q^2), \xi)}{\sigma_{\gamma^*_T p}
  (\eta (W^2, Q^2), \xi)} \Bigg|_{\eta(W^2, Q^2)\gg 1} = \frac{1}{2\rho \left(
    \epsilon = \frac{1}{6a} \right) \frac{G_T \left(
      \frac{\xi}{\eta}\right) }{G_L \left( \frac{\xi}{\eta}\right)}} .
\label{4.31a}
\ee
The ratio $R (W^2, Q^2)$ in (\ref{4.31a}), compared with (\ref{2.43J}) is
modified by the factor of $G_T (\xi / \eta) / G_L (\xi / \eta )$. 
The transverse-size enhancement of transversely polarized relative to
longitudinally polarized $(q \bar q)^{J=1}$ fluctuations from Section 3.3 must
be applied for realistic values of $m^2_1 (W^2)$, sufficiently large such that
the CDP, approximately unmodified by the finiteness of $m^2_1 (W^2)$, becomes
applicable. 
We accordingly consider $R (W^2, Q^2)$ for $\xi / \eta \ge 10$. With $\eta (W^2,
Q^2)$ in the interval of $5 < \eta (W^2 , Q^2) < 10$, this corresponds to $50
\Lambda^2_{\rm sat} (W^2) < m^2_1 (W^2) < 100 \Lambda^2_{\rm sat} (W^2)$ and $5
\Lambda^2_{\rm sat} (W^2) < Q^2 < 10 \Lambda^2_{\rm sat} (W^2)$\footnote{At
  HERA energies, we approximately have $3 {\rm GeV}^2 \le \Lambda^2_{\rm sat}
  (W^2) \le 7 {\rm GeV}^2$.}, and   
\be
m^2_1 (W^2) \gg Q^2 \gg \Lambda^2_{sat} (W^2).
\label{4.31b}
\ee
Taking into account the transverse-size enhancement in the denominator of
(\ref{4.31a}) according to (\ref{2.43J}) and (\ref{2.43I}) requires 
\be
\rho \left( \epsilon \equiv \frac{1}{6a} \right) \frac{G_T \left(
    \frac{\xi}{\eta} \cong 10 \right)}{G_L \left(
    \frac{\xi}{\eta} \cong 10 \right)} = \frac{4}{3} .
\label{4.31c}
\ee
With $\rho (\epsilon \equiv 1 / 6a)$ from (\ref{2.20}), 
and the numerical values of $G_T (\xi / \eta = 10)$ and $G_L (\xi / \eta = 10)$
from (\ref{4.29}) and (\ref{4.30}), 
$G_T (\xi / \eta \cong 10) / G_L (\xi / \eta \cong 10) \cong 0.9$,
the constraint (\ref{4.31c}) yields 
\be
a \cong 7.5 . 
\label{4.31d}
\ee
With this uniquely determined\footnote{A value of $a = 7$ is applied in the
  analysis of the experimental data in Section 5.} value of $a = 7.5$, our
ansatz (\ref{4.17}) for the dipole cross section yields a concrete realization
of the transverse-size enhancement that implies $R (W^2, Q^2) = \frac{1}{2
  \cdot \left( \frac{4}{3} \right)} = 0.375$, compare (\ref{2.43J}). 

In what follows, we will discuss the effect of a finite value of $m^2_1 (W^2) =
\xi \Lambda^2_{sat} (W^2)$ by examining the behavior of the large-$Q^2$
approximation of the cross section in
(\ref{4.28}) under variation of $\xi$. In particular, we first of all chose
the value of $\xi$ required by consistency with the experimental results in the
range of $\eta (W^2, Q^2) \ge 10$. This value of $\xi$, compare Section 5, is
given by  
\be
\xi = \xi_{exp} = 130 .
\label{4.32}
\ee
We illustrate the effect of $\xi$, by comparing the theoretical results for the
photoabsorption cross section obtained for the choice of (\ref{4.32}) with the ones for $\xi
\rightarrow \infty$ and for various values of $\xi < \xi_{\rm exp} = 130$.

\begin{figure}[h!]
\centerline{~~~~~~\epsfig{file=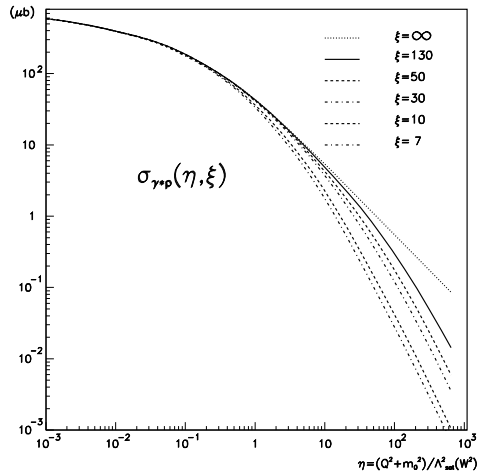,width=7cm}}
\caption{{\footnotesize The photoabsorption cross section $\sigma_{\gamma^* p} (\eta (W^2, Q^2), \xi)$
for different values of $\xi = m^2_1 (W^2) / \Lambda^2_{sat} (W^2)$.}}
\end{figure}
%Fig.8
In fig.8, we show the cross section for 
\be
\sigma_{\gamma^* p} (\eta (W^2, Q^2), \xi) = \sigma_{\gamma^*_L p} (\eta (W^2,
Q^2) , \xi) + \sigma_{\gamma^*_T p} (\eta (W^2, Q^2) , \xi)
\label{4.36a}
\ee
obtained by numerical evaluation of (\ref{4.24}) and (\ref{4.25}) together with
(\ref{B9}) and (\ref{B10}). 
The numerical input for $\Lambda^2_{sat} (W^2)$ and $m^2_0$ is identical to
what will be used in Section 5, when comparing with the experimental data. 

The main features of the behavior of $\sigma_{\gamma^* p}(\eta( W^2,Q^2),\xi$), in
fig.8 can be understood by looking at the analytic approximations in (\ref{4.28})
to (\ref{4.30}), which hold for $\eta (W^2, Q^2)$ sufficiently large compared
with unity, \\
$\eta (W^2, Q^2) > 1$:
\begin{itemize}
\item[i)]
For fixed $\xi = \xi_{exp} = 130$ and $\xi / \eta >
10$, or $\eta < \eta_{exp} = 13$, the effect of the finite upper bound of $m^2_1
(W^2)=  130 \Lambda^2_{sat} (W^2)$ becomes negligible. The corresponding
range of $Q^2$ and $W^2$ is given by
\be
Q^2 < \eta_{exp} \Lambda^2_{sat} (W^2) \cong 
\left\{ \matrix{ 39 {\rm GeV}^2, & {\rm for} \Lambda^2_{sat}(W^2)= 3{\rm GeV}^2
    , \cr
91 {\rm GeV}^2, & {\rm for} \Lambda^2_{sat}(W^2)= 7{\rm GeV}^2 .} \right.
\label{4.33}
\ee 
The result (\ref{4.33}) gives the domain, where at HERA energies 
the frequently employed approximaton of the CDP with $m^2_1 (W^2)\rightarrow\infty$
is applicable\footnote{The notation $\eta_{exp}$ for $\eta_{exp} = 13$ results
  from the choice of $\xi = \xi_{exp} = 130$ necessary for agreement with the
  experimental data for $x\le 0.1$.}. 
\item[ii)]
For fixed $\xi = \xi_{exp} = 130$, and $\xi / \eta < 10$, or $\eta > \eta_{exp}
= 13$, the approximation of $m^2_1 (W^2)\rightarrow \infty$ breaks down,
and large corrections of order $0.5$, according to (\ref{4.29}) and (\ref{4.30}),
depending of the value of $\eta (W^2 , Q^2)$, are necessary. Compare
fig.8. The finite value of $\xi = \xi_{exp} = 130$ explicitly excludes
high-mass fluctuations that have too short a lifetime to actively contribute
to the cross section.
\item[iii)]
In fig.8, we also show the theoretical results for the photoabsorption cross
section for values of $\xi$ between $\xi = 7$ and $\xi = \xi_{\rm exp} =
130$. The predicted cross sections for $\eta (W^2 , Q^2)$ sufficiently below
$\eta (W^2, Q^2) = \eta_{\rm exp} = 13$, dependent on the chosen value of $\xi$,
coincide with both the results for $\xi = \xi_{\rm exp} = 130$ and $\xi =
\infty$. This is consistent with the analytic result, $G_{T, L} (\xi / \eta )
\cong 1$ for $\xi > 10 \eta$, compare (\ref{4.29}) and (\ref{4.30}). 
The actively contributing masses $M^2_{q \bar q}$ are actually bounded by $\xi < 10
\eta$ or 
\be 
M^2_{q \bar q} < 10\eta \Lambda^2_{sat} (W^2) = 10 Q^2. ~~~(1 < \eta <
\eta_{exp} \cong 13) 
\label{4.34}
\ee
Compare Table 2. 
The upper bounds on the masses of the $q \bar q$ fluctuations, $M_{q \bar q}$,
contributing to $\sigma_{\gamma^* p} (\eta (W^2, Q^2))$ according to Table 2
approximately coincide
with the upper bounds of the $q \bar q$ masses in which the dominant contributions to
diffractive production are observed at HERA \cite{H1}. 
\begin{center}
\begin{tabular}{c|c|c|c} 
$\eta$  &  $\Lambda^2_{sat} (W^2) [ {\rm GeV}^2]$  & $Q^2 [{\rm GeV}^2]$ &  $M^2_{q \bar q} [ {\rm
  GeV}^2]$ \\
\hline

13 & 3 & 39 &390 \\
   & 7 & 91 & 910 \\
\hline
5 & 3 & 15 & 150 \\
  & 7 & 35 & 350 \\
\hline
\end{tabular}
\end{center}
\begin{itemize}
\item[Table 2] 
\footnotesize{The upper limit of the masses of the actively contributing $(q \bar q)$ fluctuations, 
$M_{q \bar q}$ for values of $\eta \cong Q^2 /
\Lambda^2_{sat} (W^2)$ and $\Lambda^2_{sat} (W^2)$ relevant for HERA energies.} 
\end{itemize}

\end{itemize}

We return to the cross section in (\ref{4.25}) and (\ref{4.16}), as well as
(\ref{4.24}) and (\ref{4.15}) and consider the approximation of 
\be
\eta (W^2, Q^2) \ll 1  
\label{4.35}
\ee
that includes the limit of (\ref{2.96}) of $W^2 \rightarrow \infty$ at fixed $Q^2$,
and specifically the limit of $Q^2 = 0$. In this limit the longitudinal cross
section vanishes, while the transverse cross section (\ref{4.16}) is given by 
\bqa
& & \sigma^{dom}_{\gamma^*_T p} (W^2, Q^2=0) = \frac{\alpha R_{e^+ e^-}}{6} \int d \vec
l_\bot^{~\prime 2} \sigma_{(q \bar q)^{J=1}_T p} (\vec l_\bot^{~\prime 2}, W^2)
\cdot \nonumber \\
& & \cdot \int^{m^2_1 (W^2)}_{m^2_0} d M^2 \left( \frac{1}{M^2} - \frac{M^2 -
    \vec l_\bot^{~\prime 2}}{M^2 | M^2 - \vec l_\bot^{~\prime 2} |} \right) = 
\label{4.36} \\
& & = \frac{\alpha R_{e^+ e^-}}{3} \int d \vec l_\bot^{~\prime 2} \bar\sigma_{(q
  \bar q)^{J=1}_T p} (\vec l_\bot^{~\prime 2} , W^2) \ln \frac{\vec
  l_\bot^{~\prime 2}}{m^2_0} \nonumber 
\eqa
Since according to (\ref{3.19}) the cross section $\bar\sigma_{(q \bar
  q)^{J=1}_T} (\vec l_\bot^{~\prime 2} , W^2)$ is non-vanishing only for $\vec
  l_\bot^{~\prime 2} < a \Lambda^2_{\rm sat} (W^2)$, the upper bound
  $m^2_1(W^2) = \xi \Lambda^2_{\rm sat} (W^2)$ in (\ref{4.36}) may be replaced
  by $m^2_1 (W^2) = a \Lambda^2_{\rm sat} (W^2)$.
With $a = 7$, and $2 {\rm GeV}^2 \le \Lambda^2_{\rm sat} (W^2) \le 7 {\rm
  GeV}^2$ at HERA energies, this implies $14 {\rm GeV}^2 \le M^2_{q \bar q} \le
  49 {\rm GeV}^2$. Only $q \bar q$ fluctuations in a strongly limited range of
  masses, bounded by approximately a value between $3.7$ GeV and $7$ GeV, dependent on $W$, are
  responsible for the photoabsorption cross section when $Q^2$ approaches the
  photoproduction limit of $Q^2 \rightarrow 0$. This analytic estimate is
  confirmed by the numerical results for $\eta = a = 7$ shown in fig.8. For
  $\eta (W^2 , Q^2) < 1$, $q \bar q$ fluctuations with masses squared larger than
  $m^2_1 (W^2) = 7 \Lambda^2_{\rm sat} (W^2)$ do not contribute to the
  interaction. 

Inserting the dipole cross section
  (\ref{3.19}) and passing to the variable $y$ according to (\ref{4.21}) and
  (\ref{4.23}), the photoproduction cross section (\ref{4.36}) becomes 
\bqa
\sigma_{\gamma p} (W^2) & = & \sigma^{dom}_{\gamma^*_T p} (W^2, Q^2 = 0) = 
\label{4.37} \\
& = & \frac{\alpha R_{e^+ e^-}}{4} \frac{\bar\sigma^{(\infty)} (W^2)}{\pi}
\int^1_{\frac{2}{3a}} dy \frac{1 - \frac{1}{2} y}{\sqrt{1-y}} \ln \frac{2
  \bar\Lambda^2_{sat} (W^2)}{3y m^2_0} . \nonumber
\eqa
The substitutions (\ref{4.26}) and (\ref{4.27}) take us back to (\ref{3.8}) and
(\ref{3.9}).

\section{Comparison with Experiment}
\renewcommand{\theequation}{\arabic{section}.\arabic{equation}}
\setcounter{equation}{0}

The total photoabsorption cross section from (\ref{4.24}) and (\ref{4.25})
together with (\ref{B9}) and (\ref{B10}) depends on the saturation scale
$\Lambda^2_{sat}(W^2)$, or rather the low-$x$ scaling variable, $\eta (W^2,
Q^2) = (Q^2 + m^2_0) / \Lambda^2_{sat} (W^2)$, the lower and the upper bounds,
$m^2_0$ and $m^2_1 (W^2) = \xi \Lambda^2_{sat} (W^2)$, on the masses of the $q
\bar q$ fluctuations, and the total $(q \bar q)p$ cross section
$\sigma^{(\infty)} (W^2)$, where from (\ref{3.23}) $\sigma^{(\infty)}(W^2)
\equiv \sigma_L^{(\infty)} (W^2) \cong  \bar\sigma^{(\infty)} (W^2)$. 

The numerical results\footnote{A computer program is available on request.} 
to be shown subsequently are based on the set of
parameters that is specified as follows. The saturation scale is parameterized
by\footnote{For the connection between $\Lambda^2_{\rm sat} (W^2)$ and
$\bar\Lambda^2_{\rm sat} (W^2)$, compare (\ref{4.31d}). The value of $C_2 =
0.27$ is taken from the previous fit in refs.\cite{DIFF2000, SCHI}. 
The difference between this value of $C_2 = 0.27$ and $C_2 = 0.29$ from 
(\ref{(p)}) is not significant in the relevant kinematic range.}
\be
\bar\Lambda^{~2}_{sat} (W^2) = \Lambda^2_{sat} (W^2) \left( 1 + \frac{1}{3a}
\right) = \bar C_1 \left( \frac{W^2}{W^2_0} + 1 \right)^{C_2}
\label{5.1}
\ee
with 
\bqa
\bar C_1 & = & 2.04 {\rm GeV}^2 , \nonumber \\
W^2_0 & = & 1081 {\rm GeV}^2 , \label{5.2} \\ 
C_2 & = & 0.27 . \nonumber 
\eqa
The lower and the upper bound on the masses of the $q \bar q$ fluctuations are
given by
\be
m^2_0 = 0.15 {\rm GeV}^2 , 
\label{5.3}
\ee
and 
\be
m^2_1 (W^2) = \xi \bar\Lambda^{~2}_{sat} (W^2) = 130 \bar\Lambda^{~2}_{sat}
(W^2).
\label{5.4}
\ee
The total cross section, $\sigma^{(\infty)}(W^2)$, is determined by requiring
\cite{DIFF2000}
consistency of the CDP at $Q^2=0$ from (\ref{4.37}) with the Regge
parameterization given by
\be
\sigma^{\rm Regge}_{\gamma p} (W^2) = A_P (W^2)^{\alpha_P - 1} + A_R
(W^2)^{\alpha_R -1} ,
\label{5.5}
\ee
where $W^2$ is to be inserted in units of ${\rm GeV}^2$, and 
\bqa
A_P & = & 63.5 \pm 0.9 \mu b , \nonumber \\
\alpha_P & = & 1.097 \pm 0.002 , \label{5.5a} \\
A_R & = & 145.0 \pm 2.0 \mu b, \nonumber \\
\alpha_R & = & 0.5 . \nonumber
\eqa
Since both the CDP and the Regge parameterization have similar (soft) energy
dependence, one finds that the variation of $\sigma^{(\infty)} (W^2)$ in the
HERA energy range is restricted to about 10\%. 
Quantitatively, since the total photoabsorption cross section is dependent on
the product of $R_{e^+ e^-} \sigma^{(\infty)} (W^2)$, we have 
\be
\sigma^{(\infty)} (W^2) \cong 
\left\{ \matrix{ 30 mb , & ({\rm for~3~active~flavors,}~~ R_{e^+ e^-} = 2) \cr
18 mb , & ({\rm for~4~active~flavors,}~~ R_{e^+ e^-} = \frac{10}{3} )} \right. 
\label{7}
\ee
Comparing the above parameters with the ones in (\ref{3.11}) to (\ref{3.16}),
from refs. \cite{DIFF2000,SCHI}, 
one notes the smaller value of $\bar C_1 = 2.04$ that is required as a
consequence in the change of the longitudinal-to-transverse ratio $R$ 
from $R=0.5$ to $R=0.375$. The magnitude of
$\xi = \xi_{\rm exp} = 130$ was determined from an eye-ball fit to the experimental data.
Compare fig.8 for the variation of the total photoabsorption cross section
under variation of $\xi$.  

\begin{figure}
~\hspace{1cm} \epsfig{file=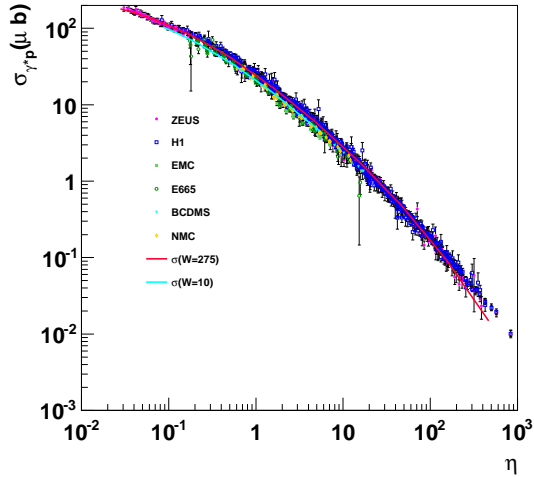,width=8cm}   
\caption{{\footnotesize The theoretical prediction for the photoabsorption cross section \hfill\break
$\sigma_{\gamma^* p} (\eta (W^2, Q^2), \xi)$ for $\xi = 130$ compared with the
experimental data on DIS.}}
\end{figure}
%Fig.9
In fig.9, we show the total cross section, 
\be
\sigma_{\gamma^*p} (W^2, Q^2) = \sigma_{\gamma^*p} \left(\eta (W^2, Q^2),
  \frac{m^2_0}{\Lambda^2_{sat} (W^2)} , \xi = \xi_{\rm exp} = 130 \right)
\label{5.7}
\ee
as a function of the low-$x$ scaling variable $\eta (W^2, Q^2)$. 
The upper and the lower theoretical curve in fig.9 refer to the variation of
$\sigma^{(\infty)} (W^2)$ under variation of the energy $W$, i.e. $\sigma (W =
275 {\rm GeV}) \equiv \sigma^{(\infty)} (W^2 = 275^2 {\rm GeV}^2)$ and $\sigma
(W = 10 {\rm GeV})
\equiv \sigma^{(\infty)} (W^2 = 100 {\rm GeV}^2)$. It is interesting to note
that the violation of scaling in $\eta (W^2, Q^2)$ of the order of about 10\%,
as a consequence of the $W$ dependence of the $(q \bar q)p$ dipole cross
section $\sigma^{(\infty)}(W^2)$, is seen in the experimental data: the
high-energy data from ZEUS and H1 lie above the data obtained at lower
energies. 
\begin{figure}
\centerline{\epsfig{file=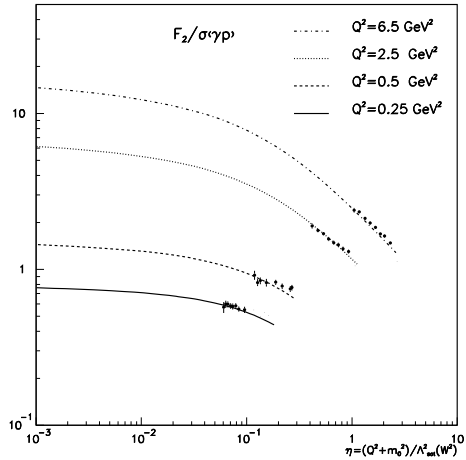,width=7cm}}
\caption{{\footnotesize The approach to the saturation limit of $F_2 (\eta (W^2, Q^2), Q^2) /
\sigma_{\gamma p} (W^2)$ for $\eta (W^2, Q^2) \ll 1$.}}
\end{figure}
%Fig.10
Figure 10 is relevant for the discussion of the limit of $W^2\!\rightarrow\!\infty$
for fixed values of $Q^2$ given in Section 2,
compare(\ref{2.96}) and Table 1. In terms of the structure function $F_2 (x \cong Q^2 / W^2
, Q^2)$ the $W^2 \rightarrow\infty$ limit in (\ref{2.96}) becomes 
\be
\lim\limits_{{W^2\rightarrow\infty}\atop{Q^2 {\rm fixed}}}
\frac{F_2 (x\cong Q^2 / W^2, Q^2)}{\sigma_{\gamma p} (W^2)} =
\frac{Q^2}{4\pi^2\alpha} .
\label{5.8}
\ee 
Higher energies are required to uniquely experimentally verify the expected saturation
property for a larger range of $\eta(W^2,\, Q^2) \ll 1$ and fixed values of $Q^2$. 
\begin{figure}
\centerline{\epsfig{file=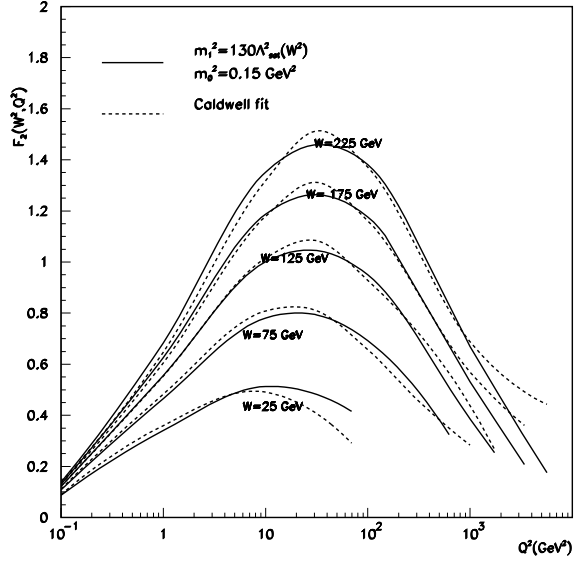,width=8.5cm}}
\caption{{\footnotesize The proton structure function $F_2 (W^2, Q^2)$ as a function of $Q^2$ for
various values of $W$. The theoretical prediction of the CDP is compared with
the Caldwell 2 P-fit as a representation of the experimental data.}}
\end{figure}
%Fig.11
\begin{figure}
\centerline{\epsfig{file=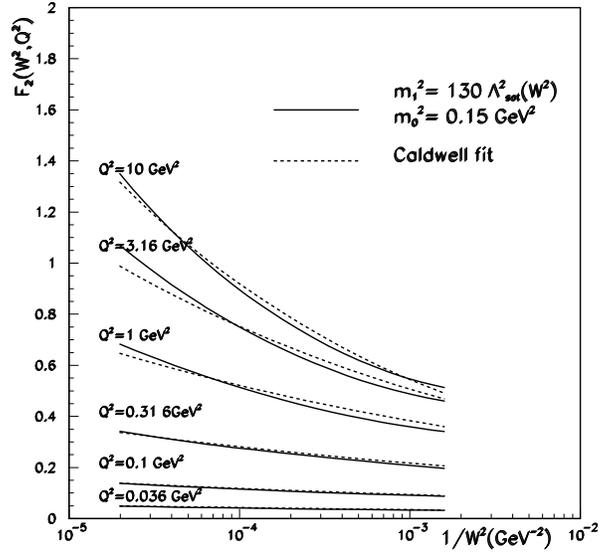,width=8.5cm}}
\caption{{\footnotesize As in Fig.11, but as a function of $1/W^2$ for various values of $Q^2 \le 10
{\rm GeV}^2$.}}
\end{figure}
%Fig.12
\begin{figure}
\centerline{\epsfig{file=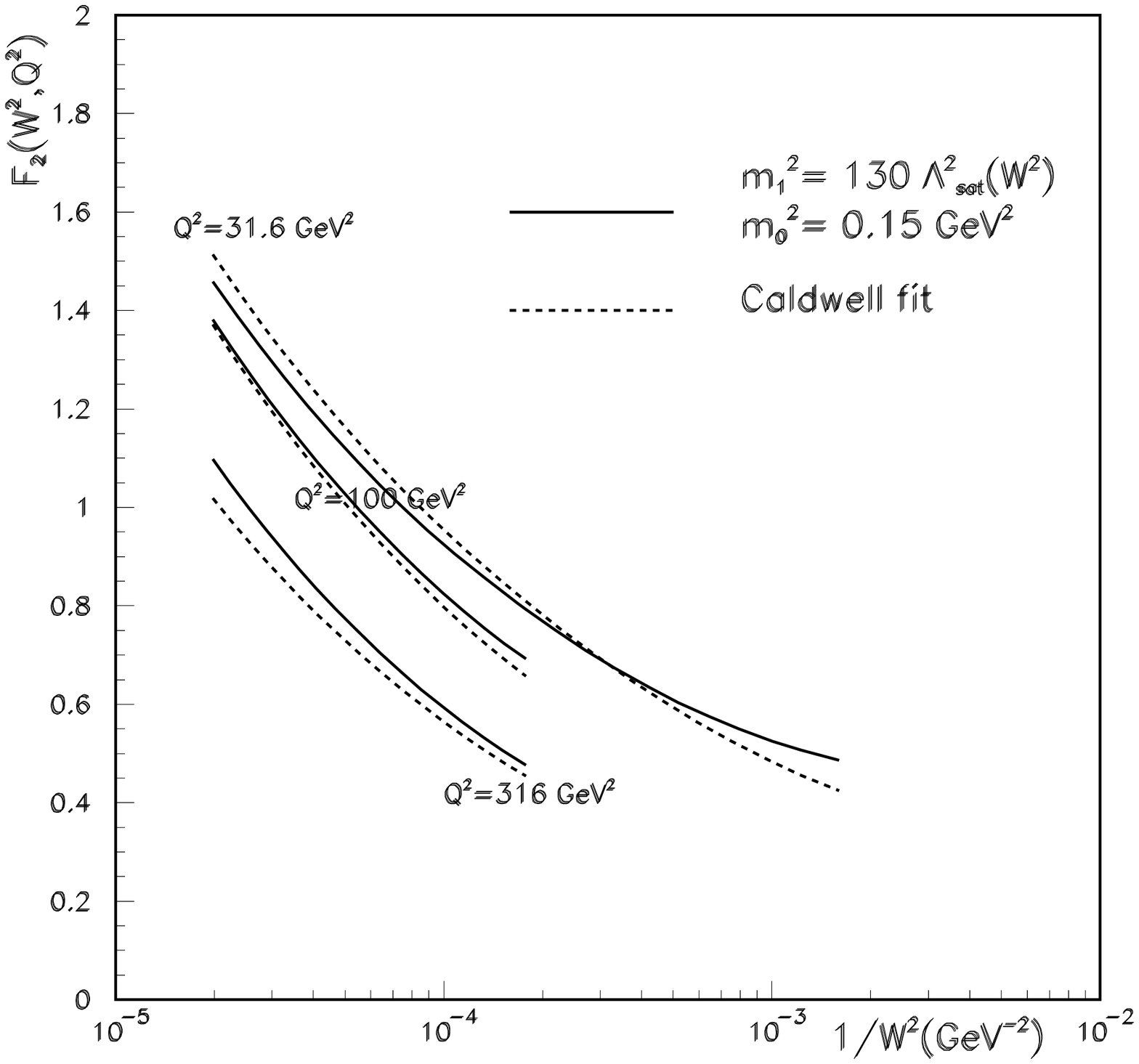,width=7cm}}
\caption{{\footnotesize As in Fig.12, but for $30 {\rm GeV}^2 < Q^2 < 316 {\rm GeV}^2$.}}
\end{figure}
%Fig.13

In figs.11 to 13, we show our predictions from the CDP for the proton structure
function $F_2 (W^2, Q^2)$ as a function of $Q^2$ for fixed values of $W^2$, and
as a function of $W^2$ for fixed values of $Q^2$. For comparison, we also show
the results of a very precise fit to the world experimental data for $F_2 (x,
Q^2)$ for $x < 0.025$ (and $Q^2 > 0$) carried out by Caldwell \cite{CAL}. 
In particular, we show the results from the so-called  2P-fit that is based on
the simple ansatz \cite{CAL}
\be
\sigma_{\gamma^*p} = \sigma_0 \frac{M^2}{Q^2 + M^2} \left( \frac{l}{l_0}
\right)^{\epsilon_0 + (\epsilon_1 - \epsilon_0) \sqrt{\frac{Q^2}{Q^2 +
      \Lambda^2}}}
\label{5.9}
\ee
where
\be
l = \frac{1}{2 x_{bj} M_p} . 
\label{5.10}
\ee
The curves in figs.11 to 13 use the mean values of the six fit parameters
$\sigma_0, M^2, l_0, \epsilon_0, \epsilon_1$ and $\Lambda^2$ given in Table 5
of ref.\cite{CAL}. 
There is acceptable agreement of the predictions of the CDP with the results of
the 2P-fit. 

In figs.14 and 15, we directly compare the theoretical results for $F_2 (W^2, Q^2)$
from the CDP (shown in figs.11 to 13) with the world experimental
data\footnote{We thank Prabhdeep Kaur for providing the plots of the
  experimental data in figs.14 to 17.} \cite{Dur}.
As expected from figs.11 to 13, there is consistency between the CDP and the
experimental data in the full range of $0.036 {\rm GeV}^2 \le Q^2 \le 316 {\rm
  GeV}^2$. The theoretical curves are restricted by the condition of $x \cong
Q^2/W^2 < 0.1$.
\begin{figure}
\hspace{-1cm}\epsfig{file=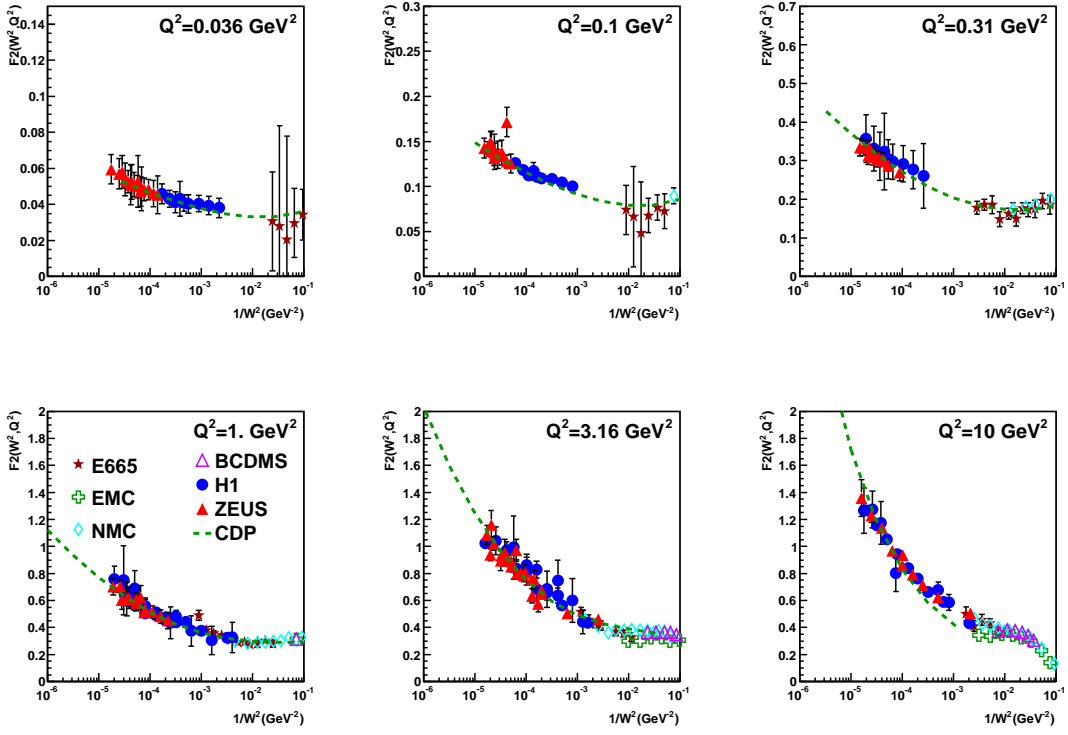,width=15cm}
\caption{{\footnotesize The predictions from the CDP for the structure functionn $F_2 (W^2, Q^2)$
compared with the experimental data for $0.036 {\rm GeV}^2 \le Q^2 \le 10 {\rm GeV}^2$.}}  
\end{figure}
%Fig.14
\begin{figure}
\hspace{-1cm} \epsfig{file=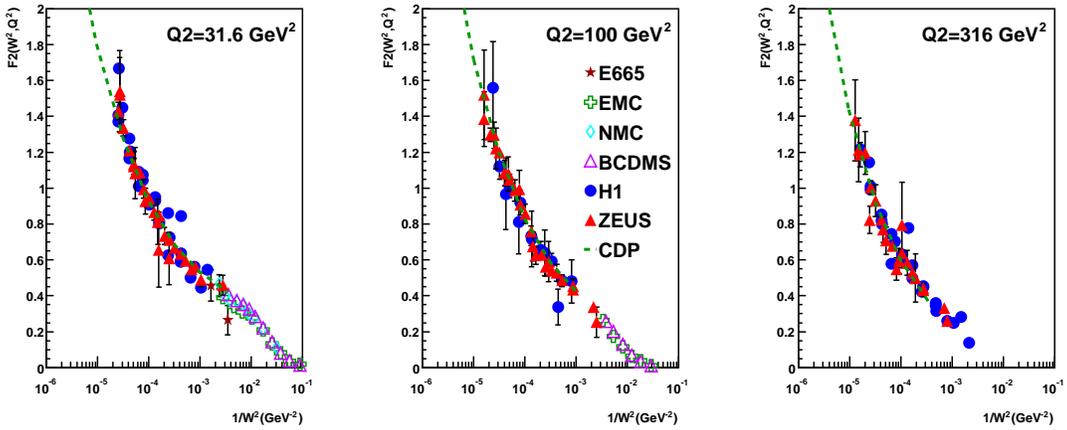,width=15cm}
\caption{{\footnotesize As in fig. 14, but for $31.6 {\rm GeV}^2 \le Q^2 \le 316 {\rm GeV}^2$.}}
\end{figure}
%Fig.15

As noted in the above discussion of the theoretical results in fig.10,
experimental data at much higher energies than available at present are needed
for a detailed verification of the approach to the saturation limit
(\ref{5.8}).
An indication of the proportionality of $F_2 (x \cong Q^2 / W^2 , Q^2)$ to
$Q^2$ according to (\ref{5.8}) becomes visible, however, when comparing the
experimental data in fig.14 for the very low values of $Q^2_1 = 0.036 {\rm
  GeV}^2$ and $Q^2_2 = 0.1 {\rm GeV}^2$ with each other. According to the
proportionality in (\ref{5.8}), for sufficiently large $W^2$ we have 
\bqa
F_2 (W^2, Q^2_2 = 0.1 {\rm GeV}^2) & = & \frac{Q^2_2}{Q^2_1} F_2 (W^2, Q^2_1 =
0.036 {\rm GeV}^2) \nonumber \\
& = & 2.78 F_2 (W^2 , Q^2_1 = 0.036 {\rm GeV}^2 ) . 
\label{5.12}
\eqa

\begin{center}
\begin{tabular}{c|c|c} 
$\frac{1}{W^2} [ {\rm GeV}^{-2} ]$  &  $F_2 (W^2, Q^2_1 = 0.036 {\rm GeV}^2)$ & 
$\frac{Q^2_2}{Q^2_1} F_2 (W^2, Q^2_1 = 0.036 {\rm GeV}^2)$ \\
\hline
$2 \cdot 10^{-5}$ & $\cong 0.055$ & 0.15 \\
$10^{-4}$ & $\cong 0.04$ & 0.11 \\
\hline
\end{tabular}
\end{center}
\begin{itemize}
\item[Table 3] 
\footnotesize{The (approximate) validity of the proportionality (\ref{5.12}). The results in
the second column were read off from fig.14. The predictions from (\ref{5.12})
in the third column (approximately) agree with the experimental results in
fig.14. }
\end{itemize}

The theoretical results for $F_2 (W^2 , Q^2_2 = 0.1 {\rm GeV}^2)$ obtained from (\ref{5.12})
and shown in Table 3 are consistent with the experimental results in fig.14. 

In figs.16 and 17, in addition to the theoretical results in figs.14 and 15, we
show the prediction (\ref{(q)}) of $F_2 (W^2) = f_2 \cdot (W^2 / 1 {\rm
  GeV}^2)^{0.29}$, where $f_2$ is the fitted normalization constant of $f_2 =
0.063$ from (\ref{(r)}), and $W^2 \cong Q^2 / x$. As expected from the
analysis in Section 2.7 and fig.4a, there is agreement between theory and experiment for
$10{\rm GeV}^2 \le Q^2 \le 100 {\rm GeV}^2$ and disagreement for values of
$Q^2$ outside of this range. 

Equation (\ref{(t)}) may be inverted and read as a prediction for $F_2
(W^2 = Q^2/x)$ from the pQCD-improved parton picture in terms of a suitable gluon
distribution i.e. as a prediction for the flavor-singlet quark distribution, according
to 
\be
F_2 (W^2 = \frac{Q^2}{x}) = \frac{5}{18} x \sum (x , Q^2)
= \frac{(2\rho + 1) \sum Q^2_q}{3\pi}
\xi^{C_2}_L \alpha_s (Q^2) G(x , Q^2) . 
\label{5.13}
\ee
In (\ref{5.13}), the numerical values for the gluon-distribution function have
to be inserted, which are obtained by evaluating the right-hand side of the
second equality in (\ref{(t)}). The resulting gluon distributions were shown in
fig.7. Since (\ref{5.13}) coincides with (\ref{(t)}),
the resulting structure function $F_2 (W^2 , Q^2)$ is identical to the one
given by (\ref{(q)}) and shown in figs.16 and 17. 

\begin{figure}
\hspace{-1cm} \epsfig{file=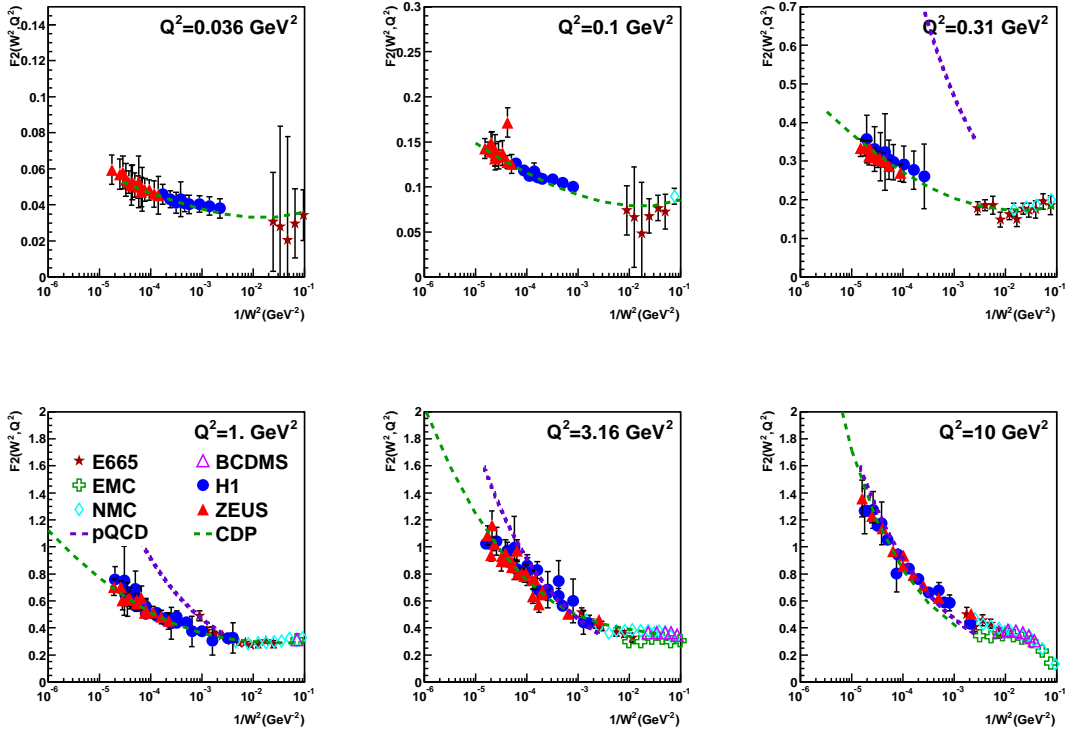,width=15cm}
\caption{{\footnotesize In addition to the prediction from the CDP, also the prediction of 
$F_2 (W^2) = f_2 \cdot (W^2 / 1 {\rm GeV}^2)^{0.29}$ from
(\ref{(q)}) and (\ref{(r)}) (valid for $10 {\rm GeV}^2 \le Q^2 < 100 {\rm GeV}^2$) for $Q^2 \le 10
{\rm GeV}^2$.}}
\end{figure}
%Fig.16
\begin{figure}
\hspace{-1cm} \epsfig{file=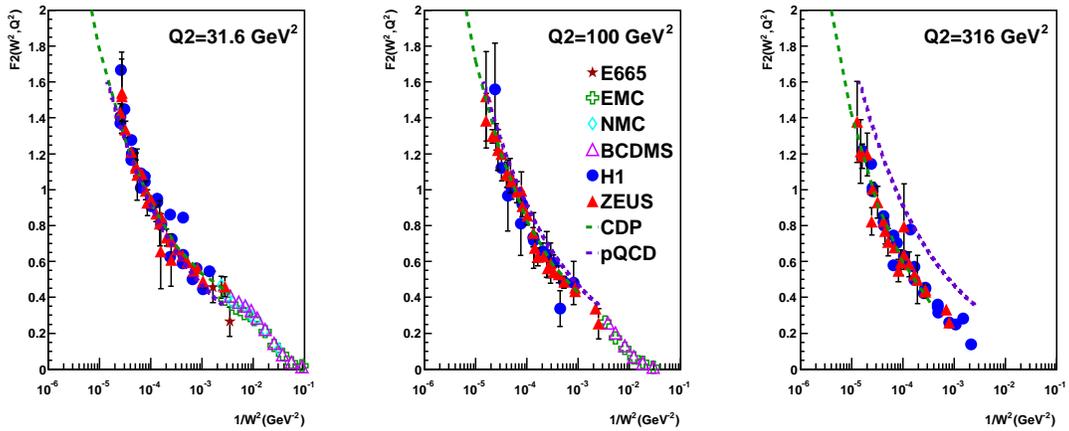,width=15cm}
\caption{{\footnotesize As in fig.16, but for $31.6 {\rm GeV}^2 \le Q^2 \le 316 {\rm GeV}^2$.}}
\end{figure}
%Fig.17
The present interpretation of the results for $F_2 (W^2, Q^2)$ is different,
however. The agreement with experiment in figs.16 and 17 shows that a suitable
choice of the gluon distribution, compare fig.7, yields agreement with
experiment for $F_2 (W^2, Q^2)$ in the relevant range of $10 {\rm GeV}^2 \le
Q^2 \le 100 {\rm GeV}^2$. The results in figs.16 and 17 thus explicitly display
the agreement with the pQCD-improved parton picture based on the gluon
distribution function of fig.7 in Section 2.7. For the ensuing discussion, we
note the proportionality of the gluon distribution function to the saturation
scale, 
\be
\alpha_s (Q^2) G (x, Q^2) \sim \left( \frac{W^2}{1{\rm GeV}^2} \right)^{C_2 =
  0.29} \sim \Lambda^2_{\rm sat} (W^2) \sigma_L^{(\infty)}
\label{5.14}
\ee
that follows from comparing (\ref{5.13}) with the representation of $F_2 (W^2,
Q^2)$ in terms of the saturation scale, $\Lambda^2_{\rm sat} (W^2)$, in 
(\ref{2.55}) with (\ref{2.72}) and $\sigma_L^{(\infty)} \cong {\rm const.}$
Compare also (\ref{2.98}) 
for the approximation of (\ref{5.1}) by the proportionality to $(W^2 / 1 {\rm
  GeV}^2)$ used in (\ref{5.14}). 

The pQCD-improved parton picture in (\ref{5.13}) with the power-like $W^2$
dependence (\ref{5.14}) fails as soon as $\eta (W^2, Q^2) < 1$, or $Q^2 < 10
  {\rm GeV}^2$, compare fig.16. The saturation behavior of the CDP sets in. For
  sufficiently large $W^2$, at any fixed value of $Q^2$, it leads to a
  logarithmic dependence of $\sigma_{\gamma^* p} (W^2, Q^2)$, and of $F_2 (W^2,
  Q^2)$, on the energy $W$, or on $\Lambda^2_{\rm sat} (W^2)$ as given in 
(\ref{2.93}), (\ref{2.96}) and (\ref{5.8}),
\bqa
F_2 (W^2, Q^2) &\sim & Q^2 \sigma_L^{(\infty)} \ln \frac{\Lambda^2_{\rm sat}
  (W^2)}{Q^2 + m^2_0} \label{5.15} \\
& \sim & Q^2 \sigma_L^{(\infty)} \ln \left( \frac{\alpha_s (Q^2) G (x ,
    Q^2)}{\sigma_L^{(\infty)} (Q^2 + m^2_0)} \right) , 
~~~~~ ( {\rm for}~ \eta (W^2, Q^2) \ll 1 ) . \nonumber
\eqa
In distinction from the pQCD-improved parton picture in (\ref{5.13}), for \\
$\eta (W^2, Q^2) < 1$, the
structure function in (\ref{5.15}) depends logarithmically on the gluon
distribution function. 

The CDP with its $W$-dependent $(q \bar q)$-dipole-proton cross section is
unique in providing a smooth transition from the region of $\eta (W^2, Q^2) >
1$, with pieceful coexistence between the CDP and the pQCD-improved parton
picture, to the saturation region of $\eta (W^2, Q^2)< 1$, exclusively governed
by the CDP. The pQCD-improved parton picture is not allowed to invade the
region of $\eta (W^2, Q^2) < 1$. The suppressed gluon distribution function at
$x < 10^{-2}$, occasionally with even negative results, from global fits
(compare fig.7) is presumably related to the inclusion of experimental
data for $F_2 (x, Q^2)$ at very low values of $Q^2$, where saturation must
actually be taken into account, compare (\ref{5.15}). 

The CDP, in distinction from the discrimination between a soft and a hard
Pomeron of the low-$x$ Regge picture \cite{Dom}, only knows of a single Pomeron
governing both the regions of $\eta (W^2, Q^2) > 1$ and of $\eta (W^2, Q^2) <
1$. The transition from $\eta (W^2, Q^2) > 1$ to $\eta (W^2, Q^2) <1$, or to
$W^2 \rightarrow \infty$ at fixed $Q^2$, is not associated with the transition
to a (first or second) soft-Pomeron exchange. The transition corresponds to
$\Lambda^2_{\rm sat} (W^2) \rightarrow \ln \Lambda^2_{\rm sat} (W^2)$, or
equivalently to $\alpha_s (Q^2) G (x , Q^2) \rightarrow \ln (\alpha_s (Q^2) G
(x, Q^2))$.
The single Pomeron of the CDP has a less strong increase of the corresponding
gluon distribution in fig.7 with decreasing $x$, when compared with the hard
Pomeron of the Regge fit. 
\begin{figure}
\centerline{\epsfig{file=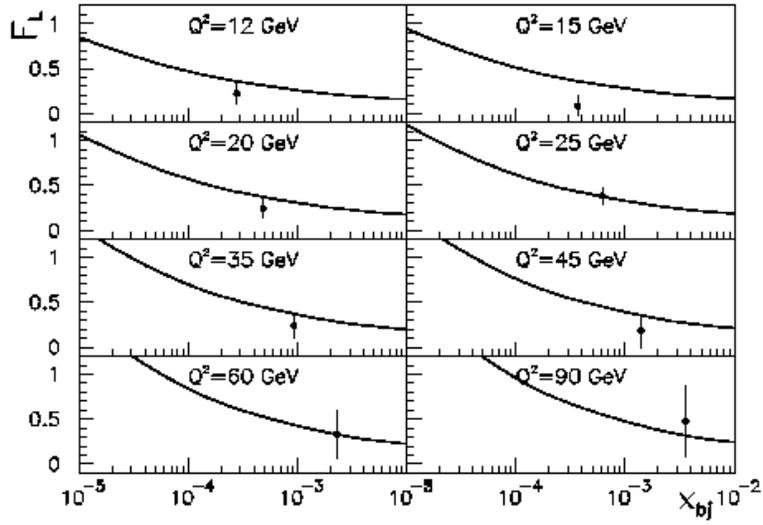,width=11cm} }
\caption{{\footnotesize The experimental results on the longitudinal structure function $F_L(x,Q^2)$
from the H1 collaboration\cite{h1data} compared with the prediction from the CDP.}}
\end{figure}                                                                     
%Fig.18
\begin{figure}
\centerline{\epsfig{file=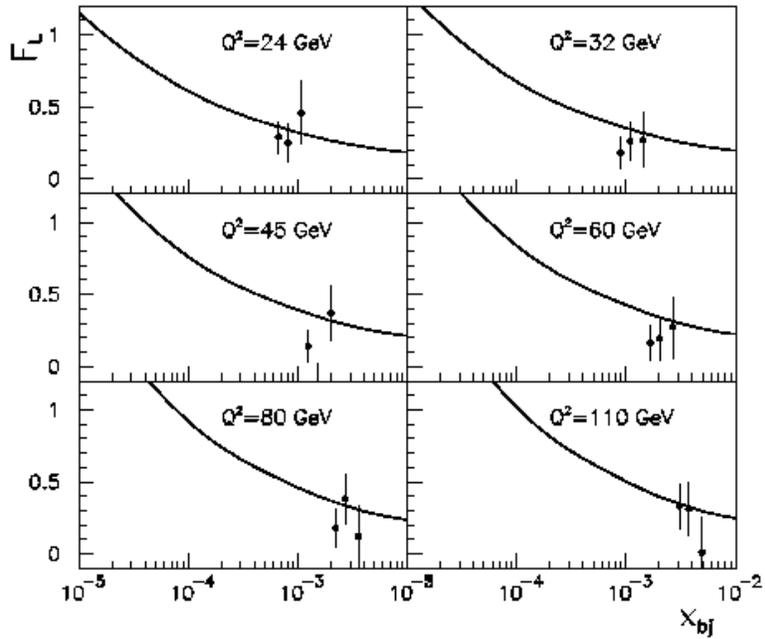,width=11cm} }
\caption{{\footnotesize As in fig.18, but showing the experimental results from the ZEUS collaboration\cite{zeusdata}.}}
\end{figure}
%Fig.19

In figs.18 and 19, we show a comparison of our predictions for the longitudinal
structure function $F_L(x,Q^2)$ with the experimental data. Since our ansatz
for the dipole cross section incorporates transverse-size enhancement, 
$\rho = {\rm const} = 4/3$, the
theoretical results in figs.18 and 19 agree with the ones in figs. 2 and 3.

\section{Conclusion}

In the present paper we reexamined and reanalysed DIS at low values of the
Bjorken scaling variable $x \cong Q^2/W^2 < 0.1$ in terms of the CDP with a
$W$-dependent color-dipole cross section. We explicitly showed that 
all essential features of the experimental data on the longitudinal and the transverse
photoabsorption cross section can be understood as a consequence of the
color-gauge-invariant $q \bar q$-dipole-proton interaction, without relying on
any specific parameterization of the dipole-proton cross section. 

We also examined the consistency between the description of the experimental
data in the CDP and the description in terms of $q \bar q$-sea and gluon
distributions of the pQCD-improved parton picture within its range of
validity. 

The resulting $(Q^2, W^2)$ plane of DIS at low $x$ consists of only two
regions, separated by the line $\eta (W^2, Q^2) \cong 1$. 

For $\eta (W^2, Q^2) \cong Q^2 / \Lambda^2_{\rm sat} (W^2) \gg 1$ i.e. for
sufficiently large $Q^2$, color transparency of the color-dipole-proton cross
section becomes relevant: the strong destructive interference among different
dipole-proton scattering amplitudes originating as a consequence of color-gauge
invariance implies a $(q \bar q)$-proton interaction that vanishes proportional
to the transverse dipole size, $\vec r^{~2}_\bot$. The photoabsorption cross
section correspondingly behaves as $\Lambda^2_{\rm sat} (W^2) / Q^2$, and the
proton structure function 
(for $10 {\rm GeV}^2 \!\! \le\!\!  Q^2\! \le\! 100 {\rm GeV}^2$) as $F_2 (x, Q^2) =
F_2 (W^2 = Q^2 / x)$. 

The experimental data for $\eta (W^2, Q^2)>1$ can alternatively be
represented in terms of the $(q \bar q)$-sea-quark and the gluon distribution
of the pQCD-improved parton picture. Consistency of the pQCD approach with the
CDP requires the gluon distribution function to be proportional to the
saturation scale, $\Lambda^2_{\rm sat} (W^2)$, and implies a definite value for
the exponent $C_2$ in the representation of the saturation scale,
$\Lambda^2_{\rm sat} (W^2) \sim (W^2)^{C_2}$. The resulting prediction, $C_2
\cong 0.27$ to $C_2 \cong 0.29$, is consistent with the experimental data. 
The formulation of the
CDP in terms of a $W$-dependent ($Q^2$-independent) color-dipole-proton cross
section is essential to arrive at this conclusion. 

With increasing energy, $W$, for any fixed dipole size, 
$\vec r_\bot^{~2}$, again due to color-gauge invariance, the destructive
interference among different amplitudes contributing to the $q \bar q$
interaction with the color field of the nucleon dies out and leads to an 
$\vec
r^{~2}_\bot$-independent limit for the $(q \bar q)$-proton cross section. The
$q \bar q$-proton cross section ``saturates'' in this high-energy limit to
become identical to a cross section of hadronic size. 

The limit of increasingly larger energy, $W$, at fixed dipole size in the
photoabsorption process is realized by $W^2 \rightarrow \infty$ at fixed $Q^2$,
or $\eta (W^2, Q^2) \ll 1$. The photoabsorption cross section increases
logarithmically with the energy according to $\ln \Lambda^2_{\rm sat} (W^2)$,
and for $W^2 \rightarrow \infty$ at any fixed value of $Q^2$,  
it reaches the limit of $(Q^2 = 0)$
photoproduction. The pQCD-improved parton picture fails, insofar as the
photoabsorption cross section in this limit depends logarithmically on the
($W$-dependent) gluon distribution function. 

A concrete parameterization of the dipole cross section is necessary for the
interpolation between the regions of $\eta (W^2, Q^2)>1$ and $\eta (W^2, Q^2)
<1$. We refined previous work in several respects, the representation of the
longitudinal-to-transverse ratio of the photoabsorption cross section by taking
into account the transverse-size enhancement of $q \bar q$ fluctuations
originating from transversely polarized photons, the extension of the CDP to
include the region of $x$ increasing to values close to $x = 0.1$,  
among others. We found
agreement with the available DIS data in the full range of $0.036 {\rm GeV}^2
\le Q^2 \le 316 {\rm GeV}^2$ for $x \le 0.1$.

\vspace{1cm}
\noindent
{\bf Acknowledgement}

\medskip
Useful discussions with Allen Caldwell and Reinhart K\"ogerler, as well as the
help of Prabdeep Kaur for providing plots of experimental data, are gratefully
acknowledged.

\clearpage

\begin{appendix}

\renewcommand{\theequation}{\Alph{section}.\arabic{equation}} 
\setcounter{section}{1}
\setcounter{equation}{0}
\section*{Appendix A. Derivation of (\ref{4.1}) and (\ref{4.2})}
In this Appendix, we derive the photoabsorption cross section 
in the momentum space (\ref{4.1}) and (\ref{4.2}) from 
the coordinate representation (\ref{2.35}).
We start with the integral representation of the modified Bessel function
\bq
  K_0(r^\prime_\perp Q) ={1\over{2\pi}}\int d^2\vec k_\perp^{~\prime} 
            {1\over{Q^2 +\vec k^{~\prime 2}_\perp }}
             e^{-i\vec r^{~\prime}_\bot \cdot\vec k_\bot^{~\prime}}
\label{a1}
\eq
where
\bq
     r^\prime_\perp =|\vec r^{~\prime}_\bot |, ~~Q=\sqrt{Q^2}.
\label{a2}
\eq
(\ref{a1}) can be easily verified from the following equations,
\bqa
   \int_0^{2\pi} d\theta \exp\left(-iz\cos\theta\right) &=&2\pi J_0(z),
\label{a3} \\
   \int_0^\infty dx{x\over{Q^2+x^2}}J_0(r^\prime_\perp x) &=&
        K_0(r_\perp^\prime Q).
\label{a4}
\eqa
We compute the following quantity
\bq
   I_L(\vec l_\perp^{~\prime 2})\equiv \int d^2\vec r^{~\prime}_\bot 
     K^2_0(r_\perp^\prime Q)
           e^{-i\vec r^{~\prime}_\bot\cdot \vec l_\bot^{~\prime}}
\label{a5}
\eq
Inserting (\ref{a1}), we find
\bqa
  I_L(\vec l_\perp^{~\prime 2})
   &=&{1\over{(2\pi)^2}}\int d^2\vec r_\perp^{~\prime} 
        \int d^2\vec k^{~\prime}_\perp   
      \nonumber \\
   &&~~~~~~~~~~     \int d^2\vec k_\perp^{~\prime\prime}
       {1\over{(Q^2+\vec k_\perp^{~\prime 2})
       (Q^2+\vec k_\perp^{~\prime\prime 2})}}
       e^{-i\vec r_\perp^{~\prime}\cdot(\vec k_\perp^{~\prime} 
          +\vec k_\perp^{~\prime\prime}
          +\vec l_\perp^{~\prime})}
     \nonumber \\
    &=& \int d^2\vec k_\perp^\prime
        {1\over{(Q^2+\vec k_\perp^{~\prime 2})(Q^2+(\vec k_\perp^\prime+
                   \vec l_\perp^{~\prime})^2 )}}
      \nonumber \\
    &=& \int d \vec k^{~\prime 2}_\bot \int_0^\pi
        d\vartheta
        {1\over{(Q^2+\vec k_\perp^{~\prime 2})(Q^2+(\vec k_\perp^\prime+
                   \vec l_\perp^{~\prime})^2 )}}
\label{a6}
\eqa
where $\vartheta$ is an angle between $\vec k_\perp^{~\prime}$ and 
$\vec l_\perp^\prime$.
Recalling (\ref{4.9}), we have
\bq
     d\vartheta = -\omega(M^2,M^{\prime 2},\vec l_\perp^{~\prime 2})
                  dM^{\prime 2}.
\label{a7}
\eq 
%Using (\ref{4.3}) and (\ref{4.4}), we can express the above
%expression in terms of $M^2$ and $M^{\prime 2}$,
%\bq
%   I_L(l_\perp^{\prime 2})
%   = \int dM^2 \int dM^{\prime 2}\omega(M^2,M^{\prime 2},
%        \vec l_\perp^{\prime 2})
%        {1\over{(Q^2+M^2)(Q^2+M^{\prime 2})}}
%\label{a7}
%\eq
Inserting (\ref{2.40}) and (\ref{a6}) and using(\ref{a7}), 
the integral in (\ref{2.35}) becomes
\bqa
  &&\int dr_\perp^{\prime 2} K_0^2(r_\perp^\prime Q)
     \sigma_{(q\bar q)_L^{J=1}p}(r_\perp^\prime,W^2)
    \nonumber \\
  &=& \int dl_\perp^{\prime 2} 
      \bar\sigma_{(q\bar q)_L^{J=1}p}(l_\perp^{\prime 2},W^2)
      \left( I_L(0)-I_L(l^{\prime 2}_\perp)\right)
    \label{a8} \\
  &=& \int dl_\perp^{\prime 2} 
      \bar\sigma_{(q\bar q)_L^{J=1}p}(l_\perp^{\prime 2},W^2)
      \nonumber \\
  &&  \int dM^2 \int dM^{\prime 2}\omega(M^2,M^{\prime 2},
         \vec l_\perp^{~\prime 2})
       \left( {1\over{(Q^2+M^2)^2}}- {1\over{(Q^2+M^2)(Q^2+M^{\prime 2})}} 
       \right), \nonumber
\eqa
which leads to (\ref{4.1}).

The transverse cross section (\ref{4.2}) is derived in a similar manner.
Differentiating (\ref{a1}) with respect to $\vec r^{~\prime}_\perp$,
one finds
\bq
  {{\vec r^{~\prime}_\perp}\over{r_\perp^\prime}}\sqrt{Q^2}K_1(r_\perp^\prime Q)
    ={i\over{2\pi}}\int d^2k_\perp^\prime
      {{\vec k_\perp^\prime }\over{Q^2+\vec k_\perp^{~\prime 2}}}
       e^{-i\vec r^{~\prime}_\perp \cdot\vec k_\perp^{~\prime}}
\label{a9}
\eq
The integral 
\bq
   I_T(l_\perp^{\prime 2})\equiv \int d^2\vec r^{~\prime}_\perp 
     K^2_1(r_\perp^\prime Q)
           e^{-i\vec r^{~\prime}_\perp \cdot\vec l_\bot^{~\prime}}
\label{a10}
\eq
can be evaluated as
\bqa
   I_T(l_\perp^{\prime 2})
   &=&{1\over{(2\pi)^2}}\int d^2\vec r_\perp^{~\prime} 
        \int d^2\vec k^\prime_\perp  
        \nonumber \\
   &&~~~~~~~~~~     \int d^2\vec k_\perp^{~\prime\prime}
       {{-\vec k^\prime_\perp\cdot\vec k_\perp^{~\prime\prime}}\over 
       {Q^2(Q^2+\vec k_\perp^{~\prime 2})
       (Q^2+\vec k_\perp^{~\prime\prime 2})}}
       e^{-i\vec r^{~\prime}_\perp\cdot(\vec k_\perp^\prime 
          +\vec k_\perp^{~\prime\prime}
          +\vec l_\perp^{~\prime})}
     \nonumber \\
    &=& {1\over{Q^2}}\int d^2\vec k_\perp^\prime
        {{\vec k_\perp^\prime \cdot(\vec k_\perp^\prime+\vec l_\perp^{~\prime})}
         \over{(Q^2+\vec k_\perp^{~\prime 2})(Q^2+(\vec k_\perp^\prime+
                   \vec l_\perp^{~\prime})^2 )}}
      \nonumber \\
    &=& {1\over{Q^2}}\int dk_\perp^{\prime 2} \int_0^\pi
        d\vartheta
        {{\vec k_\perp^\prime \cdot(\vec k_\perp^\prime+\vec l_\perp^{~\prime})}
        \over{(Q^2+\vec k_\perp^{\prime 2})(Q^2+(\vec k_\perp^\prime+
                   \vec l_\perp^{~\prime})^2 )}}
%       \nonumber \\
%   &=& {1\over{Q^2}}\int dM^2 \int dM^{\prime 2}\omega(M^2,M^{\prime 2},
%        \vec l_\perp^{\prime 2})
%        {{M^2+M^{\prime 2}-l_\perp^{\prime 2}}
%         \over{2(Q^2+M^2)(Q^2+M^{\prime 2})}}
\label{a11}
\eqa
Inserting (\ref{2.40}) and (\ref{a11}), the integral in (\ref{2.35}) becomes
\bqa
  &&\int dr_\perp^{\prime 2} K_1^2(r_\perp^\prime Q)
     \sigma_{(q\bar q)_T^{J=1}p}(r_\perp^\prime,W^2)
    \nonumber \\
  &=& \int dl_\perp^{\prime 2} 
      \bar\sigma_{(q\bar q)_T^{J=1}p}(\vec l_\perp^{~\prime 2},W^2)
      \left( I_T(0)-I_T(\vec l^{~\prime 2}_\perp)\right)
    \label{a12} \\
  &=& {1\over{Q^2}}\int d \vec l_\perp^{~\prime 2} 
      \bar\sigma_{(q\bar q)_T^{J=1}p}(\vec l_\perp^{~\prime 2},W^2)
      \nonumber \\
  &&  \int dM^2 \int dM^{\prime 2}\omega(M^2,M^{\prime 2},
         \vec l_\perp^{~\prime 2})
       \left( {{M^2}\over{(Q^2+M^2)^2}}- 
        {{M^2+M^{\prime 2}-\vec l_\perp^{~\prime 2}}
        \over{2(Q^2+M^2)(Q^2+M^{\prime 2})}} 
       \right), \nonumber
\eqa
which leads to (\ref{4.2}).

%\begin{appendix} 
 
\section*{Appendix B Correction terms}

\renewcommand{\theequation}{\Alph{section}.\arabic{equation}} 
\setcounter{section}{2} 
\setcounter{equation}{0}

In this Appendix, we will give the explicit expressions for the correction 
terms, $\Delta \sigma^{(m^2_0)}_{\gamma^*_{L,T}p} (W^2, Q^2)$ and  
$\Delta \sigma^{(m^2_1)}_{\gamma^*_{L,T}p} (W^2, Q^2)$ in (\ref{4.17}), which  
in conjunction with the dominant term guarantee the required bound on  
$M^{\prime 2}$ that is given by $m^2_0 \le M^{\prime 2} \le m^2_1 (W^2) \equiv 
m^2_1$ from (\ref{4.7}).  
 
With the splitting of the integrand (\ref{4.10}) as applied to the dominant 
term, the integrations over $d\vartheta$ in (\ref{4.12}) and (\ref{4.13})
yield the following results for the correction terms in (\ref{4.14}),
\bqa 
& & \Delta\sigma^{(m^2_0)}_{\gamma^*_Lp} (W^2, Q^2)  
   +\Delta \sigma^{(m^2_1)}_{\gamma^*_Lp} (W^2, Q^2)  
 = -\frac{\alpha R_{e^+e^-}}{3} \int d \vec l^{~\prime 2}_\bot  
   \bar\sigma_{(q \bar q)^{J=1}_Lp}(\vec l^{~\prime 2}_\bot , W^2) \cdot  
    \nonumber \\ 
& &~~~~~ \left( \int^{(\sqrt{\vec l^{~\prime 2}_\bot}+m_0)^2} 
               _{(\sqrt{\vec l^{~\prime 2}_\bot}-m_0)^2}  
                d M^2 S_{L,0} (M^2 , \vec l^{~\prime 2}_\bot , Q^2, m^2_0) 
            \right. \nonumber \\
&&~~~~~~~~~~\left.  
         + \int^{m^2_1}_{(m_1-\sqrt{\vec l^{~\prime 2}_\bot})^2}  
          d M^2 S_{L,1} (M^2 , \vec l^{~\prime 2}_\bot , Q^2, m^2_1) \right) ,  
\label{B1}  
\eqa 
and  
\bqa 
& & \Delta\sigma^{(m^2_0)}_{\gamma^*_Tp} (W^2, Q^2)  
   +\Delta \sigma^{(m^2_1)}_{\gamma^*_Tp} (W^2, Q^2)  
 = -\frac{\alpha R_{e^+e^-}}{6} \int d \vec l^{~\prime 2}_\bot  
   \bar\sigma_{(q \bar q)^{J=1}_Tp}(\vec l^{~\prime 2}_\bot , W^2) \cdot  
    \nonumber \\ 
& &~~~~~ \left( \int^{(\sqrt{\vec l^{~\prime 2}_\bot}+m_0)^2} 
               _{(\sqrt{\vec l^{~\prime 2}_\bot}-m_0)^2}  
                d M^2 S_{T,0} (M^2 , \vec l^{~\prime 2}_\bot , Q^2, m^2_0)
    \right. \nonumber \\  
&&~~~~~~~~~~\left. + \int^{m^2_1}_{(m_1-\sqrt{\vec l^{~\prime 2}_\bot})^2}  
          d M^2 S_{T,1} (M^2 , \vec l^{~\prime 2}_\bot , Q^2, m^2_1) \right) ,  
\label{B2}  
\eqa 
where 
\small
\bqa 
 S_{L,0}(M^2 , \vec l^{~\prime 2}_\bot , Q^2, m^2_0) 
 &=& {{Q^2}\over{(Q^2+M^2)^2}}{{\pi-\vartheta_0(M^2,\vec l^{~\prime 2}_\bot ,m_0^2)} 
         \over \pi}  
\label{B3} \\ 
 &&      -{{Q^2}\over{(Q^2+M^2)\sqrt X}}\left(1-{2\over \pi} 
          {\rm arctan}\sqrt{Y(M^2 , \vec l^{~\prime 2}_\bot , Q^2 , m^2_0)}\right) 
\nonumber \\ 
  S_{L,1}(M^2 , \vec l^{~\prime 2}_\bot , Q^2, m^2_1) 
  &=&{{Q^2}\over{(Q^2+M^2)^2}}{{\vartheta_1(M^2,\vec l^{~\prime 2}_\bot ,m_1^2)}\over \pi}  
\label{B4} \\ 
  &&        -{{Q^2}\over{(Q^2+M^2)\sqrt X}}{2\over \pi} 
          {\rm arctan}\sqrt{Y(M^2 , \vec l^{~\prime 2}_\bot , Q^2 , m^2_1) }  
\nonumber \\ 
 S_{T,0}(M^2 , \vec l^{~\prime 2}_\bot , Q^2, m^2_0) 
 &=& {{M^2-Q^2}\over{(Q^2+M^2)^2}}{{\pi-\vartheta_0(M^2,\vec l^{~\prime 2}_\bot ,m_0^2)} 
         \over \pi}  
\label{B5} \\ 
 &&      -{{M^2- \vec l^{~\prime 2}_\bot -Q^2}\over{(Q^2+M^2)\sqrt X}}\left(1-{2\over \pi} 
          {\rm arctan}\sqrt{Y(M^2 , \vec l^{~\prime 2}_\bot , Q^2 , m^2_0)}\right) 
\nonumber \\ 
  S_{T,1}(M^2 , \vec l^{~\prime 2}_\bot , Q^2, m^2_1) 
  &=&{{M^2-Q^2}\over{(Q^2+M^2)^2}}{{\vartheta_1(M^2,\vec l^{~\prime 2}_\bot ,m_1^2)}\over \pi}  
\label{B6} \\ 
  &&        -{{M^2- \vec l^{~\prime 2}_\bot -Q^2}\over{(Q^2+M^2)\sqrt X}}{2\over \pi} 
          {\rm arctan}\sqrt{Y(M^2 , \vec l^{~\prime 2}_\bot , Q^2 , m^2_1) }  
\nonumber 
\eqa 
\normalsize
\noindent
In (\ref{B3}) - (\ref{B6}),  
\bqa 
  \vartheta (M^2, \vec l^{~\prime 2}_\bot , m^2_{0,1}) &=&  
  \arccos \frac{m^2_{0,1} -M^2 - \vec l^{~\prime 2}_\bot} 
  {2M \sqrt{\vec l^{~\prime 2}_\bot}} , 
\label{B7} \\ 
   X(M^2 , \vec l^{~\prime 2}_\bot, Q^2) &=& (M^2 - \vec l^{~\prime 2}_\bot +  
Q^2)^2 + 4 Q^2 \vec l^{~\prime 2}_\bot , \nonumber \\ 
   Y (M^2 , \vec l^{~\prime 2}_\bot , Q^2, m^2_{0,1})  
 &=& \frac{Q^2 + (M - \sqrt{\vec l^{~\prime 2}_\bot})^2} 
          {Q^2 + (M + \sqrt{\vec l^{~\prime2}_\bot})^2} \cdot  
     \frac{1 - \cos\vartheta (M^2 , \vec l^{~\prime 2}_\bot , m^2_{0,1})} 
          {1 + \cos\vartheta (M^2 , \vec l^{~\prime 2}_\bot , m^2_{0,1})} \nonumber \\ 
 &=& - \frac{Q^2 + (M - \sqrt{\vec l^{~\prime 2}_\bot})^2} 
            {Q^2 + (M + \sqrt{\vec l^{~\prime 2}_\bot})^2} \cdot  
       \frac{(\sqrt{\vec l^{~\prime 2}_\bot}+ M)^2 - m^2_{0,1}} 
            {(\sqrt{\vec l^{~\prime 2}_\bot} - M)^2 - m^2_{0,1}} > 0 
     \nonumber \\ 
  &&~~~~~~~~~~~~~~~~~~~~~~~~{\rm for}~\sqrt{\vec l^{~\prime 2}_\bot} > 2 m_0 , \nonumber  
\eqa 
For photoproduction, $Q^2=0$, from (\ref{B2}) and (\ref{B5}), (\ref{B6}),  
we have the simplified expression 
\small
\bqa 
& & \Delta \sigma^{(m^2_0)}_{\gamma^*_T p} (W^2 , Q^2 = 0) +  
\Delta\sigma^{(m^2_1)}_{\gamma^*_T p} (W^2, Q^2=0) = \nonumber \\ 
&  & = -\frac{\alpha R_{e^+ e^-}}{6} \int d \vec l^{~\prime 2}_\bot  
\bar\sigma_{(q \bar q)^{J=1}_T p} (\vec l^{~\prime 2}_\bot , W^2) \cdot  
\label{B8} \\ 
& & \cdot \left(  \int^{(\sqrt{\vec l^{~\prime 2}_\bot} + m_0)^2}_ 
   {(\sqrt{\vec l^{~\prime 2}_\bot}-m_0)^2} {{d M^2}\over{M^2}} 
   \left( {{\pi-\vartheta_0(M^2,\vec l^{~\prime 2}_\bot ,m_0^2)}\over \pi}  
         \right.\right. \nonumber \\ 
  &&~~~~~~~~~~~~~~~~~~~~~~~~~~\left. \left.  
     -{{M^2- \vec l^{~\prime 2}_\bot }\over{|M^2- \vec l^{~\prime 2}_\bot |}} 
     \left(1-{2\over \pi} 
          {\rm arctan}\sqrt{Y(M^2 , \vec l^{~\prime 2}_\bot , Q^2 = 0 , m^2_0)}\right) 
     \right) \right. 
   \nonumber \\ 
& & \left.+ \int^{m^2_1}_{(m_1 - \sqrt{\vec l^{~\prime 2}_\bot})^2}{{dM^2}\over{M^2}} 
  \left( {{\vartheta_1(M^2,\vec l^{~\prime 2}_\bot ,m_1^2)}\over \pi} 
       -{2\over \pi} 
          {\rm arctan}\sqrt{Y(M^2 , \vec l^{~\prime 2}_\bot , Q^2 = 0 , m^2_1) } \right) 
   \right) 
\nonumber 
\eqa
\normalsize
\noindent 
for $m_1>2\sqrt{\vec l_\perp^{~\prime 2}}$. 
 
Specializing the dipole cross section in (\ref{B1}) to the ansatz (\ref{3.13})  
and its $J=1$ projections in (\ref{3.14}), the  
longitudinal cross section in (\ref{B1}) becomes 
\bqa 
& & \Delta\sigma^{(m^2_0)}_{\gamma^*_L p} (W^2, Q^2) + \Delta\sigma^ 
{(m^2_1}_{\gamma^*_L p} (W^2,Q^2) 
    = -\frac{\alpha R_{e^+ e^-}}{4} \frac{\sigma^{(\infty)}}{\pi}  
      \int^1_{\frac{2}{3a}} dy \frac{y}{\sqrt{1-y}} \nonumber\\ 
&& ~~~~~
 \left( \int^{(\sqrt{\vec l^{~\prime 2}_\bot}+m_0)^2} 
               _{(\sqrt{\vec l^{~\prime 2}_\bot}-m_0)^2}  
                d M^2 S_{L,0} (M^2 , \vec l^{~\prime 2}_\bot , Q^2, m^2_0)  
     \right. \nonumber \\
&&~~~~~~~~~~\left.
         + \int^{m^2_1}_{(m_1-\sqrt{\vec l^{~\prime 2}_\bot})^2}  
          d M^2 S_{L,1} (M^2 , \vec l^{~\prime 2}_\bot , Q^2, m^2_1) \right) ,  
\label{B9}  
\eqa 
while for the transverse cross section, we have  
\bqa 
& & \Delta\sigma^{(m^2_0)}_{\gamma^*_T p} (W^2, Q^2)  
  + \Delta\sigma^{(m^2_1}_{\gamma^*_T p} (W^2,Q^2)   
    = -\frac{\alpha R_{e^+ e^-}}{8} \frac{\sigma^{(\infty)}}{\pi}  
\int^1_{\frac{2}{3a}} dy \frac{(1-\frac{1}{2}y)}{\sqrt{1-y}} \nonumber\\ 
&& ~~~~~
   \left( \int^{(\sqrt{\vec l^{~\prime 2}_\bot}+m_0)^2} 
               _{(\sqrt{\vec l^{~\prime 2}_\bot}-m_0)^2}  
                d M^2 S_{T,0} (M^2 , \vec l^{~\prime 2}_\bot , Q^2, m^2_0)  
    \right. \nonumber \\
&&~~~~~~~~~~\left. + \int^{m^2_1}_{(m_1-\sqrt{\vec l^{~\prime 2}_\bot})^2}  
          d M^2 S_{T,1} (M^2 , \vec l^{~\prime 2}_\bot , Q^2, m^2_1) \right) ,  
\label{B10}  
\eqa 
$\vec l^{~\prime 2}_\bot$ on the right-hand side in (\ref{B9}) and (\ref{B10}) 
 is to be replaced by the integration variable $y$, 
\bq 
    \vec l^{~\prime 2}_\bot = \frac{2 \bar \Lambda^{~2}_{sat} (W^2)}{3y}  
\label{B11} 
\eq  
 
For photoproduction, from (\ref{B8}), we have  
\bqa 
& & \Delta\sigma^{(m^2_0)}_{\gamma^*_T p} (W^2, Q^2=0) + \Delta\sigma^ 
{(m^2_1}_{\gamma^*_T p} (W^2,Q^2=0) = \label{B12} \\ 
& & = -\frac{\alpha R_{e^+ e^-}}{8} \frac{\sigma^{(\infty)}(W^2)}{\pi}  
\int^1_{\frac{2}{3a}} dy \frac{1 - \frac{1}{2}y}{\sqrt{1-y}} \cdot \nonumber \\ 
& & \cdot \left(  \int^{(\sqrt{\vec l^{~\prime 2}_\bot} + m_0)^2}_ 
   {(\sqrt{\vec l^{~\prime 2}_\bot}-m_0)^2} {{d M^2}\over{M^2}} 
   \left( {{\pi-\vartheta_0(M^2,\vec l^{~\prime 2}_\bot ,m_0^2)}\over \pi}  
         \right.\right. \nonumber \\ 
  &&~~~~~~~~~~~~~~~~~~~~~~~~~~\left. \left.  
     -{{M^2- \vec l^{~\prime 2}_\bot }\over{|M^2- \vec l^{~\prime 2}_\bot |}} 
     \left(1-{2\over \pi} 
          {\rm arctan}\sqrt{Y(M^2 , \vec l^{~\prime 2}_\bot , 0 , m^2_0)}\right) 
     \right) \right. 
   \nonumber \\ 
& & \left.+ \int^{m^2_1}_{(m_1 - \sqrt{\vec l^{~\prime 2}_\bot})^2}{{dM^2}\over{M^2}} 
  \left( {{\vartheta_1(M^2,\vec l^{~\prime 2}_\bot ,m_1^2)}\over \pi} 
       -{2\over \pi} 
          {\rm arctan}\sqrt{Y(M^2 , \vec l^{~\prime 2}_\bot , 0 , m^2_1) } \right) 
   \right) 
\nonumber 
%\bigg|_{\vec l^{~\prime 2}_\bot = \frac{2\Lambda^2_{\rm sat} (W^2)}{3y}}\nonumber 
\eqa

\end{appendix}

\section*{Appendix C. Derivation of (\ref{4.28}), (\ref{4.29}) and (\ref{4.30})}

\renewcommand{\theequation}{\Alph{section}.\arabic{equation}} 
\setcounter{section}{3} 
\setcounter{equation}{0} 

In this Appendix, we derive the approximate expression for 
$\sigma^{dom}_{\gamma^*_{L/T}p}$ in the large $Q^2$ region.
we expand the integrand $I_{L/T}(\vec l_\perp^{~\prime 2}, M^2,Q^2)$
in (\ref{4.19}) and (\ref{4.20}) in terms of
\bq
  \hat x^2={{\vec l_\perp^{~\prime 2}}\over{Q^2+m_0^2}},~~~~~
  \hat y^2={{\vec l_\perp^{~\prime 2}}\over{m_1^2}},~~~~~
  \hat z^2={{m_0^2}\over{\vec l_\perp^{~\prime 2}}},
\label{c.1}
\eq
all of which are small in the limit $Q^2\gg \vec l_\perp^{~\prime 2}\gg 1$.
Each term in the integrand becomes

\bq
   -{{Q^2}\over{M^2+Q^2}}\Big|_{m_0^2}^{m_1^2}
  = {{\hat x^2 }\over{\hat x^2+\hat y^2}} +o(\hat x^2\hat z^2).
\label{c.2}
\eq
\bqa
   &&{{Q^2}\over{\sqrt{\vec l_\perp^{~\prime 2}(\vec l_\perp^{~\prime 2}+4Q^2)}}}
        \ln{{\sqrt{\vec l_\perp^{~\prime 2}
           (\vec l_\perp^{~\prime 2}+4Q^2)}\sqrt X 
          +\vec l_\perp^{~\prime 2}(3Q^2-M^2+\vec l_\perp^{~\prime 2})}\over
         {Q^2+M^2}}\Bigr]\Bigg|_{m_0^2}^{m_1^2}  \nonumber \\ 
  &=& -{{\hat x^2}\over{\hat x^2+\hat y^2}}
       +{{\hat x^4+3\hat x^2\hat y^2+6\hat y^4}\over
          {6(\hat x^2+\hat y^2)^3}}\hat x^4 +\cdots.
\label{c.3}
\eqa
\bq
   {1\over 2}\ln{{M^2+Q^2}\over{\sqrt X+M^2 -\vec l_\perp^{~\prime 2}
         +Q^2}}\Big|_{m_0^2}^{m_1^2}  \nonumber \\
  = {{\hat x^2}\over 2}\Bigl[ {{\hat x^2\hat y^2}
           \over{(\hat x^2+\hat y^2)^2}} +\cdots\Bigr],
\label{c.4}
\eq
\bqa
  && -{{2Q^2+\vec l_\perp^{~\prime 2}}\over
        {2\sqrt{\vec l_\perp^{~\prime 2}(\vec l_\perp^{~\prime 2}+4Q^2)}}}
        \ln{{\sqrt{\vec l_\perp^{~\prime 2}
            (\vec l_\perp^{~\prime 2}+4Q^2)}\sqrt X 
          +\vec l_\perp^{~\prime 2}(3Q^2-M^2+\vec l_\perp^{~\prime 2})}\over
         {Q^2+M^2}}\Bigg|_{m_0^2}^{m_1^2}   \nonumber \\
 &=& {{\hat x^2}\over{\hat x^2+\hat y^2}} 
         +{{2\hat x^4+3\hat x^2\hat y^2-3\hat y^4}\over
             {6(\hat x^2+\hat y^2)^3}}\hat x^4
           +\cdots
\label{c.5}
\eqa
Inserting (\ref{c.2})-(\ref{c.5}) into (\ref{4.18}), we find
\bq
 \sigma_{\gamma^*_Lp}^{dom}(W^2,Q^2)={{\alpha R_{e^+e^-}}\over 3}
     \int d\vec l_\perp^{~\prime 2}\bar\sigma_{(q \bar q)_L^{J=1}p}
     (\vec l_\perp^{~\prime 2},W^2)
      \hat x^2\Bigl[{{\hat x^4+3\hat x^2\hat y^2+6\hat y^4}\over 
        {6(\hat x^2+\hat y^2)^3}}\hat x^2
           +\cdots\Bigr],
\label{c.6}
\eq
and
\bq
 \sigma_{\gamma^*_Tp}^{dom}(W^2,Q^2)={{\alpha R_{e^+e^-}}\over 3}
     \int d\vec l_\perp^{~\prime 2}\bar\sigma_{(q \bar q)_L^{J=1}p}
     (\vec l_\perp^{~\prime 2},W^2)
    \hat x^2\Bigl[{{\hat x^2+3\hat y^2}\over{3(\hat x^2+\hat y^2)^3}}
          \hat x^4 +\cdots \Bigr].
\label{c.7}
\eq
Recalling
\bq
    {{\hat x^2}\over{\hat y^2}} = {\xi\over\eta},
\label{c.8}
\eq
and introducing the integration variable $y$ defined by (\ref{4.21}),
\bq
 \hat x^2={2\over {3\eta y}},
\label{c.9}
\eq
we find 
\bqa
 \sigma_{\gamma^*_Lp}^{dom}(W^2,Q^2)&=&
{{\alpha R_{e^+e^-}}\over{24}}
        \Bigl( {{\sigma^{(\infty)}(W^2)}\over\pi}\Bigr)
        \int^1_{2/(3a)} d y { y\over{\sqrt{1- y}}}   
     \hat x^2G_L({\xi\over{\eta(W^2,Q^2)}}) \nonumber \\
   &=& {{\alpha R_{e^+e^-}}\over{18}}
        \Bigl( {{\sigma^{(\infty)}(W^2)}\over\pi}\Bigr)
         {1\over\eta(W^2,Q^2)}A  
      G_L({\xi\over{\eta(W^2,Q^2)}}), 
\label{c.10} 
\eqa
and
\bqa
 \sigma_{\gamma^*_Tp}^{dom}(W^2,Q^2)
    &=&{{\alpha R_{e^+e^-}}\over {12}}
        \Bigl( {{\sigma^{(\infty)}(W^2)}\over\pi}\Bigr)  
        \int^1_{2/(3a)} d\hat y {{1-\hat y/2}\over{\sqrt{1-\hat y}}}
        \hat x^2 G_T({\xi\over{\eta(W^2,Q^2)}})  \nonumber \\
    &=&{{\alpha R_{e^+e^-}}\over {18}}
        \Bigl( {{\sigma^{(\infty)}(W^2)}\over\pi}\Bigr)  
        {1\over \eta(W^2,Q^2)} 
        \nonumber \\
     &&~~~~~~~\times \Bigl[\log{{1+A}\over{1-A}}-A\bigr]
        G_T({\xi\over{\eta(W^2,Q^2)}}).
\label{c.11}
\eqa
Here
\bq
   A\equiv \sqrt{1-{2\over{3a}}} =0.951 ~~~~{\rm for}~~a=7
\label{c.12}
\eq
and the functions $G_L({\xi\over{\eta(W^2,Q^2)}})$ and
$G_T({\xi\over{\eta(W^2,Q^2)}})$ are defined by (\ref{4.29}) 
and (\ref{4.30}).
Noting  that $A\sim 1$ and
\bq
 \ln{{1+A}\over{1-A}}-A = 2A\rho(\epsilon={1\over{6a}})
    \sim 2\rho(\epsilon={1\over{6a}})
\label{c.13}
\eq
we reach the approximate expression for the dominant parts
given in (\ref{4.28}).

\end{document}